\documentclass[a4,11pt]{article}
\usepackage{jheppub1}
\usepackage{amsmath,bm}
\usepackage{amssymb}
\usepackage{slashed}
\usepackage{graphicx}
\usepackage{color}
\usepackage[autostyle]{csquotes}
\usepackage{lscape}
\usepackage{subfig}
\usepackage{setspace}
\usepackage{bbm}
\usepackage{breqn}
\usepackage{wrapfig}
\usepackage{hyperref}
\usepackage{mathtools}
\usepackage{physics}
\usepackage{float}
\usepackage{simpler-wick}

\newcommand{\be}{\begin{equation}}
\newcommand{\ee}{\end{equation}}
\newcommand{\bea}{\begin{eqnarray}}
\newcommand{\eea}{\end{eqnarray}}
\newcommand{\beas}{\begin{eqnarray*}}
\newcommand{\eeas}{\end{eqnarray*}}

\begin{document}

\title{Reconstructing cosmological correlators via dispersion: from cutting to dressing rules}

\author[\spadesuit]{Shibam Das,}
\author[\clubsuit]{Debanjan Karan,}
\author[\clubsuit]{Babli Khatun,}
\author[\spadesuit]{ and Nilay Kundu}
\affiliation[\spadesuit]{Department of Physics \\ Indian Institute of Technology Kanpur, Kalyanpur, Kanpur 208016, India}
\affiliation[\clubsuit]{International Centre for Theoretical Sciences (ICTS-TIFR) \\ Tata Institute of Fundamental Research,
Shivakote, Hesaraghatta, Bangalore 560089, India.}

\emailAdd{shibamdas23@iitk.ac.in}
\emailAdd{debanjan.karan@icts.res.in}
\emailAdd{babli.khatun@icts.res.in}
\emailAdd{nilayhep@iitk.ac.in}

\abstract{In this work, we investigate how cosmological correlators can be reconstructed by applying the momentum-space dispersion formula to their discontinuities, treating them as functions of momentum variables associated with the corresponding de Sitter Witten diagrams. We focus on conformally coupled and massless polynomial scalar interactions (both IR-divergent and IR-convergent), and consider tree-level de Sitter Witten diagrams. We explicitly utilize the single-cut discontinuity relations, or cutting rules, involving the cosmological correlators recently constructed in \cite{Das:2025qsh} (arXiv:2512.20720). For diagrams with multiple interaction vertices, we apply the dispersion formula by cutting all internal lines in the diagram one by one, successively, thereby allowing us to reconstruct the full correlator using only lower-point contact-level objects and their discontinuity data, up to contact diagram ambiguities. We also rediscover how the cosmological correlators on the late-time slice of de Sitter space can be obtained from flat-space Feynman diagrams via a set of dressing rules. Our starting point, being the cutting rules for the cosmological correlators, also emphasizes how basic principles, such as unitarity for in-in correlators, can lead us to the dressing rules, which were previously derived in literature following a different method.
}
%\keywords{XYZ}

\maketitle
%%%%%%%%%%%%%%%%%%%%%%%%%%%%%%%%%
\section{Introduction} \label{intro}

In recent years, studying cosmological correlators has attracted widespread interest, as they can reveal theoretical insights into the primordial universe during inflation \cite{Maldacena:2002vr, Weinberg:2005vy, Chen:2010xka, Maldacena:2011nz, Wang:2013zva, Wang:2024abc, Pimentel:2025rds}. The cosmological correlators are defined as the equal-time correlation functions of quantum fields evaluated on the late-time slice of de Sitter space-time \cite{Spradlin:2001pw, Akhmedov:2013vka, Galante:2023uyf}. Traditional approaches to computing cosmological correlators include the Schwinger-Keldysh (also known as the in-in) formalism \cite{Schwinger:1960qe, Keldysh:1964ud, Weinberg:2005vy} and the method based on the wave function of the quantum fields (including the graviton) at the late-time de Sitter slice \cite{Maldacena:2002vr, Anninos_2015}. 

Given an interacting QFT in de Sitter, these in-in or wave-function-based methods are, in principle, executable. However, in practice, beyond specific, simple lower-point examples, the calculations become technically challenging. The cosmological bootstrap program, arguably a more conceptually elegant formalism, has been developed, revealing many insightful features of the analytic structure of cosmological correlators \cite{Arkani-Hamed:2018kmz, Baumann:2019oyu, Baumann:2020dch, Hogervorst:2021uvp, Baumann:2022jpr, Loparco:2023rug, SalehiVaziri:2024joi}. This method uses fundamental principles such as unitarity, analyticity \cite{Jazayeri:2021fvk, DiPietro:2021sjt, Salcedo:2022aal, Albayrak:2023hie, Goodhew:2024eup, Thavanesan:2025kyc, Thavanesan:2025wvb}, and conformal symmetries \cite{Creminelli:2011mw, Mata:2012bx, Ghosh:2014kba, Kundu:2015xta}, which follow from the isometries of de Sitter space-time \cite{Sun:2021thf} as consistency conditions, thereby imposing constraints on the mathematical structure of cosmological observables \cite{Chowdhury:2023arc, Arkani_Hamed_2025, Arkani-Hamed:2025mce, Figueiredo:2025daa, Glew:2026von} without relying on any specific underlying theory. 

The first-principles-based bootstrap methods in cosmology have borrowed the basic intuition from related developments, primarily led by the unitarity and on-shell approaches, in the study of flat-space scattering amplitudes \cite{Elvang:2013cua, Benincasa:2013faa, Henn:2014yza,cheung2017tasilecturesscatteringamplitudes}. As a result, by applying principles used to understand amplitudes in flat-space quantum field theory \cite{Weinberg:1995mt, Peskin:1995ev, Schwartz:2014sze}, one can now reimagine late-time de Sitter correlation functions beyond performing tedious time-ordered integrals in expanding space-time. In subsequent studies, this profound connection between cosmology and modern scattering amplitude techniques has been further solidified. The optical theorem \cite{Goodhew:2020hob, Goodhew:2023bcu} and the cutting rules \cite{Melville:2021lst, Goodhew:2021oqg, Chen:2025foq} in the cosmological setup provide such evidence. The cosmological optical theorem demonstrated that unitarity can be used to write down relations among wave-function coefficients in de Sitter space-time. This framework was further extended to constrain the discontinuity of wave-function coefficients, thereby establishing the cosmological cutting rules. 

One aspect which is particularly relevant to the analysis in our paper was discussed in \cite{Meltzer:2021zin}: obtaining the wave-function coefficients from the knowledge of their analytic properties when treated as functions of the complexified momenta. More specifically, it was argued that the wave-function coefficients can be reconstructed, up to contact term ambiguities, from their discontinuity by using a momentum space dispersion formula \cite{Meltzer:2021bmb}. It is worth highlighting that in \cite{Meltzer:2021zin}, the dispersion integration was primarily operating on the discontinuities of the wave-function coefficients in the complex momentum space. Once we have reconstructed the wave-function coefficients, one can, in principle, compute the late-time correlators by performing a further path integral over the late-time field variables, with the field operators inserted and weighted by the wave-function norm squared. In situations where one is interested in correlator diagrams with a large number of external legs, multiple interaction vertices, and possibly loop diagrams, computing the correlator from the wave functions via the additional path integration mentioned above becomes practically intractable. Therefore, it would be advantageous if one could reconstruct the correlator itself through the dispersion formula, without relying on the wave-function coefficients. To that end, one first needs to derive the discontinuity relations directly in terms of the cosmological correlators, which were not well formulated until recently \cite{Stefanyszyn:2024msm, Liu:2024xyi, Colipi-Marchant:2025oin, Donath:2024utn, Ema:2024hkj, Chowdhury:2026upp}.

As we mentioned above, in earlier literature on the cosmological optical theorem or cutting rules \cite{Goodhew:2020hob, Goodhew:2021oqg}, the discontinuity relations were studied for wave-function coefficients. Recently, in \cite{Das:2025qsh, Colipi-Marchant:2025oin}, working with conformally coupled and massless scalar fields with polynomial self-interactions in de Sitter, it was shown that the discontinuity due to a single cut of one internal line in the tree-level exchange diagram of a cosmological correlator can be written as a sum of products of discontinuities of lower-point data. Interestingly, in \cite{Das:2025qsh}, it was found that the lower-point discontinuity data include objects, involving the imaginary piece of the wave-function coefficients, that are not expressible as the lower-point correlator and were named as auxiliary counterparts of the correlator.

In this paper, following the idea presented in \cite{Meltzer:2021zin}, we perform dispersion integration but directly on the discontinuities of the cosmological correlators derived in \cite{Das:2025qsh}. We argue that the cosmological correlators can be reconstructed from these combined operations of dispersion and discontinuity, up to contact-term ambiguities. The resulting expression for the correlators is, of course, consistent with the expressions obtained from the standard in-in or the wave-function picture. To apply our strategy to cosmological correlators for tree-level chain diagrams with an arbitrarily large number of interaction vertices (or sites), we need to use the single-cut discontinuity, developed in \cite{Das:2025qsh}, successively to all internal lines in the diagram. As a result, the successive discontinuity of an $r$-site cosmological correlator will be decomposed into the discontinuities of lowest-point data involving a single site. Consequently, in the next step, when we apply the dispersion formula to this successive discontinuity relation, we establish that the cosmological correlator is reconstructable from the simplest elementary object available, i.e., the $1$-site objects. 

Note that at a fundamental level, the cosmological correlators are equal-time correlation functions and, hence, are inherently different from the S-matrix. Nevertheless, as mentioned above, the cosmological bootstrap program, or developments along the lines of formulating cosmological cutting rules in terms of wave-function coefficients, has successfully revealed profound connections with similar advancements in flat-space amplitudes. These intriguing analogies inspired further studies investigating more fine-grained aspects of their connections. For example, in \cite{Donath:2024utn}, this connection was explored using a new methodology, the in-out formalism.

An even more striking correspondence between cosmological correlators and flat space Feynman diagrams was revealed in \cite{Chowdhury:2023arc, Chowdhury:2025ohm}. It was shown that late-time correlators (both at the tree and loop levels) in an interacting scalar field theory in de Sitter can be obtained by implementing specific dressing rules directly to the corresponding flat-space Feynman diagrams. Their analysis is based on the shadow formalism \cite{DiPietro:2021sjt}, an alternative approach to computing late-time de Sitter correlators from an effective action in Euclidean AdS \cite{Sleight:2020obc, Sleight_2021, abhishek2025cosmologicalcorrelatorsgaugetheory, Ansari:2025fvi} obtained by introducing an extra set of auxiliary shadow fields. They also justified the consistency of these rules by comparing them with results from the conventional in-in method or wave-function methods. 

One of the primary motivations of our present work is to provide an alternative perspective and understanding of these seemingly miraculous workings of the dressing rules proposed in \cite{Chowdhury:2025ohm}. Although, as an algorithm, the dressing rules apply to a broader class of scalar field theories in de Sitter, the details, such as the explicit form of the auxiliary propagators, are theory-dependent. In other words, for each given distinct scalar interaction, one needs to go through the construction of the shadow formalism to obtain the dressing factors. As we will demonstrate, our re-derivation of the dressing rules will bypass the shadow formalism entirely. Moreover, we will propose a generalized version of the dressing rules proposed in \cite{Chowdhury:2025ohm}. Our generalized dressing rules will apply to all scalar theories with $\phi^n$ self-interactions, as the details of each interaction are built into the rules themselves. Using them, one can write the late-time de Sitter correlator involving scalar operators directly as an integral relation involving the dressing factors acting on the Feynman propagator. However, since our method reconstructs the cosmological correlator by integrating over successive discontinuities, we cannot capture the contact pieces in the final correlator. 

The rest of the paper is organized as follows: in \S\ref{review} we discuss the background material and explain the notation and conventions to set up the calculations in the following sections. Next, in \S\ref{Disp+Disc_reconstruction} we apply the dispersion integration to the discontinuity of the cosmological correlator and derive integral representations of a $2$-site and $3$-site correlator. In \S\ref{sec:rsite} we formulate the reconstruction of an $r$-site tree-level cosmological correlator via dispersion integration on the successive single cut discontinuity of the $r$-site correlator. In the following \S\ref{GeneralDressingRules}, we present a generalized set of diagrammatic dressing rules that can be applied to Feynman diagrams to evaluate cosmological correlators. We check the consistency of our generalised dressing rules in several examples presented in \S\ref{checks}. Finally, we end in \S\ref{conclusions} with a discussion of our results. Various technical details of the calculations performed in the main text are presented in the Appendices. 

\textbf{Note added:} After our work was completed, \cite{Chowdhury:2026upp} appeared, with some overlap with our results.
%%%%%%%%%%%%%%%%%%%%%%%%%%%%%%%%%%%%%%%%%%%%%%%%%%%%%%%%%%%%%%%%%%%%%%%%%%%%%%%%%%%%
%%%%%%%%%%%%%%%%%%%%%%%%%%%%%%%%%%%%%%%%%%%%%%%%%%%%%%%%%%%%%%%%%
\section{Review of the background material}
\label{review}
%%%%%%%%%%%%%%%%%%%%%%%%%%%%%%%%%%%%%%%%%%%%%%%%%%%%%%%%%%%%%%%%%%%%%%%%%%%%%%%%%%%%
In this section, we intend to describe the basic setup and review the background material that will be used in the subsequent sections. We start by noting that cosmological correlators are computed as the equal-time correlation functions of quantum fields inserted at the late-time slice of de Sitter space-time. Next, we introduce the single-cut discontinuity of the cosmological correlators. Finally, we define the dispersion formula, which can be used to reconstruct the correlator given its discontinuities, up to contact-term ambiguities. \\

\noindent \textbf{Cosmological correlators as late-time de Sitter correlators:} 
We will work in $4$-dimensional de Sitter spacetime, expressed in the expanding Poincaré patch coordinates 
\begin{equation}
ds^2 = \frac{ - d\eta^2  +  d \vec{x}^2 }{ H^2 \eta^2} ,
\end{equation}
where $\eta$ denotes the conformal time coordinate, $\eta \in (-\infty, 0)$. The limit $\eta = -\infty$ corresponds to the far past, and $\eta = 0$ represents the late-time de Sitter boundary. The cosmological correlators are evaluated on a fixed late-time slice $\eta = \eta_0$, with the understanding that the late-time limit is obtained by taking $\eta_0 \to 0^{-}$. The parameter $H$ denotes the Hubble constant, which is taken to be fixed since we work in rigid de Sitter spacetime.

The action for a scalar field with mass $m$ and $\lambda\phi^n$ interactions in the expanding Poincaré patch of de Sitter spacetime
\footnote{For a detailed discussion of this setup, see for example Refs.~\cite{Akhmedov:2013vka, Spradlin:2001pw, Galante:2023uyf}.}
is given by
\begin{equation}
S = - \frac{1}{ 2 } \int d^{4} x \, \sqrt{-g } \left(  ( \partial \phi)^2 + m^2 \phi^2 + \frac{\lambda}{n!} \phi^n\right) \, .
\end{equation}
In this paper, we focus on correlators involving conformally coupled and massless scalar fields. In rigid $dS_4$, the classical equations of motion for conformally coupled and massless scalar fields admit two independent solutions of the following form \cite{Goodhew:2020hob, Goodhew:2021oqg, Das:2025qsh}
\begin{equation} \label{mode functions}
\begin{split}
    & \text{conformally coupled scalar:} \quad f_k^-( \eta ) = \frac{ i H \eta}{\sqrt{2k}} e^{-ik\eta},~~ f_k^+(\eta) = \frac{-i H\eta}{\sqrt{ 2k }} e^{ik\eta} \, ,
    \\
    & \text{massless scalar:} \quad f_k^-( \eta ) = \frac{H}{\sqrt{2k^3}} (1+i k\eta) e^{-ik\eta}, ~~f_k^+( \eta) = \frac{H}{\sqrt{2k^3}} (1-i k\eta) e^{ik\eta } \, .
\end{split}
\end{equation}
 The canonical technique for calculating late-time de Sitter correlators is known as the in-in (also called Schwinger-Keldysh) formalism \cite{Schwinger:1960qe, Keldysh:1964ud}, which can be summarized as the following formula  
\begin{equation} \label{inincorrdef}
\begin{split}
    \langle \Omega \vert \mathcal{ O }(\eta_0) \vert \Omega \rangle &= \langle 0 \vert\,  \bar{T}  \left[ \text{exp} \left( i \int_{- \infty_+}^{\eta_0} d\eta' H_{ \text{int}}(\eta') \right)\right] \, \mathcal{O}(\eta_0 )  \, T  \left[ \text{exp} \left( - i \int_{-\infty_-}^{\eta_0} d\eta' H_{\text{int}}(\eta' ) \right)\right] \vert 0 \rangle \, ,
\end{split}
\end{equation}
where $\pm \infty_{\pm} \equiv \pm \infty \left(1 \pm i\epsilon \right)$. In eq.\eqref{inincorrdef} all the operators are in the interaction picture and $H_{\text{int}}$ denotes the interacting Hamiltonian in the interaction picture. Here, $T$ and $\bar{T}$ represent the time-ordering and anti-time-ordering, respectively. Also, the particular $ \pm i \epsilon$ principle ensures that the state at very far past is projected onto the so-called Bunch-Davies vacuum denoted by $\vert 0 \rangle$. 

Another approach to compute late-time de Sitter correlators is through the wave function of the universe, denoted by $\Psi$. Formally, the wave function is defined as a path integral over field configurations $\Phi$, subject to the initial condition in the distant past ($\eta \to -\infty$), which we take to be the Bunch-Davies initial condition. This defines the wave function as a functional of the fields at the late time slice, $\phi = \Phi(\eta = \eta_0)$. 
In a perturbative expansion in the interaction, the semi-classical wave-function can be formally expressed as 
\begin{equation} \label{Psi-wvfncoeff}
\begin{split}
    \Psi[\phi] = \exp\bigg (&-\frac{1}{2} \int\frac{ d^3 \vec{k}_1 d^3 \vec{k}_2 }{(2\pi)^6} \, \Psi_2(\vec{k}_1,\vec{k }_2) \phi(\vec{k}_1) \phi(\vec{k}_2) 
    \\
    &  +\sum_{n=3}^{\infty} \frac{1}{n!} \int\frac{ d^3 \vec{k}_1 \, d^3 \vec{k}_2 \, \cdots d^3 \vec{k}_n}{(2\pi)^{3n}} \, \Psi_n(\vec{k}_1,\vec{ k}_2,\cdots,\vec{k}_n) \, \phi(\vec{k}_1) \, \phi(\vec{k}_2)\cdots \phi(\vec{k}_n) \bigg) \, ,
\end{split}    
\end{equation}
where $\Psi_n(\vec{k}_1,\vec{k}_2,\cdots,\vec{k}_n)$ are known as the wave-function coefficients. Consequently, the late-time de Sitter correlators are defined using another path-integration as follows
\begin{equation}\label{correlator from wavefunction def}
\begin{split}
    \big{\langle} \phi_1 \phi_2 \cdots \phi_n \big{\rangle}=\int \mathcal{D} \phi~ \phi_1 \phi_2 \cdots \phi_n \, \Big{\vert}  \Psi[\phi]  \Big{\vert}^2 \, ,
\end{split}    
\end{equation}
where we have used the following short-hand notation $\phi_i = \phi(\vec{k}_i)$ for $i=1,\, 2, \cdots , n$. 

In this paper, we will always define a correlator by stripping off the three-momentum-conserving delta function. For example, any $n$-point correlator will be defined as follows 
\begin{equation}
    \big{\langle} \phi(\vec{k}_1) \, \phi(\vec{k}_2)  \cdots \phi(\vec{k}_n) \big{\rangle} = (2\pi)^3 \, \delta^3 (\vec{k}_1 + \vec{k}_2+ \cdots + \vec{k}_n) \, \mathcal{B}(\lbrace k_i \rbrace; \lbrace p_i \rbrace) \, .
\end{equation}
In $\mathcal{B}(\lbrace k_i \rbrace; \lbrace p_i \rbrace)$, the argument $\lbrace k_i \rbrace$ corresponds to the magnitude of all the external momenta, and $\lbrace p_i \rbrace$ signifies the magnitude of the internal momenta in the associated Witten diagram for the corresponding correlator. Following the convention of \cite{Das:2025qsh}, we will use a notation $\mathcal{B}^{(2)}\left( \lbrace \mathbf{k}_L, \mathbf{k}_R \rbrace ;p \right)$ where the superscript `$^{(2)}$' will be used to denote that we are looking at a $2$-site correlator with $2$ interaction vertices, with $\lbrace \mathbf{k}_L, \mathbf{k}_R \rbrace$ signifying the collection of the external momenta at the left and right vertices respectively, and $p$ denoting the modulus of the exchange momentum. Similarly, we will use $\mathcal{B}^{(r)}\left( \lbrace \mathbf{k}_L,\mathbf{k}_{M_1},...,\mathbf{k}_{M_{r-2}}, \mathbf{k}_R \rbrace ; \lbrace p_1,..,p_{r-1} \rbrace \right)$ to denote a correlator with $r$-sites or vertices, where $\lbrace \mathbf{k}_{M_1},...,\mathbf{k}_{M_{r-2}} \rbrace$ will now denote the collection of external momenta at all the intermediate sites, which are $(r-2)$ in numbers in a $r$-site correlator $\mathcal{B}^{(r)}$, and $\lbrace p_1,..,p_{r-1} \rbrace$ will denote the set of magnitude of the momenta running in the internal lines, which are $(r-1)$ in numbers.\\

\noindent \textbf{Discontinuity relation for the cosmological correlators:} 
In \cite{Das:2025qsh}, the single-cut discontinuity relation for the $2$-site correlator, $\mathcal{B}^{(2)}( \lbrace \mathbf{k}_L, \mathbf{k}_R \rbrace ;p)$ was expressed in terms of the lower-point data $\mathcal{B}^{(1)}$ and $\widetilde{\mathcal{B}}^{(1)}$ as follows
\begin{equation}\label{2siteDiscFinal Result}
\begin{split}
     { \text {Disc}}_{p } \mathcal{B}^{(2)}( \lbrace \mathbf{k}_L, \mathbf{k}_R \rbrace ;p) = \frac{1}{2 P_p(\eta_0)} \Bigg(& \text{Disc}_p \mathcal{ B}^{(1)}( \lbrace \mathbf{k}_{L} \rbrace ,p) \text{Disc}_p \mathcal{B}^{(1)}( \lbrace \mathbf{k}_{R} \rbrace ,p ) 
     \\
     & - \widetilde{\text{Disc}}_p  \widetilde{\mathcal{B}}^{(1)}( \lbrace \mathbf{ k}_{L} \rbrace ,p) \widetilde{\text{Disc}}_p  \widetilde{\mathcal{B }}^{(1)}( \lbrace \mathbf{k}_{R} \rbrace ,p) \Bigg) \, ,
\end{split}
\end{equation}
where $\mathcal{B}^{(1)}(\lbrace \mathbf{k}_{R} \rbrace ,p)$ is a one-site or contact correlator and $\widetilde{\mathcal{B}}^{(1)}( \lbrace \mathbf{k}_{L} \rbrace ,q) $ is an auxiliary counterpart of a contact correlator, and they are defined in terms of the one-site wave-function coefficients  as follows
\begin{equation} \label{defB1Btilde1}
\begin{split}
&\mathcal{B}^{(1)}(\lbrace \mathbf{k}_{L} \rbrace,q)
= \frac{2 ~\mathbb{R}e ~\psi^{ (1)} (\lbrace \mathbf{k}_L\rbrace,q )}{\mathcal{P}^{-1}_{\mathbf{ k}_L}(\eta_0) P^{-1}_{q}(\eta_0)} \,, \qquad 
 \widetilde{\mathcal{B} }^{(1)}( \lbrace \mathbf{k}_{ L} \rbrace,q) = \frac{2i ~\mathbb{I}m ~\psi^{(1)} (\lbrace \mathbf{k}_L\rbrace,q )}{\mathcal{P}^{-1}_{\mathbf{k}_L}(\eta_0) P^{-1}_{q}(\eta_0)} .
\end{split}
\end{equation}
Following \cite{Das:2025qsh}, the discontinuity operations, $\text{Disc}_{p} $, $\widetilde{\text{Disc}}_{p}$ appearing in eq.\eqref{2siteDiscFinal Result} can be defined as follows
\begin{equation} \label{defdisc}
\begin{split}
    & \text{Disc}_{p} f(p,k_1,...,k_n) = f(p,k_1,...,k_n) - f(-p,k_1,...,k_n) \, ,\\&
    \widetilde{\text{Disc}}_{p} f(p,k_1,...,k_n) = f(p,k_1,...,k_n) + f(-p,k_1,...,k_n) \, .
\end{split}
\end{equation}
In \cite{Das:2025qsh}, it was also shown that the single-cut rule for the tree-level $r$-site correlator, obtained by cutting the right-most internal line with momentum $p_{r-1}$, is as follows
\begin{equation}
\label{single cut r-site correlator}
\begin{split}
    & {\text {Disc}}_{p_{r-1}} \mathcal{B}^{(r)}( \lbrace \mathbf{k}_L,..., \mathbf{k}_R \rbrace ; \lbrace p_1, ..., p_{r-1} \rbrace ) = \frac{1}{2 P_{p_{r-1}} (\eta_0)} 
    \\
    & ~\bigg[ \text{Disc}_{p_{r-1}} \mathcal{B}^{(r-1)}\left(  \lbrace \mathbf{k}_L,...,\mathbf{k}_{M_{r-2}} \rbrace, p_{r-1} ; \lbrace p_1,..,p_{r-2} \rbrace \right) \text{Disc}_{p_{r-1}} \mathcal{B}^{(1)}( \lbrace \mathbf{k}_{R} \rbrace, p_{r-1})\\
    & ~~ - \widetilde{\text{Disc}}_{p_{r-1}} \widetilde{\mathcal{B}}^{(r-1)}\left(  \lbrace \mathbf{k}_L,...,\mathbf{k}_{M_{r-2}} \rbrace, p_{r-1} ; \lbrace p_1,..,p_{r-2} \rbrace \right) \widetilde{\text{Disc}}_{p_{r-1}} \widetilde{\mathcal{B}}^{(1)}( \lbrace \mathbf{k}_{R} \rbrace, p_{r-1}) \bigg] \, .
\end{split}
\end{equation}
Another very useful thing that we will use heavily in our paper is the successive cuts of all the internal lines in an $r$-site correlator $\mathcal{B}^{(r)}$, which can be defined as follows 
\begin{equation}
\text{Successive single cuts} = \text{Disc}_{p_{1}}\bigg[\text{Disc}_{p_{2}}\bigg[\ldots\text{Disc}_{p_{r-1}}\,\mathcal{B}^{(r)}( \lbrace \mathbf{k}_L,..., \mathbf{k}_R \rbrace ; \lbrace p_1, ..., p_{r-1} \rbrace )\bigg]\bigg] \, ,
\end{equation}
which is different from cutting all the internal lines simultaneously, leading to 
\begin{equation}
\text{Simultaneous multiple cuts} = \text{Disc}_{p_{1}, \, p_{2}, \cdots ,\,  p_{r-1}}\bigg[ \mathcal{B}^{(r)}( \lbrace \mathbf{k}_L,..., \mathbf{k}_R \rbrace ; \lbrace p_1, ..., p_{r-1} \rbrace )\bigg] \, .
\end{equation}

\noindent \textbf{Dispersive integral technique of reconstruction from discontinuity:} 
The basic idea behind the dispersive integral technique is to reconstruct a function (up to contact terms or regular pieces) from its discontinuity along a branch-cut. Let $f(z)$ be analytic on the complex plane, with its only non-analyticity arising from a branch cut along the positive real $z$-axis. Assuming the following fall-off condition $ \lim_{|z|\rightarrow \infty}\,f(z) = 0$, we can write $f(z_0)$, for a real and positive $z_0$, as 
\begin{equation}
       f(z_0) = \lim_{\epsilon\rightarrow0^+}\,f(z_0+i\epsilon) = \frac{1}{2\pi i}\int_{0}^{\infty} \frac{dz}{z-z_0-i\epsilon}\,\textbf{Disc}_{z}f(z)\,,
\end{equation}
where, where, $z_0$ is any point in complex $z$ plane and $C_{z_0}$ is the contour encircling $z =z_0$ anti-clockwise, and $\textbf{Disc}_{z}f(z)$ is the discontinuity of function $f(z)$ along the branch cut defined as \cite{Britto_2025, Baumann:2024ttn} 
\begin{equation}
    \textbf{Disc}_{z}f(z) = \lim_{\epsilon\rightarrow 0^+}\,\bigg[f(z+i\epsilon)-f(z-i\epsilon)\bigg]\,.
\end{equation}
A discussion on this is presented in Appendix~\ref{App_disc_disp}.

Using the general description given above, we can similarly apply the dispersion integral on the discontinuity of a $2$-site correlator and reconstruct it as follows 
\begin{equation}\label{disp+disc_relation}
 \begin{split}
 \mathcal{B}^{(2)}( \lbrace \mathbf{k}_L, \mathbf{k}_R \rbrace ;p)=&   \frac{1}{2\pi i}\int_{-\infty}^{+\infty} \frac{q \, dq}{q^2 - p^2} \,    {\text {Disc}}_{q} \mathcal{B}^{(2)}( \lbrace \mathbf{k}_L, \mathbf{k}_R \rbrace ;q) \, .
 \end{split}   
\end{equation}
In the presence of poles of the correlator $\mathcal{B}^{(2)}$, the relation becomes subtle because all physical poles lie on the integration contour, which makes the choice of poles important. We shift the poles of $\mathcal{B}^{(2)}(\{\mathbf{k}_L,\mathbf{k}_R\};q)$ to the lower half of the complex $q$-plane and the poles of $\mathcal{B}^{(2)}(\{\mathbf{k}_L,\mathbf{k}_R\};-q)$ to the upper half-plane. The remaining poles of the integrand are located at $q=p+i\epsilon$ and $q=p-i\epsilon$. We choose to close the contour in the upper half-plane in an anticlockwise direction. With this pole-prescription, and using the fact that $\text{Disc}_q\,\mathcal{B}^{(2)}$ is antisymmetric under $q\to -q$, the integral in eq.\eqref{disp+disc_relation} reduces to a sum over residues. This approach is simpler than performing contour integrals in the presence of branch cuts. The simplification follows from our definition of the discontinuity $\text{Disc}_q$ for cosmological correlators, which coincides with the standard discontinuity $\textbf{Disc}_{q^2}$, as shown in Appendix~\ref{App_disc_disp}. The idea of dispersive integration on discontinuities to reconstruct the wave-function coefficients was previously used in \cite{Meltzer:2021zin}, where a different discontinuity operation (with respect to $k^2$) compared to ours in eq.\eqref{defdisc} was used. Our discussion in Appendix \ref{App_disc_disp} explains why both definitions yield the same answers. 

Similarly, an $r$-site correlator
$\mathcal{B}^{(r)}(\mathbf{k}_L,\ldots,\mathbf{k}_R\};\{p_1,\ldots,p_{r-1}\})$ can be reconstructed by successively applying the $\text{Disc}$ operation to all internal lines, followed by iterative dispersion relations
\begin{equation}
\label{r site disc+disp ansatz substituted}
\begin{split}
    \mathcal{B}^{(r)}( \lbrace \mathbf{k}_L,..., \mathbf{k}_R \rbrace ;& \lbrace p_1, ..., p_{r-1} \rbrace ) = \bigg(\frac{1}{2\pi i}\bigg) ^{(r-1)}\left(\prod_{i = 1}^{r-1}\int\frac{q_{i}\,dq_{i}}{q_{i}^2-p_{i}^2} \right) \times \\
    & \text{Disc}_{q_{1}}\bigg[\text{Disc}_{q_{2}}\bigg[\ldots\text{Disc}_{q_{r-1}}\,\mathcal{B}^{(r)}( \lbrace \mathbf{k}_L,..., \mathbf{k}_R \rbrace ; \lbrace q_1, ..., q_{r-1} \rbrace )\bigg]\bigg] \, .
\end{split}
\end{equation}
In general, this is not true with $\textbf{Disc}$ of a multivariable function with complex branch-cut structures. In our case, however, the simplification is possible because the $\text{Disc}$ operator is defined as in eq.\eqref{defdisc}. As an explicit example, in \S~\ref{check 3-site phi5}, we consider the $3$-site $\phi^5$ cosmological correlator. We reconstructed the full correlator by performing iterative dispersion integrals over its successive cut discontinuities, after choosing the appropriate poles as discussed above, and the final result matched the expected one from the standard in-in calculation.

It should be noted that the dispersive integration technique cannot reconstruct the contact pieces in the correlator, since the ${\text {Disc}}_{q}$ operation is insensitive to any analytic pieces in the correlator. Also, while applying the dispersive integration to discontinuities of cosmological correlators, it is important that the latter falls off sufficiently fast as $p \to \infty$, so that the deformation of the contour to pick up the discontinuity across the real line is justified. From the fall-off behavior of the bulk-to-bulk propagator with Bunch-Davies initial conditions, it can be checked that the correlator falls off sufficiently rapidly as the magnitude of the exchanged momenta increases, i.e., $|k^2|\to \infty$. \\

\noindent \textbf{Cosmological dressing rule:} In \cite{Chowdhury:2025ohm, Chowdhury:2025nnk}, it was shown that the late-time de Sitter correlator (involving scalar field operators) can be obtained from the corresponding flat space Feynman diagram by dressing the latter with certain auxiliary propagators. This result was established using the shadow formalism \cite{DiPietro:2021sjt}, an alternative method for computing the de Sitter correlator. The idea is to derive these correlators in Euclidean anti-de Sitter space (EAdS) \cite{ Sleight:2020obc, Sleight:2021plv} and map them back to de Sitter via analytic continuation. For a single scalar field with polynomial interactions in de Sitter spacetime, the Wick-rotated Feynman rules are generated by the EAdS shadow effective action containing two fields, $\phi^+$ and $\phi^-$. The dressing rules were derived from that shadow effective action and the spectral representation of EAdS Green's function \cite{Liu:1998ty, Raju:2012zr, Chowdhury:2025ohm}. Furthermore, it was shown that the late-time de Sitter correlator can be obtained from the corresponding flat space Feynman diagram by dressing it with certain auxiliary propagators \cite{Chowdhury:2025ohm, Chowdhury:2025nnk}. 
It was argued that there can be two types of auxiliary propagators: dashed and dotted propagators. The expressions for these propagators are theory dependent, and their exact forms for some specific theories are written in Table-$3$ of \cite{Chowdhury:2025ohm}.

For example, the $4$-point (or $2$-site) correlator involving the $\phi^3$ interaction of conformally coupled scalars can be expressed as follows \cite{Chowdhury:2025ohm}
\begin{equation}
    \mathcal{B}^{(2)}( \lbrace \mathbf{k}_L, \mathbf{k}_R \rbrace ;y_{12}) =  \mathcal{A}_4^{(1)}+\mathcal{A}_4^{(2)}\,,
\end{equation}
where $\mathcal{A}_4^{(1)}$ denotes the dressed Feynman diagram with two dashed auxiliary propagator dressings, and $\mathcal{A}_4^{(2)}$ denotes the dressed Feynman diagram with two dotted auxiliary propagator dressings (see Figure-$4$ of \cite{Chowdhury:2025ohm}). The explicit expressions of $\mathcal{A}_4^{(1)}, \mathcal{A}_4^{(2)}$ are 
\begin{equation}
\label{dressingphi32sitecor}
\begin{split}
&\mathcal{A}_4^{(1)}
=
\lambda^2
\int_{-\infty}^{\infty} dp
\left(
-2i \int_0^{\infty} ds_1 \,
\frac{p}{p^2 + (s_1 + k_{12})^2}
\right)
\left(
-2i \int_0^{\infty} ds_2 \,
\frac{p}{p^2 + (s_2 + k_{34})^2}
\right) \times \\
& \qquad \qquad  \qquad \qquad 
\left(
\frac{1}{\pi} \frac{1}{p^2 + y_{12}^2}
\right)\,,
\end{split}
\end{equation}
and 
\begin{equation}
\label{dressingphi32sitecor1}
\begin{split}
\mathcal{A}_4^{(2)}
=
\lambda^2 \pi
\int_{-\infty}^{\infty} dp \,
\frac{1}{p^2 + y_{12}^2},
\end{split}
\end{equation}
with $k_{ij} = |\vec{k}_i|+|\vec{k}_j|\,,\, y_{ij} = |\vec{k}_i+\vec{k}_j|$.

From eq.\eqref{dressingphi32sitecor}, eq.\eqref{dressingphi32sitecor1}, one can identify the Euclidean flat-space massless propagator carrying four-momentum $P^{\mu} = (p, \vec{y}_{12})$ appearing explicitly in the integrands. All the remaining factors, apart from this flat-space massless propagator, are absorbed into the definition of the dressing factors.

%%%%%%%%%%%%%%%%%%%%%%%%%%%%%%%%%%%%%%%%%%%%%%%%%%%%%%%%%%%%%%%%%%%%%%%%%%%%%%%%%%%%
\section{Reconstructing cosmological correlators via dispersion integration} \label{Disp+Disc_reconstruction}

In this section, we argue that a $2$-site and $3$-site cosmological correlator involving scalar fields with polynomial interactions can be reconstructed from the dispersion-integration of its discontinuities. 

It will be crucial for us to implement the dispersion integration across the discontinuities of the correlator by successively cutting all internal lines in the tree-level diagram. Therefore, our reconstruction formula via dispersion integration will only involve, on the RHS, discontinuities of the $1$-site contact level data, which are essentially $\mathcal{B}^{(1)}$ and $\widetilde{\mathcal{B}}^{(1)}$. For the $2$-site correlator, this can be trivially achieved by cutting a single internal line. Next, for a $3$-site or higher correlator, we will have to cut multiple internal lines one by one to obtain successive discontinuities that reduce to only $1$-site objects at each site. 

Although our analysis in this section will be generically valid to, e.g., conformally coupled or massless theories, with IR-divergent or convergent interactions, we will particularly consider the conformally coupled IR convergent theories to justify the usefulness of our calculations. In the following sections, we will demonstrate the applicability of the formulation to other cases.  
%%%%%%%%%%%%%%%%%%%%%%%%%%%%%%%%%%%%%%%%%%%%%%%%%%%%%%%%%%%%%%%%%%%%%%%%%%%%%%%%%%%%%%%%%%%%%%%%%%%%%%%%%%%%%%%%%%%%%%%%%%%%%%%%%%%%

\subsection{$2$-site correlator} \label{subsec2siteReconstruct}

For the $2$-site correlator $\mathcal{B}^{(2)}$, one can reconstruct it from knowledge of its discontinuity and, further, using the dispersion relation. For a $2$-site correlator, the single-cut discontinuity obtained by cutting the only available internal line will suffice to express it in terms of $1$-site contact-level data. Therefore, using eq.\eqref{2siteDiscFinal Result}, one can write the reconstruction formula in this case as follows
\begin{equation}\label{2-site correlator}
\begin{split}
\mathcal{B}^{(2)}&( \lbrace \mathbf{k}_L,  \mathbf{k}_R \rbrace ; \, p) =  -\frac{1}{2\pi i}\int_{-\infty}^{+\infty} \frac{q dq}{q^2 - p^2} \frac{1}{2P_q(\eta_0)} \times \\
 & \Bigg( \text{Disc}_p \mathcal{ B}^{(1)}( \lbrace \mathbf{k}_{L} \rbrace ,p) \text{Disc}_p \mathcal{B}^{(1)}( \lbrace \mathbf{k}_{R} \rbrace ,p ) 
      - \widetilde{\text{Disc}}_p  \widetilde{\mathcal{B}}^{(1)}( \lbrace \mathbf{ k}_{L} \rbrace ,p) \widetilde{\text{Disc}}_p  \widetilde{\mathcal{B }}^{(1)}( \lbrace \mathbf{k}_{R} \rbrace ,p) \Bigg) \, .
\end{split}
\end{equation}

Next, as a proof of concept, we explicitly work out how a $2$-site correlator can be reconstructed using eq.\eqref{2-site correlator} by focusing on the conformally coupled polynomial IR convergent interaction (i.e., $\phi^n$ interactions with $n \ge 4$). In this theory, the $1$-site wave-function coefficient is given as 
\begin{equation} \label{1-site_wvfn_coeff}
\begin{split}
    \psi^{(1)}(\lbrace \mathbf{k}_L\rbrace,q )=  
    %\frac{\lambda i^{n} (n-4)!}{H^4 \eta_0^n} \frac{1}{(k_L+q)^{n-3}} \sum_{k=0}^{n-4} \frac{1}{k!}\Big(-i (k_L+q) \eta_0 \Big)^k ,
    \frac{\lambda (n-4)!}{H^4 \eta_0^n} \frac{1}{(k_L+q)^{n-3}} \Bigg[ i^n  e^{- i \eta_0 (k_L+q)}\Big{\vert}_{n-4} \Bigg] \, , ~~~~\text{for}~~n \geq 4 \, , 
\end{split}
\end{equation}
where we have used the compact notation
 \begin{equation}
 e^{- i \eta_0 (k_L+q)}\Big{\vert}_{n-4} = \sum_{k=0}^{n-4} \frac{1}{k!}\Big(-i (k_L+q) \eta_0 \Big)^k \, .
 \end{equation}
Following the definitions given in eq.\eqref{defB1Btilde1}, the $1$-site correlator $\mathcal{B}^{(1)}$ and the corresponding auxiliary counterpart $\widetilde{\mathcal{B}}^{(1)}$ can be obtained from the real and imagianry pieces of $\psi^{(1)}$ given in eq.\eqref{1-site_wvfn_coeff}. 
We will be interested in extracting out the leading pieces in $\mathcal{B}^{(1)}$ and $\widetilde{\mathcal{B}}^{(1)}$ in the late-time limit as $\eta_0 \to 0$. In what follows, we will distinguish the cases for $n=$ even and odd in the polynomial interaction $\lambda\phi^n$ by denoting them as $n=2m$ and $n=2m+1$ respectively for $m$ being an integer. The leading order expressions in $\eta_0 \to 0$ can be written as follows 
\begin{equation}\label{corcompact_sec}
\begin{split}
    & \mathcal{B}^{(1)}(\lbrace \mathbf{k}_L\rbrace,q ) =  
\frac{\lambda (-1)^m (2m-4)! H^{4m-4} \eta_0^{2m}}{q \lbrace 2k_i \rbrace^L~(k_L+q)^{2m-3}} \, , ~~~~~~~~~~\text{for}~n=2m \, ,
    \\
    & \mathcal{B}^{(1)}(\lbrace \mathbf{k}_L\rbrace,q ) 
    =\frac{\lambda (-1)^m (2m-3)! H^{4m-2} \eta_0^{2m+2}}{q \lbrace 2k_i \rbrace^L ~ (k_L+q)^{2m-3}} \, , ~~~~~~~\text{for}~n=2m+1 \,.
\end{split}
\end{equation}
and 
\begin{equation}\label{1 site B tilde_sec}
\begin{split}
    & \widetilde{\mathcal{B}}^{(1)}(\lbrace \mathbf{k}_L\rbrace,q ) = 
 \frac{i\lambda (-1)^{m+1} (2m-4)! H^{4m-4} \eta_0^{2m+1}}{q \lbrace 2k_i \rbrace^L ~ (k_L+q)^{2m-4}} \, , ~~~~~~\text{for}~n=2m \,, 
    \\
    & \widetilde{\mathcal{B}}^{(1)}(\lbrace \mathbf{k}_L\rbrace,q ) 
    = \frac{i \lambda (-1)^m  (2m-3)! H^{4m-2} \eta_0^{2m+1} }{q \lbrace 2 k_i \rbrace^L (k_L+q)^{2m-2}} \, , ~~~~~~~~~ \text{for}~n=2m+1\,.
\end{split}
\end{equation}
where we have used the following notation 
\begin{equation} \label{notation1}
\begin{split}
\lbrace 2 k_i \rbrace^L = \prod_{i=1}^{n-1} 2 k_i .
\end{split}
\end{equation}
In the next step, we use eq.\eqref{corcompact_sec} and eq.\eqref{1 site B tilde_sec} in eq.\eqref{2-site correlator} for both the cases $n= 2m$ and $n=2m+1$. \\

\noindent \textbf{For $n=2m$ cases:} 
From eq.\eqref{corcompact_sec} and eq.\eqref{1 site B tilde_sec} we learn that for a given $n=2m$ case, in the late-time $\eta_0 \to 0$ limit, $\mathcal{B}^{(1)} \sim \eta_0^{2m}$ but $\widetilde{\mathcal{B}}^{(1)} \sim \eta_0^{2m+1}$. Therefore, the leading contribution to the RHS of eq.\eqref{2-site correlator} will come from $\mathcal{B}^{(1)}$, and the auxiliary object $\widetilde{\mathcal{B}}^{(1)}$ will not contribute to this order. This is also justified by the fact that in the $\eta_0 \to 0$ limit, to the leading order, $\psi^{(1)}$ is fully real. Consequently, for $n=2m$ cases, the discontinuity of the one-site correlator can be written as
\begin{equation}\label{discleft1}
 \begin{split}
     \text{Disc}_q \mathcal{B}^{(1)}( \lbrace \mathbf{k}_{L} \rbrace ,q) 
     = \frac{\lambda (-1)^m H^{4m-4} \eta_0^{2m}}{q \lbrace 2k_i \rbrace^L} \partial_{k_L}^{2m-4} \left( \frac{1}{k_L + q } +  \frac{1}{k_L - q } \right) \, .
 \end{split}   
\end{equation}
We can express eq.\eqref{discleft1} more compactly as 
\begin{equation}\label{discleft_sec}
 \begin{split}
     \text{Disc}_q \mathcal{B}^{(1)}( \lbrace \mathbf{k}_{L} \rbrace ,q) = \frac{-\lambda_e^c}{q \lbrace 2k_i \rbrace^L} \mathcal{D}_{k_L} \left( \frac{2k_L}{q^2-k_L^2} \right) \, . 
 \end{split}   
\end{equation}
Similarly, we get
\begin{equation}\label{discright_sec}
 \begin{split}
     \text{Disc}_q \mathcal{B}^{(1)}( \lbrace \mathbf{k}_{R} \rbrace ,q)= \frac{-\lambda_e^c}{q \lbrace 2 k_i \rbrace^R} \mathcal{D}_{k_R} \left( \frac{2k_R}{q^2-k_R^2} \right) \, ,
 \end{split}   
\end{equation}
where we introduced the following notation for a  differential operator and an effective coupling constant as 
\begin{equation} \label{lambda c e}
   \lbrace 2 k_i \rbrace^R = \prod_{i=n}^{2n-2} 2 k_i,~~~~\mathcal{D}_{k_L} \equiv \partial_{k_L}^{2m-4},~~~~~ \lambda_e^c = \lambda (-1)^m H^{4m-4} \eta_0^{2m} \, . 
\end{equation}
Substituting eq.\eqref{discleft_sec} and eq.\eqref{discright_sec} in eq.\eqref{2-site correlator} we obtain
\begin{equation} \label{CC even 2site reconstruction}
\begin{split}
     \mathcal{B}^{(2)}( \lbrace \mathbf{k}_L, \mathbf{k}_R \rbrace ;p) 
     = \frac{\lambda^2 H^{4n-10} \eta_0^{2n-2}}{\lbrace 2 k_i \rbrace^L \lbrace 2 k_i \rbrace^R}  \mathcal{D}_{k_L}  \mathcal{D}_{k_R} \left( \frac{-1}{2\pi i}\int_{-\infty}^{+\infty} \frac{dq}{q^2 - p^2}  \frac{2k_L}{q^2-k_L^2}  \frac{2k_R}{q^2-k_R^2} \right) \, .
\end{split} 
\end{equation}
Note that in the last step, we have restored $n$ by writing $2m=n$. \\

\noindent \textbf{For $n=2m+1$ cases:}
Next, for a given $n=2m+1$ case, from eq.\eqref{corcompact_sec} and eq.\eqref{1 site B tilde_sec} one can see that $\mathcal{B}^{(1)} \sim \eta_0^{2m+2}$ but $\widetilde{\mathcal{B}}^{(1)} \sim \eta_0^{2m+1}$ in the late-time $\eta_0 \to 0$ limit. Therefore, to the RHS of eq.\eqref{2-site correlator} the contribution from $\widetilde{\mathcal{B}}^{(1)}$ dominates the contribution from $\mathcal{B}^{(1)}$ in the leading order in $\eta_0 \to 0$ limit, which is also consistent with the fact that to this order $\psi^{(1)}$ is fully imaginary. Hence, using the definition of $\widetilde{\text{Disc}}_p$ from eq.\eqref{defdisc}, for the auxiliary object we get
\begin{equation}\label{disctildeleft_sec}
\begin{split}
\widetilde{\text{Disc}}_p  \widetilde{\mathcal{B}}^{(1)}( \lbrace &\mathbf{k}_{L} \rbrace ,p)
= \frac{-i \lambda_o^{\psi}}{q \lbrace 2k_i \rbrace^L} \widetilde{\mathcal{D}}_{k_L} \left( \frac{2q}{q^2-k_L^2} \right) \, ,
\end{split}    
\end{equation}
where we have defined 
\begin{equation} \label{lambda psi o}
 \widetilde{\mathcal{D}}_{k_L} \equiv \partial_{k_L}^{2m-3},~~~~~ \lambda_o^{\psi} = \lambda (-1)^{m} H^{4m-2} \eta_0^{2m+1} \, .
\end{equation}
Similarly,
\begin{equation}\label{disctilderight_sec}
\begin{split}
\widetilde{\text{Disc}}_p  \widetilde{\mathcal{B}}^{(1)}( \lbrace \mathbf{k}_{R} \rbrace ,p) = \frac{- i \lambda_o^{\psi}}{q \lbrace 2k_i \rbrace^R}  \widetilde{\mathcal{D}}_{k_L} \left(  \frac{2 q}{q^2 - k_R^2} \right) \, .
\end{split}
\end{equation}
Substituting eq.\eqref{disctildeleft_sec} and eq.\eqref{disctilderight_sec} in eq.\eqref{2-site correlator} we get the following
\begin{equation} \label{CC odd 2site reconstruction}
\begin{split}
     \mathcal{B}^{(2)}( \lbrace \mathbf{k}_L, \mathbf{k}_R \rbrace ;p) =  \frac{\lambda^2 H^{2n-10} \eta_0^{2n-2}}{\lbrace 2 k_i \rbrace^L \lbrace 2 k_i \rbrace^R}  \widetilde{\mathcal{D}}_{k_L} \widetilde{\mathcal{D}}_{k_R} \left( \frac{-1}{2\pi i}\int_{-\infty}^{+\infty} \frac{dq}{q^2 - p^2}  \frac{2 q}{q^2 - k_L^2} \frac{2 q}{q^2 - k_R^2} \right)\, ,
\end{split} 
\end{equation}
where, again, we have restored $n$ by writing $2m+1=n$. One can verify that eq.\eqref{CC even 2site reconstruction} and eq.\eqref{CC odd 2site reconstruction} are consistent with the expected results for the $2$-site cosmological correlator in conformally coupled IR-convergent theories, as, e.g., they match with \cite{Chowdhury:2025ohm}. We also explicitly verified that the method based on dispersion integration on the discontinuities can indeed reconstruct the $2$-site correlators in such theories. 

%%%%%%%%%%%%%%%%%%%%%%%%%%%%%%%%%%%%%%%%%%%%%%%%%%%%%%%%%%%%%%%%%%%%%
%%%%%%%%%%%%%%%%%%%%%%%%%%%%%%%%%%%%%%%%%%%%%%%%%%%%%%%%%%%%%%%%%%%%%%%%%%%%%%%%%%%%%%%%%%%%%%%%%%%%%%%%%%%%%%%%%%%%%%%%%%%%%%%%%%%%%%%%%%%%%%%%%%%%%%%%%%%%%%%%%%%%%%%%%%%%%%%%%%%%%%%%%%%%%%%%%%%%%%%%%%%%%%%%%%%%%%%%%%%%%%
\subsection{$3$-site correlator} \label{sec:3site}

In this sub-section, we will argue how a $3$-site correlator $\mathcal{B}^{(3)}$ can be reconstructed by applying the dispersion formula to its discontinuity. In \cite{Das:2025qsh}, a single-cut discontinuity relation for the $3$-site correlator was written, where one of the two internal lines (or bulk-to-bulk propagators) was cut. This was explicitly derived as eq.(D.18) in Appendix D in \cite{Das:2025qsh}, which we write here again for convenience
\begin{equation}\label{3 site single cut}
	\begin{split}
		{\text {Disc}}_{p_2} \mathcal{B}^{(3)}( \lbrace \mathbf{k}_L, \mathbf{k}_M , \mathbf{k}_R \rbrace ;p_1;p_2) =&  \frac{1}{2 P_{p_2}(\eta_0)} \Big( \text{Disc}_{p_2} \mathcal{B}^{(2)}( \lbrace \mathbf{k}_{L}, \mathbf{k}_M \rbrace , p_2; p_1) \text{Disc}_{p_2} \mathcal{B}^{(1)}( \lbrace \mathbf{k}_{R} \rbrace ,p_2) 
		\\
		& - \widetilde{\text{Disc}}_{p_2} \widetilde{\mathcal{B}}^{(2)}( \lbrace \mathbf{k}_{L}, \mathbf{k}_M \rbrace , p_2; p_1) \widetilde{\text{Disc}}_{p_2} \widetilde{\mathcal{B}}^{(1)}( \lbrace \mathbf{k}_{R} \rbrace ,p_2)\Bigg) \, .
	\end{split}
\end{equation}
From the RHS of eq.\eqref{3 site single cut} it is clear that the single-cut discontinuity for $\mathcal{B}^{(3)}$ was obtained by cutting the bulk-to-bulk propagator associated with $p_2$ and thus it leaves behind $\text{Disc}_{p_2} \mathcal{B}^{(2)}$ and $\widetilde{\text{Disc}}_{p_2} \widetilde{\mathcal{B}}^{(2)}$. In order to express the discontinuity of $\mathcal{B}^{(3)}$ completely in terms of discontinuities of the lower-most-point objects, i.e., $\mathcal{B}^{(1)}$ and $\widetilde{\mathcal{B}}^{(1)}$, associated with only a single site, we need to take a further discontinuity by cutting the remaining bulk-to-bulk propagator corresponding to $p_2$. Thus, to express the discontinuity of $\mathcal{B}^{(3)}$ in terms of discontinuities of lower-point objects associated with single sites, we apply the single-cut rule with respect to the energy of each internal propagator successively. In this way the full discontinuity is obtained completely from the lower-most-point data. 
Following this, taking discontinuity with respect to $p_1$ in eq.\eqref{3 site single cut}, we can write
\begin{equation}\label{3 site successive cut}
	\begin{split}
		& {\text {Disc}}_{p_1} \Big[ {\text {Disc}}_{p_2} \mathcal{B}^{(3)}( \lbrace \mathbf{k}_L, \mathbf{k}_M , \mathbf{k}_R \rbrace ;p_1;p_2) \Big] \\
		& =  \frac{1}{2 P_{p_2}(\eta_0)} \Big( {\text {Disc}}_{p_1} \Big[ \text{Disc}_{p_2} \mathcal{B}^{(2)}( \lbrace \mathbf{k}_{L}, \mathbf{k}_M \rbrace , p_2; p_1) \Big]  \times  \text{Disc}_{p_2} \mathcal{B}^{(1)}( \lbrace \mathbf{k}_{R} \rbrace ,p_2) 
		\\
		&\qquad - {\text {Disc}}_{p_1} \Big[ \widetilde{\text{Disc}}_{p_2} \widetilde{\mathcal{B}}^{(2)}( \lbrace \mathbf{k}_{L}, \mathbf{k}_M \rbrace , p_2; p_1) \Big] \times \widetilde{\text{Disc}}_{p_2} \widetilde{\mathcal{B}}^{(1)}( \lbrace \mathbf{k}_{R} \rbrace ,p_2)\Bigg) \, .
	\end{split}
\end{equation}
By definition, see eq.\eqref{defdisc}, the ${\text {Disc}}_{p}$ operation is antisymmetric in $p$. Therefore, eq.\eqref{3 site successive cut} can be written in the following form to make the antisymmetry in $p_2$ manifest, 
\begin{equation}\label{3 site successive cut second eq}
	\begin{split}
		& {\text {Disc}}_{p_1} \Big[ {\text {Disc}}_{p_2} \mathcal{B}^{(3)}( \lbrace \mathbf{k}_L, \mathbf{k}_M , \mathbf{k}_R \rbrace ;p_1;p_2) \Big] \\
		& =  \frac{1}{2 P_{p_2}(\eta_0)} \Big[\text{Disc}_{p_1} \mathcal{B}^{(2)}( \lbrace \mathbf{k}_{L}, \mathbf{k}_M \rbrace , p_2; p_1) \times  \text{Disc}_{p_2} \mathcal{B}^{(1)}( \lbrace \mathbf{k}_{R} \rbrace ,p_2) 
		\\
		&\qquad\qquad-\text{Disc}_{p_1} \widetilde{\mathcal{B}}^{(2)}( \lbrace \mathbf{k}_{L}, \mathbf{k}_M \rbrace , p_2; p_1) \times \widetilde{\text{Disc}}_{p_2} \widetilde{\mathcal{B}}^{(1)}( \lbrace \mathbf{k}_{R} \rbrace ,p_2)\Bigg]-\bigg(p_2\leftrightarrow -p_2\bigg) \, .
	\end{split}
\end{equation} 
To simplify $\text{Disc}_{p_1} \widetilde{\mathcal{B}}^{(2)}$ appearing on the RHS of eq.\eqref{3 site successive cut second eq} we will make use of the following relation
 \begin{equation}
 \label{single cut auxiliary correlator def2} 
 \begin{split}
     {\text {Disc}}_{p_{1}} \widetilde{\mathcal{B}}^{(r)}( \lbrace \mathbf{k}_L, \mathbf{k}_R \rbrace ; p_1 )  = {\text {Disc}}_{p_{1}} \mathcal{B}^{(r)}( \lbrace \mathbf{k}_L, \mathbf{k}_R \rbrace ; p_1 ) \bigg|_{\mathcal{B}^{(1)}( \lbrace \mathbf{k}_{R} \rbrace, p_{1})\leftrightarrow\widetilde{\mathcal{B}}^{(1)}( \lbrace \mathbf{k}_{R} \rbrace, p_{1})}\,, 
\end{split}
\end{equation}
with $r=2$. This relation will also be used in our analysis to reconstruct the $r$-site correlator in \S \ref{sec:3site} and will be proven for generic $r$ in Appendix \ref{app_r_site_subsec1}. Next, in eq.\eqref{3 site successive cut second eq} we substitute $\text{Disc}_{p_1} \widetilde{\mathcal{B}}^{(2)}$ from eq.\eqref{single cut auxiliary correlator def2} with $r=2$ and making further use of eq.\eqref{2-site correlator}. Subsequently, some algebraic manipulations will allow us to the successive discontinuity relation for $\mathcal{B}^{(3)} \left( \lbrace \mathbf{k}_L, \mathbf{k}_M , \mathbf{k}_R \rbrace ;p_1;p_2 \right)$, such that the RHS of eq.\eqref{3 site successive cut second eq} gets decomposed into a summation of four terms 
\begin{equation}\label{3sitetermstotal}
\begin{split}
  & \text{Disc}_{p_1}\Big[ \text{Disc}_{p_2} \mathcal{B}^{(3)} \left( \lbrace \mathbf{k}_L, \mathbf{k}_M , \mathbf{k}_R \rbrace ;p_1;p_2 \right) \Big] 
   =\text{Disc}_{p_1}\Big[ \text{Disc}_{p_2} \mathcal{B}^{(3)}  \Big] \Bigg{\vert}_{\text{term 1}} \\
   & +\text{Disc}_{p_1}\Big[ \text{Disc}_{p_2} \mathcal{B}^{(3)} \Big] \Bigg{\vert}_{\text{term 2}} +\text{Disc}_{p_1}\Big[ \text{Disc}_{p_2} \mathcal{B}^{(3)}  \Big] \Bigg{\vert}_{\text{term 3}}+\text{Disc}_{p_1}\Big[ \text{Disc}_{p_2} \mathcal{B}^{(3)} \Big] \Bigg{\vert}_{\text{term 4}} \, ,
 \end{split}
 \end{equation}
where each of the four terms on the RHS of eq.\eqref{3sitetermstotal} can be expressed as given below
\begin{equation}\label{3sitetrem1final mt} 
\begin{split}
    & \text{Disc}_{p_1} \Big[ {\text {Disc}}_{p_2} \mathcal{B}^{(3)}\Big] \Bigg{\vert}_{\text{term-1}} = \frac{1}{4 {{P_{p_1}(\eta_0)}}{{P_{p_2}(\eta_0)}}} \Big[ \text{Disc}_{p_1}  \mathcal{B}^{(1)}(\lbrace \mathbf{k}_{L} \rbrace,p_1) \times \\
    & \qquad \qquad  \widetilde{\text{Disc}}_{p_1,p_2}\,\mathcal{B}^{(1)} (\lbrace \mathbf{k}_{M} \rbrace, p_1,  p_2) \times \text{Disc}_{p_2} \mathcal{B}^{(1)}(\lbrace \mathbf{k}_{R} \rbrace,p_2)\Big]-\bigg(p_2\leftrightarrow -p_2\bigg) \, ,
\end{split}    
\end{equation}
\begin{equation}\label{3sitetrem2final mt} 
\begin{split}
    & \text{Disc}_{p_1}\Big[ \text{Disc}_{p_2} \mathcal{B}^{(3)} \Big] \Bigg{\vert}_{\text{term 2}} =-\frac{1}{4 {{P_{p_1}(\eta_0)}}{{P_{p_2}(\eta_0)}}} \Big[\text{Disc}_{p_1} \mathcal{B}^{(1)}(\lbrace \mathbf{k}_{L} \rbrace,p_1) \times \\
    &  \qquad \qquad   \text{Disc}_{p_1,p_2}\,\widetilde{\mathcal{B}}^{(1)} (\lbrace \mathbf{k}_{M} \rbrace, p_1,  p_2)  \times \widetilde{\text{Disc}}_{p_2}\widetilde{\mathcal{B}}^{(1)}(\lbrace \mathbf{k}_{R} \rbrace ,p_2) \Big] -\bigg(p_2\leftrightarrow -p_2\bigg)\, ,
    \end{split}
\end{equation}
\begin{equation}\label{3sitetrem3final mt} 
\begin{split}
    & \text{Disc}_{p_1}\Big[ \text{Disc}_{p_2} \mathcal{B}^{(3)} \Big] \Bigg{\vert}_{\text{term 3}} = -\frac{1}{4 {{P_{p_1}(\eta_0)}}{{P_{p_2}(\eta_0)}}} \Big[ \widetilde{\text{Disc}}_{p_1}\widetilde{\mathcal{B}}^{(1)}(\lbrace \mathbf{k}_{L} \rbrace ,p_1) \times \\
    & \qquad \qquad     \text{Disc}_{p_1,p_2}\,\widetilde{\mathcal{B}}^{(1)} (\lbrace \mathbf{k}_{M} \rbrace, p_1,  p_2) \times \text{Disc}_{p_2} \mathcal{B}^{(1)}(\lbrace \mathbf{k}_{R} \rbrace,p_2)  \Big] -\bigg(p_2\leftrightarrow -p_2\bigg)\, ,
    \end{split}
\end{equation}
and 
\begin{equation}\label{3sitetrem4final mt} 
 \begin{split}
      & \text{Disc}_{p_1} \Big[ \text{Disc}_{p_2} \mathcal{B}^{(3)} \Big] \Bigg{\vert}_{\text{term 4}}=\frac{1}{4 {{P_{p_1}(\eta_0)}}{{P_{p_2}(\eta_0)}}} \Big[ \widetilde{\text{Disc}}_{p_1}\widetilde{\mathcal{B}}^{(1)}(\lbrace \mathbf{k}_{L} \rbrace ,p_1) \times \\
    &  \qquad \qquad   \widetilde{\text{Disc}}_{p_1,p_2}\, \mathcal{B}^{(1)} (\lbrace \mathbf{k}_{M} \rbrace, p_1,  p_2) \times  \widetilde{\text{Disc}}_{p_2}\widetilde{\mathcal{B}}^{(1)}(\lbrace \mathbf{k}_{R} \rbrace ,p_2) \Big] -\bigg(p_2\leftrightarrow -p_2\bigg) \,,
 \end{split}   
\end{equation}
where we have defined
\begin{equation} \label{discndisctBBt_sec}
\begin{split}
    &\widetilde{\text{Disc}}_{p_i,p_{i+1}}\,
    \mathcal{B}^{(1)}(\{\mathbf{k}_{M_i}\},p_i,p_{i+1})
    = \mathcal{B}^{(1)}(\{\mathbf{k}_{M_i}\},p_i,p_{i+1})
    + \mathcal{B}^{(1)}(\{\mathbf{k}_{M_i}\},-p_i,-p_{i+1})  \, , \\[6pt]
    &\text{Disc}_{p_i,p_{i+1}}\,
    \widetilde{\mathcal{B}}^{(1)}(\{\mathbf{k}_{M_i}\},p_i,p_{i+1})
    = \widetilde{\mathcal{B}}^{(1)}(\{\mathbf{k}_{M_i}\},p_i,p_{i+1})
    - \widetilde{\mathcal{B}}^{(1)}(\{\mathbf{k}_{M_i}\},-p_i,-p_{i+1})\,.
\end{split}
\end{equation}
Note that the definitions of discontinuities given above in eq.\eqref{discndisctBBt_sec} correspond to simultaneous cutting of two internal lines attached to a vertex in the middle.

Finally, the $3$-site correlator will be obtained by applying the dispersion relation on eq.\eqref{3sitetermstotal} 
 \begin{equation}\label{dispersion 3 site mt}
\begin{split}
   & \mathcal{B}^{(3)} \left( \lbrace \mathbf{k}_L, \mathbf{k}_M , \mathbf{k}_R \rbrace ;p_1;p_2 \right)\\
   & = \bigg(\frac{-1}{2\pi i}\bigg)^2\int_{-\infty}^{+\infty}\frac{q_1\,dq_1}{q_1^2-p_1^2}\,\int_{-\infty}^{+\infty}\frac{q_2\,dq_2}{q_2^2-p_2^2} ~ \text{Disc}_{q_1}\Big[ \text{Disc}_{q_2} \mathcal{B}^{(3)} \left( \lbrace \mathbf{k}_L, \mathbf{k}_M , \mathbf{k}_R \rbrace ;q_1;q_2 \right) \Big]\\
  & =\sum_{i=1}^4 \mathcal{B}_i^{(3)} \left( \lbrace \mathbf{k}_L, \mathbf{k}_M , \mathbf{k}_R \rbrace ;p_1;p_2 \right) \, ,
   \end{split}
\end{equation} 
where the summation takes care of the four terms written in eq.\eqref{3sitetrem1final mt}, \eqref{3sitetrem2final mt}, \eqref{3sitetrem3final mt}, \eqref{3sitetrem4final mt}. Individually, each of the four terms $\mathcal{B}_i^{(3)} \left( \lbrace \mathbf{k}_L, \mathbf{k}_M , \mathbf{k}_R \rbrace ;p_1;p_2 \right)$ will have the following expressions 
\begin{equation}
\label{expterm1 final mt}
\begin{split}
  &  \mathcal{B}_1^{(3)} \left( \lbrace \mathbf{k}_L, \mathbf{k}_M , \mathbf{k}_R \rbrace ;p_1;p_2 \right)=  \bigg(\frac{1}{2\pi i}\bigg)^2\int_{-\infty}^{\infty}\frac{q_1\,dq_1}{p_1^2-q_1^2}\,\int_{-\infty}^{\infty}\frac{q_2\,dq_2}{p_2^2-q_2^2}~ \frac{2}{4 {{P_{q_1}(\eta_0)}}{{P_{q_2}(\eta_0)}}}  \times  \\
  &  \quad \quad \text{Disc}_{q_1} \mathcal{B}^{(1)}(\lbrace \mathbf{k}_{L} \rbrace,q_1) \times \widetilde{\text{Disc}}_{q_1,q_2}\,\mathcal{B}^{(1)} (\lbrace \mathbf{k}_{M} \rbrace, q_1,  q_2) \times \text{Disc}_{q_2} \mathcal{B}^{(1)}(\lbrace \mathbf{k}_{R} \rbrace,q_2) \, ,
  \end{split}
  \end{equation}
\begin{equation}
\label{expterm2 final mt}
\begin{split}
    & \mathcal{B}_2^{(3)} \left( \lbrace \mathbf{k}_L, \mathbf{k}_M , \mathbf{k}_R \rbrace ;p_1;p_2 \right)= -\bigg(\frac{1}{2\pi i}\bigg)^2\int_{-\infty}^{\infty}\frac{q_1\,dq_1}{p_1^2-q_1^2}\,\int_{-\infty}^{\infty}\frac{q_2\,dq_2}{p_2^2-q_2^2} ~ \frac{2}{4 {{P_{q_1}(\eta_0)}}{{P_{q_2}(\eta_0)}}} \times \\
    &  \quad \quad  \text{Disc}_{q_1} \mathcal{B}^{(1)}(\lbrace \mathbf{k}_{L} \rbrace,q_1) \times \text{Disc}_{q_1,q_2}\,\widetilde{\mathcal{B}}^{(1)} (\lbrace \mathbf{k}_{M} \rbrace, q_1,  q_2) \times \widetilde{\text{Disc}}_{q_2}\widetilde{\mathcal{B}}^{(1)}(\lbrace \mathbf{k}_{R} \rbrace ,q_2)  \, ,
    \end{split}
    \end{equation}
\begin{equation}
\label{expterm3 final mt}
 \begin{split} 
  & \mathcal{B}_3^{(3)} \left( \lbrace \mathbf{k}_L, \mathbf{k}_M , \mathbf{k}_R \rbrace ;p_1;p_2 \right)= -\bigg(\frac{1}{2\pi i}\bigg)^2\int_{-\infty}^{\infty}\frac{q_1\,dq_1}{p_1^2-q_1^2}\,\int_{-\infty}^{\infty}\frac{q_2\,dq_2}{p_2^2-q_2^2} ~ \frac{2}{4 {{P_{q_1}(\eta_0)}}{{P_{q_2}(\eta_0)}}} \times  \\
    &  \quad \quad  \widetilde{\text{Disc}}_{q_1}\widetilde{\mathcal{B}}^{(1)}(\lbrace \mathbf{k}_{L} \rbrace ,q_1) \times \text{Disc}_{q_1,q_2}\,\widetilde{\mathcal{B}}^{(1)} (\lbrace \mathbf{k}_{M} \rbrace, q_1,  q_2) \times \text{Disc}_{q_2} \mathcal{B}^{(1)}(\lbrace \mathbf{k}_{R} \rbrace,q_2)  \, ,
 \end{split}
 \end{equation}
 and 
 \begin{equation}
 \label{expterm4 final mt}
 \begin{split} 
  & \mathcal{B}_4^{(3)} \left( \lbrace \mathbf{k}_L, \mathbf{k}_M , \mathbf{k}_R \rbrace ;p_1;p_2 \right)= \bigg(\frac{1}{2\pi i}\bigg)^2\int_{-\infty}^{\infty}\frac{q_1\,dq_1}{p_1^2-q_1^2}\,\int_{-\infty}^{\infty}\frac{q_2\,dq_2}{p_2^2-q_2^2} ~ \frac{2}{4 {{P_{q_1}(\eta_0)}}{{P_{q_2}(\eta_0)}}}  \times \\
    & \quad \quad \widetilde{\text{Disc}}_{q_1}\widetilde{\mathcal{B}}^{(1)}(\lbrace \mathbf{k}_{L} \rbrace ,q_1) \times \widetilde{\text{Disc}}_{q_1,q_2}\,\mathcal{B}^{(1)} (\lbrace \mathbf{k}_{M} \rbrace, q_1,  q_2) \times \widetilde{\text{Disc}}_{q_2}\widetilde{\mathcal{B}}^{(1)}(\lbrace \mathbf{k}_{R} \rbrace ,q_2) \, ,
 \end{split}
 \end{equation}
 where $\text{Disc}_{q_1,q_2}$ and $\widetilde{\text{Disc}}_{q_1,q_2}$ are defined in eq.\eqref{discndisctBBt_sec}. 

Thus, in eq.\eqref{dispersion 3 site mt}, we have obtained the full $3$-site correlator $\mathcal{B}^{(3)} \left( \lbrace \mathbf{k}_L, \mathbf{k}_M , \mathbf{k}_R \rbrace ;p_1;p_2 \right)$ completely in terms of the discontinuities ($ \text{Disc}_{q}$ and $\widetilde{\text{Disc}}_{q}$) of $1$-site objects $ \mathcal{B}^{(1)}$ and $\widetilde{\mathcal{B}}^{(1)}$, which explicitly appear in eq.\eqref{expterm1 final mt}-eq.\eqref{expterm4 final mt}. 

These general expressions can further be simplified using explicit forms of $\mathcal{B}^{(1)}$ and $\widetilde{\mathcal{B}}^{(1)}$. To this end, let us remind ourselves that for a $3$-site correlator with conformally coupled $\lambda \phi^n$ interactions, we are actually computing a $(3n-4)$-point correlator. Also, there will be two internal lines or bulk-to-bulk propagators, which, after being successively cut and stuck to the late-time boundary, will give us three $1$-site objects, each with $n$ bulk-to-boundary propagators. Therefore, considering, for example, conformally coupled IR convergent scalar field theories, the objects $\mathcal{B}^{(1)}( \lbrace \mathbf{k}_{L/R} \rbrace ,q)$, $\widetilde{\mathcal{B}}^{(1)}( \lbrace \mathbf{k}_{L/R} \rbrace ,p)$, $\mathcal{B}^{(1)} (\lbrace \mathbf{k}_{M} \rbrace, q_1,  q_2)$, and $\widetilde{\mathcal{B}}^{(1)} (\lbrace \mathbf{k}_{M} \rbrace, q_1,  q_2)$ can all be straightforwardly computed using the relations given in eq.\eqref{corcompact_sec} and eq.\eqref{1 site B tilde_sec}. Furthermore, using them in eq.\eqref{dispersion 3 site mt}, and eq.\eqref{expterm1 final mt}-eq.\eqref{expterm4 final mt}, we can reconstruct the $3$-site correlator completely in such theories. 

For conformally coupled IR convergent theories (with the interaction term being $\lambda \phi^n$ with $n>4$), and for $n=\text{even}=2m$, one can obtain the explicit expressions for the $3$-site correlator, to the leading order in $\eta_0 \to 0$, as
\begin{equation} \label{B3 n=2m CC}
\begin{split}
   & \mathcal{B}^{(3)} \left( \lbrace \mathbf{k}_L, \mathbf{k}_M , \mathbf{k}_R \rbrace ;p_1;p_2 \right)  = \mathcal{B}_1^{(3)} \left( \lbrace \mathbf{k}_L, \mathbf{k}_M , \mathbf{k}_R \rbrace ;p_1;p_2 \right)  \\
   & = \frac{\left( - \lambda_e^c \right)^3}{\prod_{i=1}^{3(n-2)+2} 2 k_i} \left( \frac{1}{H^2 \eta_0^2} \right)^{2} \times \left( \mathcal{D}_{k_L}\mathcal{D}_{k_M} \mathcal{D}_{k_R} \right) \prod_{m=1}^{2} \left(  \frac{1}{2 \pi i} \int_{-\infty}^{+\infty} \frac{ dq_m}{p_m^2 - q_m^2}\right) \times
   \\
   &~~~~~~~~~~~~~~~~~~~~~~~~~~~~~~~~~~~~~ \Bigg(  \frac{2 k_L}{q_1^2 - k_L^2}   \frac{2 k_{M}}{( q_1 + q_2)^2 - k_{M_j}^2} \frac{2 k_R}{q_2^2 - k_R^2} \Bigg).
   \end{split}
\end{equation} 
Similarly, for $n=\text{odd}=2m+1$, we can rewrite eq.\eqref{expterm1 final mt}-\eqref{expterm4 final mt} for conformally coupled theories with odd polynomial interaction as follows
\begin{equation} \label{expterm1 final mt CC}
\begin{split}
    & \mathcal{B}_1^{(3)} \left( \lbrace \mathbf{k}_L, \mathbf{k}_M , \mathbf{k}_R \rbrace ;p_1;p_2 \right)= \frac{\left( - \lambda_o^c \right)^3}{\prod_{i=1}^{3(n-2)+2} 2 k_i} \left( \frac{1}{H^2 \eta_0^2} \right)^{2} \times     \\
     & \qquad  \left( \mathcal{D}_{k_L}\mathcal{D}_{k_M} \mathcal{D}_{k_R} \right) \prod_{m=1}^{2} \left(  \frac{1}{2 \pi i} \int_{-\infty}^{+\infty} \frac{ dq_m}{p_m^2 - q_m^2}\right) \Bigg(  \frac{2 k_L}{q_1^2 - k_L^2}   \frac{2 k_{M}}{( q_1 + q_2)^2 - k_{M_j}^2} \frac{2 k_R}{q_2^2 - k_R^2} \Bigg) \, ,
\end{split}
\end{equation}
\begin{equation}\label{expterm2 final mt CC}
\begin{split}
    & \mathcal{B}_2^{(3)} \left( \lbrace \mathbf{k}_L, \mathbf{k}_M , \mathbf{k}_R \rbrace ;p_1;p_2 \right)= - \frac{(-\lambda_o^c)(-i \lambda_o^\psi) (-i \lambda_o^\psi) }{\prod_{i=1}^{3(n-2)+2} 2 k_i} \left( \frac{1}{H^2 \eta_0^2} \right)^{2}  \times     \\
     & \qquad   \left( \mathcal{D}_{k_L} \widetilde{\mathcal{D}}_{k_M} \widetilde{\mathcal{D}}_{k_R} \right) \prod_{m=1}^{2} \left(  \frac{ 1}{2 \pi i} \int_{-\infty}^{+\infty} \frac{ dq_m}{p_m^2 - q_m^2}\right) \left( \frac{2 k_L}{q_1^2 - k_L^2} \frac{2( q_1 + q_2)}{(q_1 + q_2)^2 - k_M^2}  \frac{2 q_2}{q_2^2 - k_R^2} \right) \, ,
\end{split}
\end{equation}
\begin{equation}\label{expterm3 final mt CC}
\begin{split}
    & \mathcal{B}_3^{(3)} \left( \lbrace \mathbf{k}_L, \mathbf{k}_M , \mathbf{k}_R \rbrace ;p_1;p_2 \right)= - \frac{( -i \lambda_o^\psi) (- i \lambda_o^\psi) (- \lambda_o^c)}{\prod_{i=1}^{3(n-2)+2} 2 k_i} \left( \frac{1}{H^2 \eta_0^2} \right)^{2}  \times     \\
     & \qquad   \left( \widetilde{\mathcal{D}}_{k_L} \widetilde{\mathcal{D}}_{k_M} \mathcal{D}_{k_R} \right) \prod_{m=1}^{2} \left(  \frac{ 1}{2 \pi i} \int_{-\infty}^{+\infty} \frac{ dq_m}{p_m^2 - q_m^2}\right) \left( \frac{2 q_1}{q_1^2 - k_L^2} \frac{2( q_1 + q_2)}{(q_1 + q_2)^2 - k_M^2}  \frac{2 k_R}{q_2^2 - k_R^2} \right) \, ,
\end{split}
\end{equation}
\begin{equation}\label{expterm4 final mt CC}
\begin{split}
    & \mathcal{B}_4^{(3)} \left( \lbrace \mathbf{k}_L, \mathbf{k}_M , \mathbf{k}_R \rbrace ;p_1;p_2 \right)= \frac{(- i \lambda_o^\psi) (- \lambda_o^c) (- i \lambda_o^\psi) }{\prod_{i=1}^{3(n-2)+2} 2 k_i} \left( \frac{1}{H^2 \eta_0^2} \right)^{2}  \times     \\
     & \qquad   \left( \widetilde{\mathcal{D}}_{k_L}\mathcal{D}_{k_M} \widetilde{\mathcal{D}}_{k_R} \right) \prod_{m=1}^{2} \left(  \frac{ 1}{2 \pi i} \int_{-\infty}^{+\infty} \frac{ dq_m}{p_m^2 - q_m^2}\right) \left( \frac{2 q_1}{q_1^2 - k_L^2} \frac{2 k_M}{(q_1 + q_2)^2 - k_M^2}  \frac{2 q_2}{q_2^2 - k_R^2} \right) \, .
\end{split}
\end{equation}
The intermediate steps to arrive at these expressions are given in Appendix \ref{app - detail 3 site disp+disc}. 

From these expressions, it is clear that the leading order expressions for the $n=\text{even}$
and $n=\text{odd}$ polynomial interactions are different, which will have crucial implications for the diagrammatic dressing rules in \S \ref{GeneralDressingRules}. Also, it must be noted that for odd polynomial interactions, the leading $\eta_0$ contribution can be achieved by adding up only the contributions given in eq.\eqref{expterm2 final mt CC}, \eqref{expterm3 final mt CC}, \eqref{expterm4 final mt CC}, whereas the contribution in eq.\eqref{expterm4 final mt CC} is sub-dominant compared to them $\eta_0 \to 0$. This can be verified by noting that 
\begin{equation}
    \left( \lambda_o^c \right)^3 \sim \eta_0^{6m+6}, \qquad  \lambda_o^c \left( \lambda_o^\psi \right)^2 \sim \eta_0^{6m+4}\, ,
\end{equation} 
which follows from the fact that $\lambda_o^c \sim \eta_0^{2m+2}, ~ \lambda_o^\psi \sim \eta_0^{2m+1}$, see eq.\eqref{lambda c e} and eq.\eqref{lambda psi o}. 

%%%%%%%%%%%%%%%%%%%%%%%%%%%%%%%%%%%%%%%%%%%%%%%%%%%%%%%%%%%%%%%%%%%%%
%%%%%%%%%%%%%%%%%%%%%%%%%%%%%%%%%%%%%%%%%%%%%%%%%%%%%%%%%%%%%
\section{Reconstructing $r$-site correlator via induction method}\label{sec:rsite}

In this section, we aim to generalize our formalism to reconstruct an $r$-site tree-level correlator by applying dispersion integration on the successive discontinuity of the $r$-site correlator obtained by cutting all its $(r-1)$ internal lines one by one. For that, we first need to obtain the successive single cut discontinuity of the $r$-site correlator expressed in terms of the discontinuities (both $ \text{Disc}_{q}$ and $\widetilde{\text{Disc}}_{q}$) of $r$ numbers of $1$-site, $n$-point objects (i.e., $\mathcal{B}^{(1)}$ and $\widetilde{\mathcal{B}}^{(1)}$). This is exactly what we have achieved for the $3$-site correlator in eq.\eqref{3sitetermstotal} with eq.\eqref{3sitetrem1final mt}-\eqref{3sitetrem4final mt}.  

We start with the single cut discontinuity obtained by cutting the right-most internal line for an $r$-site correlator as written in eq.\eqref{single cut r-site correlator}. Since ${\text {Disc}}_{p_{r-1}} \mathcal{B}^{(r)}$ is antisymmetric in the exchange: $(p_{r-1}\leftrightarrow -p_{r-1})$, eq.\eqref{single cut r-site correlator} can also be written as
\begin{equation} \label{discdiscBr0}
\begin{split}
    & {\text {Disc}}_{p_{r-1}} \mathcal{B}^{(r)}( \lbrace \mathbf{k}_L,..., \mathbf{k}_R \rbrace ; \lbrace p_1, ..., p_{r-1} \rbrace )  = \frac{1}{2 P_{p_{r-1}} (\eta_0)}\times 
    \\
    & ~~\bigg[ \bigg(\mathcal{ B}^{(r-1)}\left(  \lbrace \mathbf{k}_L,...,\mathbf{k}_{M_{r-2}} \rbrace, p_{r-1} ; \lbrace p_1,..,p_{r-2} \rbrace \right)  \times  \text{Disc}_{p_{r-1}} \mathcal{B}^{(1)}( \lbrace \mathbf{k}_{R} \rbrace, p_{r-1})\\
    & ~~ -\widetilde{\mathcal{B}}^{(r-1)}\left(  \lbrace \mathbf{k }_L,...,\mathbf{k}_{M_{r-2}} \rbrace, p_{r-1} ; \lbrace p_1,..,p_{r-2} \rbrace \right)  \times  \widetilde{\text{Disc}}_{p_{r-1}} \widetilde{\mathcal{B}}^{(1)}( \lbrace \mathbf{k}_{ R} \rbrace, p_{r-1}) \bigg)\\
    & ~~-\bigg(p_{r-1}\leftrightarrow -p_{r-1}\bigg)\bigg] \, . 
\end{split}
\end{equation}
Next, we take another discontinuity by cutting the second internal line with energy $p_{r-2}$ from the right end of the $r$-site correlator. 
Furthermore, using the identity in eq.\eqref{single cut auxiliary correlator def2}, which we write here again for convenience, \footnote{We have derived eq.\eqref{single cut auxiliary correlator def3} in Appendix \ref{app_r_site_subsec1}.}
 \begin{equation}
 \label{single cut auxiliary correlator def3} 
 \begin{split}
     &{\text {Disc}}_{p_{r-1}} \widetilde{\mathcal{B}}^{(r)}( \lbrace \mathbf{k}_L,..., \mathbf{k}_R \rbrace ; \lbrace p_1, ..., p_{r-1} \rbrace )  \\&= {\text {Disc}}_{p_{r-1}} \mathcal{B}^{(r)}( \lbrace \mathbf{k}_L,..., \mathbf{k}_R \rbrace ; \lbrace p_1, ..., p_{r-1} \rbrace ) \bigg|_{\mathcal{B}^{(1)}( \lbrace \mathbf{k}_{R} \rbrace, p_{r-1})\leftrightarrow\widetilde{\mathcal{B}}^{(1)}( \lbrace \mathbf{k}_{R} \rbrace, p_{r-1})}\,, 
\end{split}
\end{equation}
we can write eq.\eqref{discdiscBr0} as 
\begin{equation}
\begin{split}
    & {\text {Disc}}_{p_{r-2}}\bigg[{\text {Disc}}_{p_{r-1}} \mathcal{B}^{(r)}( \lbrace \mathbf{k}_L,..., \mathbf{k}_R \rbrace ; \lbrace p_1, ..., p_{r-1} \rbrace )\bigg] = \frac{1}{2 P_{p_{r-1}} (\eta_0)} \times \\
    &\bigg[ {\text {Disc}}_{p_{r-2}}\mathcal{ B}^{(r-1)}\left(  \lbrace \mathbf{k}_L,...,\mathbf{k}_{M_{r-2}} \rbrace, p_{r-1} ; \lbrace p_1,..,p_{r-2} \rbrace \right)  \times  \text{Disc}_{p_{r-1}} \mathcal{B}^{(1)}( \lbrace \mathbf{k}_{R} \rbrace, p_{r-1})\\
    & -\text {Disc}_{p_{r-2}}\mathcal{ B} ^{(r-1)}\left(  \lbrace \mathbf{k}_L,...,\mathbf{k}_{M_{r-2}} \rbrace, p_{r-1} ; \lbrace p_1,..,p_{r-2} \rbrace \right)\bigg|_{\mathcal{B}^{(1)}( \lbrace \mathbf{k}_{M_{r-2}} \rbrace, p_{r-2},p_{r-1})\leftrightarrow\widetilde{\mathcal{B}}^{(1)}( \lbrace \mathbf{k}_{M_{r-2}} \rbrace, p_{r-2},p_{r-1})}\\&~~  \times  \widetilde{\text{Disc}}_{p_{r-1}} \widetilde{\mathcal{B}}^{(1)}( \lbrace \mathbf{k}_{ R} \rbrace, p_{r-1}) -\bigg(p_{r-1}\leftrightarrow -p_{r-1}\bigg)\bigg] \, .
\end{split}
\end{equation}
Now, performing successive cut discontinuity with respect to all internal energies, we will obtain 
\begin{equation}
\label{r-site successive dis with r-1 site}
\begin{split}
  &{\text {Disc}}_{p_{1}}\ldots\bigg[{\text {Disc}}_{p_{r-1}} \mathcal{B}^{(r)}( \lbrace \mathbf{k}_L,..., \mathbf{k}_R \rbrace ; \lbrace p_1, ..., p_{r-1} \rbrace )\bigg] = \widehat{\mathcal{I}}_{r}( \lbrace \mathbf{k}_L,..., \mathbf{k}_R \rbrace ; \lbrace p_1, ..., p_{r-1} \rbrace )
    \\
    & = \frac{1}{2 P_{p_{r-1}} (\eta_0)} \bigg[\widehat{\mathcal{I}}_{(r-1)}\left(  \lbrace \mathbf{k}_L,...,\mathbf{k}_{M_{r-2}} \rbrace, p_{r-1} ; \lbrace p_1,..,p_{r-2} \rbrace \right) \times  \text{Disc}_{p_{r-1}} \mathcal{B}^{(1)}( \lbrace \mathbf{k}_{R} \rbrace, p_{r-1})\\
    & -\widehat{\mathcal{I}}_{(r-1)}\left(  \lbrace \mathbf{k}_L,...,\mathbf{k}_{M_{r-2}} \rbrace, p_{r-1} ; \lbrace p_1,..,p_{r-2} \rbrace \right) \bigg|_{\mathcal{B}^{(1)}( \lbrace \mathbf{k}_{M_{r-2}} \rbrace, p_{r-2},p_{r-1})\leftrightarrow\widetilde{\mathcal{B}}^{(1)}( \lbrace \mathbf{k}_{M_{r-2}} \rbrace, p_{r-2},p_{r-1})}\\&  \times  \widetilde{\text{Disc}}_{p_{r-1}} \widetilde{\mathcal{B}}^{(1)}( \lbrace \mathbf{k}_{ R} \rbrace, p_{r-1}) -\bigg(p_{r-1}\leftrightarrow -p_{r-1}\bigg)\bigg] \, ,
\end{split}
\end{equation}
where we have defined $\widehat{\mathcal{I}}_s$ as the successive cut discontinuity of the $r$-site correlator 
\begin{equation}
\label{successive cut discontinuity def1 with I}
    \widehat{\mathcal{I}}_r( \lbrace \mathbf{k}_L,..., \mathbf{k}_R \rbrace ; \lbrace q_1, ..., q_{r-1} \rbrace )  \equiv  \text{Disc}_{q_{1}}\bigg[\ldots\bigg[ \text{Disc}_{q_{r-1}}\,\mathcal{B}^{(s)}( \lbrace \mathbf{k}_L,..., \mathbf{k}_R \rbrace ; \lbrace q_1, ..., q_{r-1} \rbrace )\bigg] \bigg] \,.
\end{equation}
It can be anticipated that following the iterative process of cutting all the internal lines one by one and writing an expression for $\widehat{\mathcal{I}}_r$ in terms of the $\text{Disc}_{q}$ and $\widetilde{\text{Disc}}_{q}$ of $\mathcal{B}^{(1)}$ and $\widetilde{\mathcal{B}}^{(1)}$ will be too complicated. Instead, we will use an inductive method. We will first propose an algorithm to write down an expression for the successive discontinuity of an $s$-site correlator in terms of $\text{Disc}_{q}$ and $\widetilde{\text{Disc}}_{q}$ of $\mathcal{B}^{(1)}$ and $\widetilde{\mathcal{B}}^{(1)}$. Next, we will show that this algorithm consistently reproduces the successive discontinuities for the $2$-site ($\mathcal{B}^{(2)}$) and $3$-site ($\mathcal{B}^{(3)}$) correlators, obtained in the previous two subsections. Assuming that this algorithm holds for $\widehat{\mathcal{I}}_{r-1}$, we will then argue that the same algorithm will generate $\widehat{\mathcal{I}}_{r}$, the successive discontinuity for an $r$-site correlator from lower-point discontinuity data using the same algorithm.

From the iterative structure of taking successive discontinuities, one learns that $\widehat{\mathcal{I}}_s $ can be reorganized as 
\begin{equation}
\label{r site successive cut disc ansatz}
\begin{split}
    &\widehat{\mathcal{I}}_s( \lbrace \mathbf{k}_L,..., \mathbf{k}_R \rbrace ; \lbrace q_1, ..., q_{s-1} \rbrace ) = \text{Disc}_{q_{1}}\bigg[\ldots\text{Disc}_{q_{s-1}}\,\mathcal{B}^{(s)}( \lbrace \mathbf{k}_L,..., \mathbf{k}_R \rbrace ; \lbrace q_1, ..., q_{s-1} \rbrace)\bigg] \\&=  \mathcal{I}_s( \lbrace \mathbf{k}_L,..., \mathbf{k}_R \rbrace ; \lbrace q_1, ..., q_{s-1} \rbrace )\\&~~~~~~~~~ - \sum_{\substack{\sigma_2,\dots,\sigma_{s-1}=\pm 1 \\ (\sigma_2,\dots,\sigma_{s-1})\neq (+,\dots,+)}} \left(\prod_{i=2}^{s-1}\sigma_i\right) \times \, \mathcal{I}_s(\{\mathbf{k}_L,\dots,\mathbf{k}_R\}; \{q_1,\sigma_2 q_2,\dots,\sigma_{s-1} q_{s-1}\}) \, ,
\end{split}
\end{equation}
where in the RHS above, for an $s$-site correlator, there will be a linear combination of terms, each expressed as products of 
\begin{equation} \label{lists_elements}
\begin{split}
& \text{Disc}_{q_i} \mathcal{B}^{(1)}(\lbrace \mathbf{k} \rbrace ;q_i)\,, ~~\widetilde{\text{Disc}}_{q_i} \mathcal{B}^{(1)}(\lbrace \mathbf{k} \rbrace ;q_i)\,, ~~\text{Disc}_{q_i} \widetilde{\mathcal{B}^{(1)}}(\lbrace \mathbf{k} \rbrace ;q_i)\,, ~~ \widetilde{\text{Disc}}_{q_i} \widetilde{\mathcal{B}^{(1)}}(\lbrace \mathbf{k} \rbrace ;q_i) \, ,
\end{split}
\end{equation}
coming from each of the vertices. In eq.\eqref{r site successive cut disc ansatz}, the LHS is a linear combination of the objects $\mathcal{B}^{(s)}$ with sign flips of the internal energies. However, the RHS is a combination of terms, each expressed in terms of the lowest-point $1$-site data listed in eq.\eqref{lists_elements}. One can, in principle, obtain this relation by following the successive discontinuity operation in a brute force manner. Also, the RHS of eq.\eqref{r site successive cut disc ansatz} has been organized in the particular way such that the second term on the RHS signifies the fact that $\widehat{\mathcal{I}}_s( \lbrace \mathbf{k}_L,..., \mathbf{k}_R \rbrace ; \lbrace q_1, ..., q_{s-1} \rbrace )$ is antisymmetric in sign flips of the internal energies: $q_2, q_3, ..., q_{s-1}$ individually. Furthermore, it is easy to verify that this reorganization is in sync with what we obtained in eq.\eqref{3 site successive cut second eq} for the successive discontinuity of the $3$-site correlator. \\

\noindent \textbf{The algorithm to obtain successive discontinuity:} We propose the following algorithm to evaluate $\mathcal{I}_s$ 
\begin{itemize}
    \item By successively cutting $(s-1)$ internal lines of an $s$-site correlator, we obtain a product of discontinuities 
    ($\text{Disc}$, $\widetilde{\text{Disc}}$) of the contact correlator $\mathcal{B}^{(1)}$ or auxiliary object $\widetilde{\mathcal{B}}^{(1)}$ at each site, after attaching the internal lines to the boundary. 
    
    \item For the two sites (or vertices) present at the two edges (left and right ends of the corresponding diagram), one receives contributions of the following form
    \[
        \text{Disc}_{p_i}\,\mathcal{B}^{(1)}(\{\mathbf{k}_{L/R}\},p_i)
        \qquad\text{or}\qquad
        \widetilde{\text{Disc}}_{p_i}\,\widetilde{\mathcal{B}}^{(1)}(\{\mathbf{k}_{L/R}\},p_i).
    \]
\item For the vertices in the middle of the diagram, the contributions are
    \[
       \widetilde{\text{Disc}}_{p_i,p_{i+1}}\,
    \mathcal{B}^{(1)}(\{\mathbf{k}_{M_i}\},p_i,p_{i+1})\, , \quad \text{or} \quad \text{Disc}_{p_i,p_{i+1}}\,
    \widetilde{\mathcal{B}}^{(1)}(\{\mathbf{k}_{M_i}\},p_i,p_{i+1}) \, ,
    \]
which are defined in eq.\eqref{discndisctBBt_sec} . 
\item Since $\widetilde{\mathcal{B}}_1$ is purely imaginary, but the correlators are real quantities, in each candidate term representing the full correlator, always an even number of $\widetilde{\mathcal{B}}_1$ insertions must be present. One must consider all possible combinations in which an even number of sites contribute $\widetilde{\mathcal{B}}_1$, which will be summed over. 
    
   \item Along with the vertex contributions, we must also multiply the following contribution coming from all the internal lines:
\[
    \frac{1}{2 \!\left(\prod_{i=\text{all internal lines}} P_{p_i}(\eta_0)\right)}.
\]
\item All possible combinations of contributions from each vertex must be multiplied and summed to form the series. One also has to decide the relative sign of each contribution in the summation by multiplying the following factor whenever $\widetilde{\mathcal{B}}_1$ occurs:
    \[
        (-1)^{\text{position of }\widetilde{\mathcal{B}}_1\text{ (counted from left or right)}}.
    \]
\end{itemize}

\noindent \textbf{The algorithm is consistent with $2$ and $3$-site successive discontinuities:} 
For a $2$-site correlator, using the algorithm, we get the following expression for $\widehat{\mathcal{I}}_{2}( \lbrace \mathbf{k}_L, \mathbf{k}_R \rbrace ; p_1) $
\begin{equation}
\begin{split}
    &\widehat{\mathcal{I}}_{2}( \lbrace \mathbf{k}_L, \mathbf{k}_R \rbrace ; p_1 ) = \text{Disc}_{p_1}\mathcal{B}^{(2)}( \lbrace \mathbf{k}_L, \mathbf{k}_R \rbrace ; p_1 )\\
    & = \frac{1}{2P_{q_1}(\eta_0)}\bigg(\text{Disc}_{q_1}\,\mathcal{B}^{(1)}(\{\mathbf{k}_{L}\},q_1)\,\text{Disc}_{q_1}\,\mathcal{B}^{(1)}(\{\mathbf{k}_{R}\},q_1) \\
    & + (-1)^{1+2}\,\widetilde{\text{Disc}}_{q_1}\,\widetilde{\mathcal{B}}^{(1)}(\{\mathbf{k}_{L}\},q_1)\,\widetilde{\text{Disc}}_{q_1}\,\widetilde{\mathcal{B}}^{(1)}(\{\mathbf{k}_{R}\},q_1)\bigg) \, ,
\end{split}
\end{equation}
which is consistent with eq.\eqref{2siteDiscFinal Result}. Note that, in eq.\eqref{r site successive cut disc ansatz}, the second term on the RHS vanishes for $r=2$. Now, substituting the above expression into the dispersion integral of eq.\eqref{disp+disc_relation}, we obtain the $2$-site correlator, which is consistent with eq.\eqref{2-site correlator}.

Similarly, for a $3$-site correlator, using the algorithm, we get the following expression
\begin{equation}
\label{3site_rule_total}
    \widehat{\mathcal{I}}_{3}( \lbrace \mathbf{k}_L, \mathbf{k}_M,  \mathbf{k}_R \rbrace ; p_1, p_2 )= 
    \mathcal{I}_{3}( \lbrace \mathbf{k}_L, \mathbf{k}_M,  \mathbf{k}_R \rbrace ; p_1, p_2 )-\mathcal{I}_{3}( \lbrace \mathbf{k}_L, \mathbf{k}_M,  \mathbf{k}_R \rbrace ; p_1, -p_2 )\,,
\end{equation}
where, 
\begin{equation} \label{3site_rule}
\begin{split}
    &\mathcal{I}_{3}( \lbrace \mathbf{k}_L, \mathbf{k}_M,  \mathbf{k}_R \rbrace ; p_1, p_2 )= \frac{1}{4P_{p_1}(\eta_0)\,P_{p_2}(\eta_0)}\times\\
    &~\bigg(\text{Disc}_{p_1}\,\mathcal{B}^{(1)}(\{\mathbf{k}_{L}\},p_1)\,\widetilde{\text{Disc}}_{\{p_1,p_2\}}\,\mathcal{B}^{(1)}(\{\mathbf{k}_{M}\},p_1, p_2)\,\text{Disc}_{p_2}\,\mathcal{B}^{(1)}(\{\mathbf{k}_{R}\},p_2)\\
    & ~ +(-1)^{(2+3)}\text{Disc}_{p_1}\,\mathcal{B}^{(1)}(\{\mathbf{k}_{L}\},p_1)\,\text{Disc}_{\{p_1,p_2\}}\,\widetilde{\mathcal{B}}^{(1)}(\{\mathbf{k}_{M}\},p_1, p_2)\,\widetilde{\text{Disc}}_{p_2}\,\widetilde{\mathcal{B}}^{(1)}(\{\mathbf{k}_{R}\},p_2)\\& ~+(-1)^{(1+2)}\widetilde{\text{Disc}}_{p_1}\,\widetilde{\mathcal{B}}^{(1)}(\{\mathbf{k}_{L}\},p_1)\,\text{Disc}_{\{p_1,p_2\}}\,\widetilde{\mathcal{B}}^{(1)}(\{\mathbf{k}_{M}\},p_1, p_2)\,\text{Disc}_{p_2}\,\mathcal{B}^{(1)}(\{\mathbf{k}_{R}\},p_2)\\&~+(-1)^{(1+3)}\widetilde{\text{Disc}}_{p_1}\,\widetilde{\mathcal{B}}^{(1)}(\{\mathbf{k}_{L}\},p_1)\,
    \widetilde{\text{Disc}}_{\{p_1,p_2\}}\,\mathcal{B}^{(1)}(\{\mathbf{k}_{M}\},p_1, p_2)\,\widetilde{\text{Disc}}_{p_2}\,\widetilde{\mathcal{B}}^{(1)}(\{\mathbf{k}_{R}\},p_2)\bigg) \, ,
\end{split}
\end{equation}
such that $\text{Disc}_{\{p_1,p_2\}}\,\widetilde{\mathcal{B}}^{(1)}$ and $\text{Disc}_{\{p_1,p_2\}}\,\widetilde{\mathcal{B}}^{(1)}$ are given in eq.\eqref{discndisctBBt_sec}. Also, note that on the RHS in eq.\eqref{3site_rule}, each term contains an even number of $\widetilde{\mathcal{B}}^{(1)}$s, and the relative signs of each of the four terms are consistent with the rule stated above, depending on the positions of the $\widetilde{\mathcal{B}}^{(1)}$s in each of them. Finally, one can verify that eq.\eqref{3site_rule_total} after substituting eq.\eqref{3site_rule} agrees with the successive cut discontinuity for $3$-site correlator written in eq.\eqref{3sitetermstotal}.  \\

\noindent \textbf{Inductive proof for an $r$-site correlator:} Let us start by focusing on eq.\eqref{r-site successive dis with r-1 site}, obtained by performing successive cut discontinuity with respect to all internal energies on an $r$-site correlator. We write the expression here again for convenience 
\begin{equation}
\label{r-site successive dis with r-1 site repeat}
\begin{split}
  &{\text {Disc}}_{p_{1}}\ldots\bigg[{\text {Disc}}_{p_{r-1}} \mathcal{B}^{(r)}( \lbrace \mathbf{k}_L,..., \mathbf{k}_R \rbrace ; \lbrace p_1, ..., p_{r-1} \rbrace )\bigg] = \widehat{\mathcal{I}}_{r}( \lbrace \mathbf{k}_L,..., \mathbf{k}_R \rbrace ; \lbrace p_1, ..., p_{r-1} \rbrace )
    \\
    & = \frac{1}{2 P_{p_{r-1}} (\eta_0)} \bigg[\widehat{\mathcal{I}}_{(r-1)}\left(  \lbrace \mathbf{k}_L,...,\mathbf{k}_{M_{r-2}} \rbrace, p_{r-1} ; \lbrace p_1,..,p_{r-2} \rbrace \right) \times  \text{Disc}_{p_{r-1}} \mathcal{B}^{(1)}( \lbrace \mathbf{k}_{R} \rbrace, p_{r-1})\\
    & ~~ -\widehat{\mathcal{I}}_{(r-1)}\left(  \lbrace \mathbf{k}_L,...,\mathbf{k}_{M_{r-2}} \rbrace, p_{r-1} ; \lbrace p_1,..,p_{r-2} \rbrace \right) \bigg|_{\mathcal{B}^{(1)}( \lbrace \mathbf{k}_{M_{r-2}} \rbrace, p_{r-2},p_{r-1})\leftrightarrow\widetilde{\mathcal{B}}^{(1)}( \lbrace \mathbf{k}_{M_{r-2}} \rbrace, p_{r-2},p_{r-1})}\\&~~  \times  \widetilde{\text{Disc}}_{p_{r-1}} \widetilde{\mathcal{B}}^{(1)}( \lbrace \mathbf{k}_{ R} \rbrace, p_{r-1}) -\bigg(p_{r-1}\leftrightarrow -p_{r-1}\bigg)\bigg] \, ,
\end{split}
\end{equation}
where the function $\widehat{\mathcal{I}}_{s}$ for any integer $s$ is given in eq.\eqref{r site successive cut disc ansatz}.

Now, we will assume that $\widehat{\mathcal{I}}_{(r-1)}$ follows from our algorithm mentioned above and, hence, can be written explicitly in terms of the objects listed in eq.\eqref{lists_elements}. With this assumption, following the principle of the method of induction, our final goal is to argue that $\widehat{\mathcal{I}}_{r}$ is also consistent with our algorithm. In other words, to show that the RHS of eq.\eqref{r-site successive dis with r-1 site repeat} can also be constructed, as an expression written in terms of the elements in eq.\eqref{lists_elements}, following our algorithm. 

In order to achieve that, we need to resolve two potential problems as mentioned below:

\begin{itemize}
\item The left vertex is an edge vertex for both the $(r-1)$-site correlator and the $r$-site correlator. But the right edge vertex for the $(r-1)$-site correlator is now one of the middle vertices, which is adjacent to the right-most edge vertex for the $r$-site correlator. So, following our algorithm, we may get a different vertex factor for the $(r-1)$-th position of the $r$-site correlator from eq.\eqref{r-site successive dis with r-1 site repeat} compared to the right edge vertex contribution of an $(r-1)$-site vertex. Regarding the contribution from the right most edge vertex for an $r$-site correlator, we see from eq.\eqref{r-site successive dis with r-1 site repeat} that the contribution in the first term on the RHS goes as $\text{Disc}_{p_{r-1}} \mathcal{B}^{(1)}( \lbrace \mathbf{k}_{R} \rbrace, p_{r-1})$, which is consistent with our algorithm. So, the problem is only with the contribution from the last but one vertex from the right edge of the $r$-site correlator. 

\item To understand the second potential problem, let us focus on the second term of the RHS of eq.\eqref{r-site successive dis with r-1 site repeat}. The contribution from the right-most vertex of the $r$-site correlator here involves the auxiliary counterpart of the $1$-site correlator, i.e., $\widetilde{\text{Disc}}_{p_{r-1}} \widetilde{\mathcal{B}}^{(1)}( \lbrace \mathbf{k}_{ R} \rbrace, p_{r-1})$. However, following our algorithm, there is a particular relative sign to be taken care of whenever we encounter $\widetilde{\mathcal{B}}^{(1)}$, depending on the position of the vertex that produces it. So, there might be concern that the second term of the RHS of eq.\eqref{r-site successive dis with r-1 site repeat} will not follow the sign rule of our algorithm. 
\end{itemize}

Interestingly, it can be argued that both of the above-mentioned concerns can be satisfactorily addressed, so that eq.\eqref{r-site successive dis with r-1 site repeat} is indeed consistent with our algorithm for obtaining the successive discontinuity of an $r$-site correlator. Intuitively, the reason that these two problematic terms can be handled properly can be traced back to our starting assumption that the successive discontinuities of the $(r-1)$-site correlator $\widehat{\mathcal{I}}_{(r-1)}\left(  \lbrace \mathbf{k}_L,...,\mathbf{k}_{M_{r-2}} \rbrace, p_{r-1}; \lbrace p_1,..,p_{r-2} \rbrace \right)$ can be consistently produced following our algorithm, and hence will have a reorganization as in eq.\eqref{r site successive cut disc ansatz} such as 
\begin{equation}
\label{r-1-site successive disc}
\begin{split}
    &\widehat{\mathcal{I}}_{(r-1)}\left(  \lbrace \mathbf{k}_L,...,\mathbf{k}_{M_{r-2}} \rbrace, p_{r-1} ; \lbrace p_1,..,p_{r-2} \rbrace \right) \\&=  \mathcal{I}_{(r-1)}\left(  \lbrace \mathbf{k}_L,...,\mathbf{k}_{M_{r-2}} \rbrace, p_{r-1} ; \lbrace p_1,..,p_{r-2} \rbrace \right) \\& - \sum_{\substack{\sigma_2,\dots,\sigma_{r-2}=\pm 1 \\ (\sigma_2,\dots,\sigma_{r-2})\neq (+1,\dots,+1)}} \left(\prod_{i=2}^{s-1}\sigma_i\right) \times \, \mathcal{I}_{(r-1)}\left(  \lbrace \mathbf{k}_L,...,\mathbf{k}_{M_{r-2}} \rbrace, p_{r-1} ; \lbrace p_1,\sigma_2p_2,..,\sigma_{r-2}p_{r-2} \rbrace \right) \,,
\end{split}
\end{equation}
where $\mathcal{I}_{(r-1)}$ can be consistently constructed using our algorithm. Now we must note that all possible combinations of terms involving products of the basic elements written in eq.\eqref{lists_elements} for an $(r-1)$-site correlator, with the sign flips of the internal energies, are present on the RHS of eq.\eqref{r-1-site successive disc}. Therefore, it is always possible to swap terms between the first and the second terms on the RHS of eq.\eqref{r-1-site successive disc} such that the two potential problems mentioned above get resolved. We present an elaborated demonstration of this in Appendix \ref{app induction_details r site}. 

Following the arguments presented above, we conclude that the successive discontinuity of the $r$-site correlator, as written in eq.\eqref{r site successive cut disc ansatz}, can be constructed in terms of the basic $1$-site elements listed in eq.\eqref{lists_elements} via the algorithm we mentioned above. 

Finally, after constructing $ \widehat{\mathcal{I}}_r( \lbrace \mathbf{k}_L,..., \mathbf{k}_R \rbrace ; \lbrace q_1, ..., q_{s-1} \rbrace )$ for an $r$-site correlator, we substitute it from eq.\eqref{r site successive cut disc ansatz} (with $s=r$), into eq.\eqref{r site disc+disp ansatz substituted}, and get the $r$-site correlator reconstructed as  
\begin{equation}
\label{r site disc+disp ansatz substituted_ssec}
\begin{split}
   & \mathcal{B}^{(r)}( \lbrace \mathbf{k}_L,..., \mathbf{k}_R \rbrace ; \lbrace p_1, ..., p_{r-1} \rbrace ) = \bigg(\frac{-1}{2\pi i}\bigg) ^{(r-1)}\times 2^{r-2} \times \\
   & \qquad \qquad \prod_{i = 1}^{r-1}\int\frac{q_{i}\,dq_{i}}{q_{i}^2-p_{i}^2} \,~ \mathcal{I}_r( \lbrace \mathbf{k}_L,..., \mathbf{k}_R \rbrace ; \lbrace q_1, ..., q_{r-1} \rbrace )\,.
\end{split}
\end{equation}
It is important to note that on the RHS above, we only get $\mathcal{I}_r$. It is easy to verify that the second term, including the summation over possible sign flips of $q_2, q_3, ..., q_{r-1}$, on the RHS of eq.\eqref{r site successive cut disc ansatz} contributes the same as the first term on the RHS, i.e., $\mathcal{I}_r$. This is due to the fact that in the dispersion integration, we are integrating over $-\infty < q_i < + \infty$, and a variable change of the form $q_i \to -q_i$ will make those integrations over the second term on the RHS of eq.\eqref{r site successive cut disc ansatz} same as that of the first term on the RHS. We have noticed similar behavior for the $3$-site correlator while going from eq.\eqref{3 site successive cut} to eq.\eqref{dispersion 3 site mt}. 

%%%%%%%%%%%%%%%%%%%%%%%%%%%%%%%%%%%%%%%%%%%%%%%%%%%%%%%%%%%%%%%%
\section{Diagrammatic dressing rules for cosmological correlators} \label{GeneralDressingRules}
%%%%%%%%%%%%%%%%%%%%%%%%%%%%%%%%%%%%%%%%%%%%%%%%%%%%%%%%%%%%%%%%%%%%%%%%%%%%%%%%%%%%
In this section, we present a set of diagrammatic rules that can be directly followed to reconstruct a tree-level $r$-site scalar correlator. These rules are justified by the dispersive reconstruction method involving the successive cut discontinuities of the correlator that we discussed in detail in the previous section \S\ref{Disp+Disc_reconstruction}, particularly for the $r$-site correlator in \S\ref{sec:rsite}. The diagrammatic rules are useful because they allow us to write down the correlator directly (in integral form), bypassing the complicated steps of implementing successive single cuts for all internal lines; see \S\ref{Disp+Disc_reconstruction}. Starting from a Feynman diagram in Minkowski spacetime, these rules prescribe how to \textit{`dress'} it with additional factors to obtain the expression for the $r$-site late-time de Sitter correlator. 

A similar set of dressing rules for cosmological correlators was written down recently in \cite{Chowdhury:2025ohm}; however, a different method, namely the Shadow formalism, was used. Moreover, the specifics and detailed expressions of the dressing rules depended on the interaction vertices in the corresponding theory, as well as on whether one considered IR-divergent or IR-finite cases. However, our diagrammatic rules are universally applicable to all such situations. Like \cite{Chowdhury:2025ohm}, we will restrict ourselves to scalar field theories in de Sitter, including both conformally coupled and massless scalars.

\subsection{General statement of the diagrammatic dressing rules} \label{genericrules}
To obtain the full result of an $r$-site tree level correlator, $ \mathcal{B}^{(r)}( \lbrace \mathbf{k}_L,..., \mathbf{k}_R \rbrace ; \lbrace p_1, ..., p_{r-1} \rbrace )$ we propose the following set of rules: \\

\noindent $\bullet$ \textbf{Rule-1: Color-coded Feynman diagrams:} For a tree-level late-time de Sitter correlator involving scalars, draw the corresponding Feynman diagram at tree level with a massless scalar scattering in the flat space, but without demanding energy conservation at any vertex. We will denote the external legs attached to a vertex with their corresponding energies $k_i$. The energy associated with an internal line between the $i$-th vertex and $(i+1)$-th vertex of the Feynman diagram will be denoted by $q_i$. Next, we encircle each vertex of the Feynman diagram using either blue or red. We consider all possible diagrams constructed using this color-coding rule, subject to the constraint that the number of red vertices must be even. Therefore, we will get 
    \[
    \binom{r}{0} + \binom{r}{2} + \binom{r}{4} + \ldots + \binom{r}{2\lfloor r/2 \rfloor} = 2^{r-1},
    \]
numbers of color-coded Feynman diagrams for an $r$-site Feynman diagram. Note that $\lfloor j/2\rfloor $ represents the greatest integer less than or equal to $\frac{j}{2}$. \\

\noindent $\bullet$ \textbf{Rule-2:} \textbf{Contribution from colored vertices:} The contributions from the two types of color-coded vertices are as follows:\\
\textbf{Blue vertex:} The contribution from the blue vertices will depend on whether they are encircling edge vertices or middle vertices. Every edge blue vertex will contribute the following
\begin{equation}\label{firsteq of diag blue}
\begin{split}
    & \text{Left-most vertex:}~~~\text{Disc}_{q_1} \mathcal{B}^{(1)}(\lbrace \mathbf{k}_L \rbrace, q_1) \, ,
    \\
    & \text{Right-most vertex:}~~~\text{Disc}_{q_{r-1}} \mathcal{B}^{(1)}(\lbrace \mathbf{k}_R \rbrace, q_{r-1}) \, .
\end{split}
\end{equation}
Likewise, any middle blue vertex will contribute the following
\begin{equation}\label{secondeq of diag blue}
\begin{split}
   \text{Middle vertices:} ~~~ & 2~\widetilde{\text{Disc}}_{q_j, q_{j+1}} \mathcal{B}^{(1)}(\lbrace \mathbf{k}_{M_j} \rbrace, q_j, q_{j+1})  \, , \quad \text{where}~~j=1,\cdots, r-2 \, .
\end{split}
\end{equation}
\textbf{Red vertex:} Similarly, the contribution from the red vertices will be different depending on whether they are edge vertices or middle vertices. From each red vertex on the edges we will get 
\begin{equation}\label{firsteq of diag red}
\begin{split}
    & \text{Left-most vertex:}~~~\widetilde{\text{Disc}}_{q_1} \widetilde{\mathcal{B}}^{(1)}(\lbrace \mathbf{k}_L \rbrace, q_1) \, ,
    \\
    & \text{Right-most vertex:}~~~\widetilde{\text{Disc}}_{q_{r-1}} \widetilde{\mathcal{B}}^{(1)}(\lbrace \mathbf{k}_R \rbrace, q_{r-1}) \, .
\end{split}
\end{equation}
Any middle red vertex will contribute
\begin{equation}\label{secondeq of diag red}
\begin{split}
   \text{Middle vertices:} ~~~ &2~\text{Disc}_{q_j, q_{j+1}} \widetilde{\mathcal{B}}^{(1)}(\lbrace \mathbf{k}_{M_j} \rbrace, q_j, q_{j+1})  \, , \quad \text{where}~~j=1,\cdots, r-2 \, .
\end{split}
\end{equation}
\\
\noindent $\bullet$ \textbf{Rule-3:} \textbf{Contribution from internal propagators:} For any Feynman propagator corresponding to the flat space Feynman diagrams, each internal line carrying energy $q_i$, will contribute, after being dressed, as follows 
\begin{equation} \label{rule-propagator}
    \frac{1}{2\pi i} \int_{-\infty}^{+\infty} \frac{q_i~d q_i}{2 P_{q_i}(\eta_0)} \frac{1}{p_i^2-q_i^2} \, .
\end{equation}
It should be noted that the Feynman propagator is associated with a $4$-momentum $p_i^{\mu} = (q_i,\vec{p}_i)$, such that we have only $3$-momentum conservation, and we will be denoting $p_i = |\vec{p}_i|$. \\

\noindent $\bullet$ \textbf{Rule-4:} \textbf{Relative sign rule:}
Among various possibilities, for each given distribution of red and blue vertices in a color-coded Feynman diagram, we will need to decide a relative sign factor as follows
\begin{equation}
\label{sign rule}
     (-1)^{\text{sum of positions of all red vertices}}.
\end{equation}
Note that our convention for counting the position of the vertex is from left to right.\\
\\
\noindent $\bullet$ \textbf{Rule-5: Summing all contributions:} Finally, we need to sum over all the color-coded Feynman diagrams, each with a different arrangement of red and blue vertices consistent with the rules prescribed above, after integrating with respect to energies of the internal lines.

%%%%%%%%%%%%%%%%%%%%%%%%%%%%%%%%%%%%%%%%%%%%%%%%%%%%
\subsection{Applying dressing rules: Conformally coupled scalars with $\lambda \phi^n$ interaction} \label{ssec- CC dressing rules}
%%%%%%%%%%%%%%%%%%%%%%%%%%%%%%%%%%%%%%%%%%%%%%%%%%%%
In this sub-section, we consider conformally coupled scalar fields with $\lambda \phi^n$ interaction and discuss the form of the diagrammatic dressing rules while the detail derivation is given in Appendix \ref{appCC_IRdiv}. These rules will allow us to directly write down the expression for the $r$-site cosmological correlators in such theories by dressing the corresponding flat space Feynman diagram.

We divide our analysis for conformally coupled scalars into two parts: the IR-convergent case (i.e., $\phi^n$ type of interaction with $n\geq 4$) and the IR-divergent case (e.g., $\phi^3$ interactions). Note that for the IR-divergent case, we need to regularize the integrations. For both cases, from eq.\eqref{rule-propagator}, we know that the contribution from the internal Feynman propagators (the $i$-th one) (with the dressing factors) is the same, i.e., 
\begin{equation}
    \frac{1}{2\pi i} \int_{-\infty}^{+\infty} \frac{q_i^2~d q_i}{H^2 \eta_0^2} \frac{1}{p_i^2 - q_i^2}
\end{equation}
However, the contribution from the vertices (both at the edges and in the middle) will differ between the two cases, as we demonstrate below.
%%%%%%%%%%%%%%%%%%%%%%%%%%%%%%%%%%%%%%%%%%%%%%%%%%%%%
\subsubsection{IR convergent cases: conformally coupled $\lambda \phi^n$  with $n \ge 4$}\label{CC IR convergent dressing}

In the following, we list down the contributions from the blue vertices appearing anywhere in the $r$-site diagram.
\begin{equation}\label{discleft}
 \begin{split}
     &\text{Left-most blue vertex:}~~~\text{Disc}_{q_1} \mathcal{B}^{(1)}( \lbrace \mathbf{k}_{L} \rbrace ,q_1) = \frac{-\lambda_1}{q_1 \lbrace 2k_i \rbrace^L} \mathcal{D}_{k_L} \left( \frac{2k_L}{q_1^2-k_L^2} \right) \,,\\&
     \text{Right-most blue vertex:}~~~\text{Disc}_{q_{r-1}} \mathcal{B}^{(1)}( \lbrace \mathbf{k}_{R} \rbrace ,q_{r-1})= \frac{-\lambda_1}{q_{r-1} \lbrace 2 k_i \rbrace^R} \mathcal{D}_{k_R} \left( \frac{2k_R}{q_{r-1}^2-k_R^2} \right)\,,\\&
     \text{Middle blue vertex:}~~~\\
     & \qquad 2~\widetilde{\text{Disc}}_{q_{j+1},q_j}\mathcal{B}^{(1)}(\lbrace \mathbf{k}_{M_{j}} \rbrace, q_{j+1}, q_{j})
    =  \frac{-\lambda_1}{q_j q_{j+1} \lbrace 2k_i \rbrace^{M_j}}  \mathcal{D}_{k_{M_j}} \left( \frac{2\,k_{M_j}}{(q_{j+1} + q_j )^2 - k_{M_j}^2} \right) \,,
 \end{split}   
\end{equation} 
where we have introduced the notation
 \begin{equation}
 \mathcal{D}_{k_L} \equiv \partial_{k_L}^{2m-4} \, ,
 \end{equation}
and the modified coupling $\lambda_1$ is defined as 
\begin{equation}
\begin{split}
   \lambda_1 = \lambda_e^c =& \lambda (-1)^m H^{4m-4} \eta_0^{2m} ~~~~~~~~~~~~~~~~~~~~~ (\text{for } n= \text{even}=2m) \, ,\\
   \lambda_1 = \lambda_o^{c} =& \lambda (2m-3)(-1)^{m} H^{4m-2} \eta_0^{2m+2}~~~~~~ (\text{for } n= \text{odd} = 2m+1) \, .
   \end{split}
\end{equation}
Similarly, for the red vertices, the contribution will be the following
\begin{equation} \label{red for odd}
 \begin{split}
     &\text{Left-most red vertex:}~~~\widetilde{\text{Disc}}_{q_1}  \widetilde{\mathcal{B}}^{(1)}( \lbrace \mathbf{k}_{L} \rbrace ,q_1) = \frac{-i \lambda_o^{\psi}}{q_1 \lbrace 2k_i \rbrace^L} \widetilde{\mathcal{D}}_{k_L} \left( \frac{2q_1}{q_1^2-k_L^2} \right)\,,\\&
     \text{Right-most red vertex:}~~~\widetilde{\text{Disc}}_{q_{r-1}}  \widetilde{\mathcal{B}}^{(1)}( \lbrace \mathbf{k}_{R} \rbrace ,q_{r-1}) 
= \frac{-i \lambda_o^{\psi}}{q_{r-1} \lbrace 2k_i \rbrace^L} \widetilde{\mathcal{D}}_{k_R} \left( \frac{2q_{r-1}}{q_{r-1}^2-k_R^2} \right)\,, \\
&\text{Middle red vertex:}~~~\\
     & \qquad 2 \, \text{Disc}_{q_j,q_{j+1}}\, \widetilde{\mathcal{B}}^{(1)} (\lbrace \mathbf{k}_{M_j} \rbrace, q_j,  q_{j+1})
    = \frac{-i \lambda_o^\psi}{q_j q_{j+1} \lbrace 2 k_i \rbrace^M} \widetilde{\mathcal{D}}_{k_{M_j}} \left(\frac{2( q_j +q_{j+1})}{(q_j +q_{j+1})^2 - k_{M_j}^2} \right) \, , 
 \end{split}   
\end{equation}
where, now, we have another notation $\widetilde{\mathcal{D}}_{k_L}$, and the modified coupling as
\begin{equation}
  \widetilde{\mathcal{D}}_{k_L} \equiv \partial_{k_L}^{2m-3} \, , \qquad \qquad \lambda_o^{\psi} = \lambda (-1)^{m} H^{4m-2} \eta_0^{2m+1} \quad \quad \, .
\end{equation}

%%%%%%%%%%%%%%%%%%%%%%%%%%%%%%%%%%%%%%%%%%%%%%%%%%%%%%
\subsubsection{IR divergent case: conformally coupled $\lambda\phi^3$ interaction }\label{phi3dressing}

In this section, we provide the explicit form of the vertex rules for the conformally coupled $\phi^3$ interaction. In Appendix \ref{detail vertex IR div cc} we have derived the contribution of each vertex explicitly. Here, we only list down the expression of each vertex type.

The blue vertices have the following expressions 
\begin{equation}\label{discleftphi3cc}
 \begin{split}
     &\text{Left-most blue vertex:}~~~\text{Disc}_{q_1} \mathcal{B}^{(1)}( \lbrace \mathbf{k}_{L} \rbrace ,q_1) = \frac{\pi \lambda H^2 \eta_0^3}{\lbrace 2 k_i \rbrace^L q_1} \,,\\&
     \text{Right-most blue vertex:}~~~\text{Disc}_{q_{r-1}} \mathcal{B}^{(1)}( \lbrace \mathbf{k}_{R} \rbrace ,q_{r-1})= \frac{\pi \lambda H^2 \eta_0^3}{\lbrace 2 k_i \rbrace^R q_{r-1}} \,,\\&
     \text{Middle blue vertex:}~~~2~\widetilde{\text{Disc}}_{q_{j+1},q_j}\mathcal{B}^{(1)}(k_{M_{j}}, q_{j+1}, q_{j})
    =  \frac{\pi \lambda H^2 \eta_0^3}{\lbrace 2 k_i \rbrace^{M_j} q_j q_{j+1}} \, .
 \end{split}   
\end{equation}
Similarly, for the red vertex, we will have following contributions
\begin{equation}
\label{discrightphi3cc}
 \begin{split}
     &\text{Left-most red vertex:}~~~\widetilde{\text{Disc}}_{q_1}  \widetilde{\mathcal{B}}^{(1)}( \lbrace \mathbf{k}_{L} \rbrace ,q_1) = \frac{i \lambda H^2 \eta_0^3}{\lbrace 2k_i \rbrace^L  q_1}  \int_{0}^{\infty} ds_L \frac{2q_1}{(s_L+k_L)^2-q_1^2}\,,\\&
     \text{Right-most red vertex:}~~~\\
     & \qquad \widetilde{\text{Disc}}_{q_{r-1}}  \widetilde{\mathcal{B}}^{(1)}( \lbrace \mathbf{k}_{R} \rbrace ,q_{r-1}) 
   = \frac{i \lambda H^2 \eta_0^3}{\lbrace 2k_i \rbrace^R  q_{r-1}}  \int_{0}^{\infty} ds_R \frac{2q_{r-1}}{(s_R+k_R)^2-q_{r-1}^2}\,,
\\&
     \text{Middle red vertex:}~~~ \\
     & \quad 2 ~ \text{Disc}_{q_j,q_{j+1}}\, \widetilde{\mathcal{B}}^{(1)} (\lbrace \mathbf{k}_{M_j} \rbrace, q_j,  q_{j+1})
    = \frac{i \lambda H^2 \eta_0^3}{\lbrace 2 k_i \rbrace^{M_j} q_{j} q_{j+1}}  \int_{0}^{\infty} ds_j \frac{2(q_j+q_{j+1})}{(s_j+k_{M_{j}})^2-(q_j+q_{j+1})^2} .
 \end{split}   
\end{equation}

%%%%%%%%%%%%%%%%%%%%%%%%%%%%%%%%%%%%%%%%%%%%%%%%%%%%%%%%%%%%%%%%%%%%%%
\subsection{Applying dressing rules: Massless scalars with $ \lambda\chi^3$ interaction}  \label{rules massless}
%(\textcolor{blue}{pending})}
In this subsection, we will consider massless scalar fields (denoted by $\chi$). Restricting ourselves to cases with a cubic interaction $ \lambda \chi^3$. We will derive explicit expressions for the diagrammatic dressing rules by following the generic rules of section~\ref{genericrules}. For detailed derivation of this section, we refer to Appendix \ref{app-massless-details}.

Firstly, the contribution from the internal lines (say the $i$-th one) of a corresponding Feynmann diagram, see eq.\eqref{rule-propagator}, after being dressed by the factors involving the power spectrum $P_{q_i}(\eta_0)$ for this case, will look like \footnote{We can drop the factor $\big(1 +  q_i^2 \eta_0^2 \big)^{-1} \sim 1$ in the denominator in eq.\eqref{massless prop dressing}, in the $\eta_0 \to 0$ limit.}
\begin{equation} \label{massless prop dressing}
   \frac{1}{2\pi i}\int_{-\infty}^{\infty} \frac{dq_i}{\vec{p_i}^2 - q_i^2}\times\bigg(\frac{q_i^4}{H^2\,\eta_0^2\big(1 + q_i^2\eta_0^2\big)}\bigg)\,.
\end{equation}
Now, we list the vertex contributions, starting with the blue vertices.
\begin{equation}\label{discbluemasslesschi3}
 \begin{split}
     &\text{Left-most blue vertex:}~~~\\
     &\quad \text{Disc}_{q_1} \mathcal{B}^{(1)}( \lbrace \mathbf{k}_{L} \rbrace ,q_1) = \lim_{\epsilon\rightarrow 0^+}\frac{\lambda\,H^2\,\cos(\pi\epsilon/2)}{\lbrace{2\,k_i^3\rbrace}^Lq_1^3}\int_{0}^{\infty}ds_L\frac{\mathcal{Q}_3(k_L,q_1,s_L)}{(s_L+k_{L})^2-q_1^2} \,,\\&
     \text{Right-most blue vertex:}~~~\\
     &\quad \text{Disc}_{q_{r-1}} \mathcal{B}^{(1)}( \lbrace \mathbf{k}_{R} \rbrace ,q_{r-1})= \lim_{\epsilon\rightarrow 0^+}\frac{\lambda\,H^2\,\cos(\pi\epsilon/2)}{\lbrace{2\,k_i^3\rbrace}^Rq_{r-1}^3}\int_{0}^{\infty}ds_R\frac{\mathcal{Q}_3(k_R,q_{r-1},s_R)}{(s_R+k_{R})^2-q_{r-1}^2} \,,\\&
     \text{Middle blue vertex:}~~~\\
     &\quad 2\,\widetilde{\text{Disc}}_{q_{j+1},q_j}\mathcal{B}^{(1)}(k_{M_{j}}, q_{j},  q_{j+1})
    =  \lim_{\epsilon\rightarrow 0^+}\frac{\lambda\,H^2\,\cos(\pi\epsilon/2)}{\lbrace{2\,k_i^3\rbrace}^{M_j}q_j^3q_{j+1}^3}\int_{0}^{\infty}ds_j\frac{\mathcal{Q}_3(k_{M_j},q_j+q_{j+1},s_j)}{(s_j+k_{M_j})^2-(q_j+q_{j+1})^2}  \,,
 \end{split}   
\end{equation}
where the function $\mathcal{Q}_3$ is defined in eq.\eqref{defQ3_app}.

Similarly, for the red vertex, we have the following contributions
\begin{equation}\label{discredmasslesschi3}
 \begin{split}
     &\text{Left-most red vertex:}~~~\\
     &\quad \widetilde{\text{Disc}}_{q_1}  \widetilde{\mathcal{B}}^{(1)}( \lbrace \mathbf{k}_{L} \rbrace ,q_1) = \lim_{\epsilon\rightarrow 0^+}\frac{i \lambda\,H^2\,\sin(\pi\epsilon/2)}{\lbrace{2\,k_i^3\rbrace}^Lq_1^3}\int_{0}^{\infty}ds_L\frac{\widetilde{\mathcal{Q}}_3(k_L,q_1,s_L)}{(s_L+k_{L})^2-q_1^2}\,,\\&
     \text{Right-most red vertex:}~~~\\
     &\quad \widetilde{\text{Disc}}_{q_{r-1}}  \widetilde{\mathcal{B}}^{(1)}( \lbrace \mathbf{k}_{R} \rbrace ,q_{r-1}) 
= \lim_{\epsilon\rightarrow 0^+}\frac{i \lambda\,H^2\,\sin(\pi\epsilon/2)}{\lbrace{2\,k_i^3\rbrace}^Rq_{r-1}^3}\int_{0}^{\infty}ds_R\frac{\widetilde{\mathcal{Q}}_3(k_R, q_{r-1}, s_R)}{(s_R+k_{R})^2-q_{r-1}^2}\,,
\\&
     \text{Middle red vertex:}~~~\\
     &\quad 2\,\text{Disc}_{q_j,q_{j+1}}\, \widetilde{\mathcal{B}}^{(1)} (\lbrace \mathbf{k}_{M_j} \rbrace, q_j,  q_{j+1})
    = \lim_{\epsilon\rightarrow 0^+}\frac{i \lambda\,H^2\,\sin(\pi\epsilon/2)}{\lbrace{2\,k_i^3\rbrace}^{M_j}q_j^3q_{j+1}^3}\int_{0}^{\infty}ds_j\frac{\widetilde{\mathcal{Q}}_3(k_{M_j}, q_{j}+q_{j+1},s_j)}{(s_j+k_{M_j})^2-(q_j+q_{j+1})^2}\,,
 \end{split}   
\end{equation}
where the function $\widetilde{\mathcal{Q}}_3$ is defined in eq.\eqref{defQ3tilde_app}.

Before concluding this section, let us compare our results with the ones in \cite{Chowdhury:2025ohm}. It is easy to check that eq.\eqref{discbluemasslesschi3} and eq.\eqref{discredmasslesschi3} are equivalent to eq.(3.34) and eq.(3.35), respectively, in \cite{Chowdhury:2025ohm}. From this comparison, we arrive at the following conclusions for the massless correlator dressing factor with $\lambda\chi^3$ interactions 
\begin{itemize}
\item The dashed auxiliary propagator attached to different vertices of the massless correlator in \cite{Chowdhury:2025ohm} (see table 3) corresponds to the Euclideanized version of our blue vertex contributions in eq. \eqref{discbluemasslesschi3}, up to an overall normalization. Similarly, the dotted auxiliary propagator for the massless correlator in \cite{Chowdhury:2025ohm} (see table 3) is equivalent (up to an overall normalization) to the Euclideanized version of our red vertex contributions in eq. \eqref{discredmasslesschi3}.
\item The Euclideanization of our result is implemented by replacing the cut momentum magnitude as $p_1 \rightarrow i p_1$.
\end{itemize}
For dressing factors of correlators with conformally coupled theories we learn that
\begin{itemize}
\item For conformally coupled $\lambda\phi^4$ interaction, only admits the dashed auxiliary propagator in \cite{Chowdhury:2025ohm} (see table 3). Likewise, in our rules, we have the corresponding Euclideanized version (up to an overall normalization) of only the blue vertex contributions in eq. \item For the conformally coupled $ \lambda \phi^3$ interaction, we see that the dashed auxiliary propagator attached to different vertices of the correlator in \cite{Chowdhury:2025ohm} (see table 3) corresponds to the Euclideanized version of our red vertex contributions in eq. \eqref{discleftphi3cc}, up to an overall normalization. And the dotted auxiliary propagator in \cite{Chowdhury:2025ohm} (see table 3) is identical to the Euclideanized version (up to an overall normalization) of our blue vertex contributions in eq. \eqref{discrightphi3cc}.
\end{itemize}
%%%%%%%%%%%%%%%%%%%%%%%%%%%%%%%%%%%%%%%%%%%%%%%%%%%%%%%%%%%%%%%%%%%%
%%%%%%%%%%%%%%%%%%%%%%%%%%%%%%%%%%%%%%%%%%%%%%%%%%%%%%%%%%%%%%%%%
\section{Explicit checks of the diagrammatic dressing rules}\label{checks}

In this section, we will pick up some examples of the conformally coupled and massless interacting scalar field theories and explicitly reconstruct cosmological correlators using the set of diagrammatic dressing rules that we have written down for these theories in \S\ref{GeneralDressingRules}. In each example, our starting point will be a Feynman diagram in Minkowski space-time. Our purpose will be to validate the effectiveness and the utility of reconstructing the correlators using these diagrammatic dressing rules. We want to convey that, rather than performing the momentum integrations in the dispersive reconstruction formula, calculating the correlator using the set of rules we proposed is more economical. However, we want to reiterate that our reconstruction method, based on dispersion integration over discontinuities, will not capture the contact-term pieces if they are present in the correlator. 

We will also work out the example of a $3$-site correlator in $\phi^5$ theory from the dressing rules, which, to the best of our knowledge, has not been calculated. This example also shows that one need not resort to the shadow formalism-based method of \cite{Chowdhury:2025ohm} to obtain the dressing rules. Our generic rules presented in \S\ref{genericrules} will apply straightforwardly, and the final expression for the correlator is consistent with the result obtained via the in-in method. 

%%%%%%%%%%%%%%%%%%%%%%%%%%%%%%%%%%%%%%%%%%%%%%%%%%%%%%%%%%%%%%%%%%%%
\subsection{Explicit checks for conformally coupled $\lambda\phi^4$ theory}

\begin{figure}[h]
    \centering
   \includegraphics[width=0.4\textwidth]{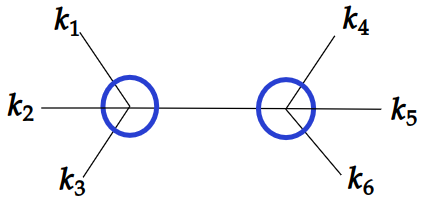}
    \caption{Diagrammatic representation of a $2$-site dS correlator for conformally coupled $\phi^4$}
    \label{dS phi4 dressing}
\end{figure}

For a $2$-site correlator in $\lambda\,\phi^4$ theory (an IR-convergent example), we will use our diagrammatic dressing rules from \S\ref{CC IR convergent dressing}. 
As mentioned earlier (see discussions around eq.\eqref{discleft1}), for $\lambda \phi^{n=\text{ even}}$ interactions involving conformally coupled scalars and $n \geq 4$, we obtain diagrams with only blue circles. The color-coded Feynman diagram for the $2$-site correlator with the $\lambda\phi^4$ interaction is shown in Fig.-\ref{dS phi4 dressing}.
\\

Using the diagrammatic rules from \S~\ref{GeneralDressingRules} and \S~\ref{CC IR convergent dressing}, we directly get the following
\begin{equation}
    \mathcal{B}^{(2)}_{\phi^4}(\lbrace \mathbf{k}_L, \mathbf{k}_R \rbrace; p) = \text{Fig.1} = \left( \frac{\lambda^2 H   ^6 \eta_0^6}{\prod_{i=1}^6 2k_i} \right)\bigg(\frac{1}{2\pi i}\bigg)\int_{-\infty}^{+\infty} \frac{ dq}{p^2 - q^2}  \bigg(\frac{2k_L}{q^2-k_L^2}\bigg)\bigg(\frac{2k_R}{q^2-k_R^2}\bigg)\,.
\end{equation}
\subsection{Explicit checks for conformally coupled $\lambda\phi^3$ theory}
In this section, we will reconstruct a cosmological correlator involving conformally coupled scalars with $\lambda\phi^3$ interaction in de Sitter, using our diagrammatic dressing rules. We will consider examples of $2$-site and $3$-site correlators and show that our result is consistent with that of \cite{Chowdhury:2025ohm}.
\subsubsection{$2$-site correlator}
\begin{figure}[h]
    \centering
   \includegraphics[width=0.6\textwidth]{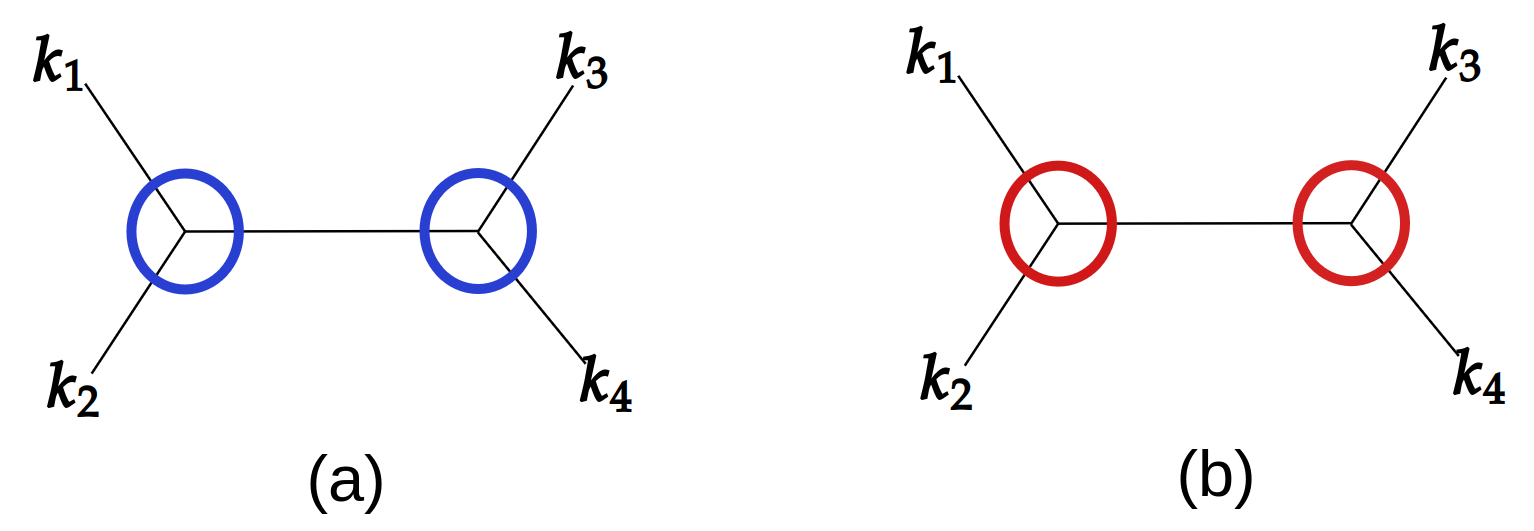}
    \caption{Diagrammatic representation of a $2$-site dS correlator for conformally coupled $\phi^3$}
    \label{dS phi3 dressing}
\end{figure}

First, using the first rule in \S~\ref{genericrules}, we need to draw all possible color-coded Feynman diagrams for the $2$-site correlator, which are given in Fig.-\ref{dS phi3 dressing}.
Next, using the propagator and vertex-dressing rules of \S~\ref{phi3dressing}, we obtain the contributions from the color-coded Feynman diagrams. Next, we consider the contributions from the internal line following the rule-4 of section \S~\ref{genericrules}. Then, assigning the appropriate relative sign factor corresponding to a diagram we get the following
\begin{equation}\label{B2 phi3 diagrams}
\begin{split}
    & \text{Fig.2(a)} = \left( \frac{\lambda^2 H    ^2 \eta_0^4}{\prod_{i=1}^4 2k_i} \right)\bigg(\frac{1}{2\pi i}\bigg)\int_{-\infty}^{+\infty} \frac{ dq}{p^2 - q^2}   \pi^2 \,,\\&
     \text{Fig.2(b)} = \left( \frac{\lambda^2 H    ^2 \eta_0^4}{\prod_{i=1}^4 2k_i} \right)\bigg(\frac{1}{2\pi i}\bigg)\int_{-\infty}^{+\infty} \frac{ dq}{p^2 - q^2}  \int_{0}^{\infty} ds_1  \int_{0}^{\infty} ds_2 ~\times\\
     & \qquad \qquad \qquad  \bigg(\frac{2q}{(s_1+k_1+k_2)^2-q^2} \frac{2q}{(s_2+k_3+k_4)^2-q^2} \bigg)\,.
\end{split}    
\end{equation}
Then, using rule-5 of \S~\ref{genericrules}, we add the two contributions to get 
\begin{equation}\label{B2 phi3 final}
\begin{split}
    & \mathcal{B}^{(2)}( k_1,k_2,k_3,k_4 ;p) 
     = \left( \frac{\lambda^2 H    ^2 \eta_0^4}{\prod_{i=1}^4 2k_i} \right)\bigg(\frac{1}{2\pi i}\bigg)\int_{-\infty}^{+\infty} \frac{ dq}{p^2 - q^2}~ \times \\
     & \qquad \qquad \qquad  \bigg( \pi^2 + \int_{0}^{\infty} ds_1  \int_{0}^{\infty} ds_2 \frac{2q}{(s_1+k_1+k_2)^2-q^2} \frac{2q}{(s_2+k_3+k_4)^2-q^2} \bigg)\,.
\end{split}    
\end{equation}

\subsubsection{$3$-site correlator}
\begin{figure}[h]
    \centering
   \includegraphics[width=0.6 \textwidth]{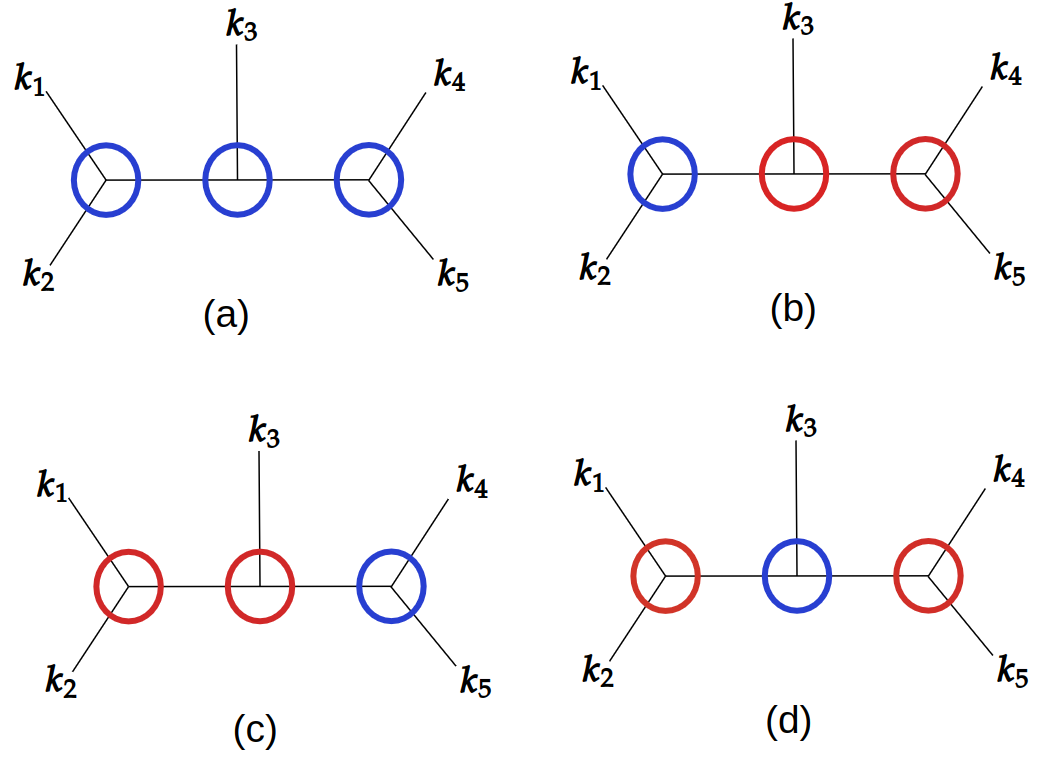}
    \caption{Diagrammatic representation of a $3$-site dS correlator for conformally coupled $\phi^3$}
    \label{dS 3site phi3 dressing}
\end{figure}

Similarly, using the rules of \S~\ref{genericrules}, we have the following color-coded Feynman diagrams, as in Fig.-\ref{dS 3site phi3 dressing}, corresponding to a $3$-site correlator. 

Using the vertex dressing factors of \S~\ref{phi3dressing} and using the rule of \S~\ref{genericrules}, we have the following contributions from each diagrams
\begin{equation}
\label{term1phi3}
\begin{split}
  \text{Fig.3(a)}= \frac{\pi^3 \lambda^3 H^2 \eta_0^5}{\prod_{i=1}^5 2k_i}
 \bigg(\frac{1}{2\pi i}\bigg)^2
 \int_{-\infty}^{\infty}\frac{dq_1}{p_1^2-q_1^2}\,
 \int_{-\infty}^{\infty}\frac{dq_2}{p_2^2-q_2^2}
\end{split}
\end{equation}

\begin{equation}
\label{term2phi3}
\begin{split}
 \text{Fig. 3(b)} =\;&
 \frac{-4\pi \lambda^3 H^2 \eta_0^5}{\prod_{i=1}^5 2k_i}
 \bigg(\frac{1}{2\pi i}\bigg)^2
 \int_{0}^{\infty} ds_2
 \int_{0}^{\infty} ds_3
 \int_{-\infty}^{\infty}\frac{dq_1}{p_1^2-q_1^2}\,
 \int_{-\infty}^{\infty}\frac{dq_2}{p_2^2-q_2^2} \\
 & \times
 \frac{ (q_1+q_2)}{(s_2+k_3)^2-(q_1+q_2)^2}
 \frac{q_2}{(s_3+k_4+k_5)^2-q_2^2}
\end{split}
\end{equation}

\begin{equation}
\begin{split}
 \text{Fig.3(c)} =\;&
 \frac{-4\pi\lambda^3 H^2 \eta_0^5}{\prod_{i=1}^5 2k_i}
 \bigg(\frac{1}{2\pi i}\bigg)^2
 \int_{0}^{\infty} ds_1
 \int_{0}^{\infty} ds_2 
 \int_{-\infty}^{\infty}\frac{dq_1}{p_1^2-q_1^2}\,
 \int_{-\infty}^{\infty}\frac{dq_2}{p_2^2-q_2^2} \\
 & \times
 \frac{ q_1}{(s_1+k_1+k_2)^2-q_1^2}
 \frac{(q_2+q_1)}{(s_2+k_3)^2-(q_2+q_1)^2}
\end{split}
\end{equation}

\begin{equation}
\begin{split}
 \text{Fig. 3(d)} =\;&
 \frac{4\pi\lambda^3 H^2 \eta_0^5}{\prod_{i=1}^5 2k_i}
 \bigg(\frac{1}{2\pi i}\bigg)^2
 \int_{0}^{\infty} ds_1
 \int_{0}^{\infty} ds_3
 \int_{-\infty}^{\infty}\frac{dq_1}{p_1^2-q_1^2}\,
 \int_{-\infty}^{\infty}\frac{dq_2}{p_2^2-q_2^2} \\
 & \times
 \frac{ q_1}{(s_1+k_1+k_2)^2-q_1^2}
 \frac{q_2}{(s_3+k_4+k_5)^2-q_2^2}
\end{split}
\end{equation}
The final answer, therefore, will be a sum of all four contributions written above. 

\subsection{Explicit checks for massless $\lambda\chi^3$ theory}
In this section, we will reconstruct the $2$-site correlator for massless scalar fields (denoted by $\chi$) with the $\chi^3$ interaction using our diagrammatic dressing rules. 
Following the rules laid out in \S\ref{genericrules}, there will be two color-coded Feynman diagrams for a $2$-site massless correlator, same as shown in Fig.-\ref{dS phi3 dressing}; however, the diagrammatic dressing rules will differ.  Using the dressing factors of \S\ref{rules massless} and using the rules of \S\ref{genericrules}, we obtain the following contributions from each of the diagrams 
\begin{equation}
\begin{split}
    \text{Fig. 4(a)} =&\lim_{\epsilon\rightarrow 0^+}\frac{\lambda^2\,H^4\,\cos^2(\pi\epsilon/2)}{\prod_{i = 1}^{4}4k_i^3}\bigg(\frac{1}{2\pi i}\bigg)\int_{0}^{\infty}ds_L\int_{0}^{\infty}ds_R\int_{-\infty}^{\infty} \frac{dq_1}{p_1^2 - q_1^2}\\
    &\times\bigg(\frac{1}{H^2\,\eta_0^2\big(1 + q_1^2\eta_0^2\big)q_1^4}\bigg) \bigg(\frac{\mathcal{Q}_3(k_1+k_2, q_1, s_L)}{(s_L+k_1+k_2)^2-q_1^2}\frac{\mathcal{Q}_3(k_3+k_4, q_1, s_R)}{(s_R+k_3+k_4)^2-q_1^2}\bigg) \, ,
\end{split}    
\end{equation}
\begin{equation}
\begin{split}
    \text{Fig. 4(b)} =&\lim_{\epsilon\rightarrow 0^+}\frac{-\lambda^2\,H^4\,\sin^2(\pi\epsilon/2)}{\prod_{i = 1}^{4}4k_i^3}\bigg(\frac{1}{2\pi i}\bigg)\int_{0}^{\infty}ds_L\int_{0}^{\infty}ds_R\int_{-\infty}^{\infty} \frac{dq_1}{p_1^2 - q_1^2}\\&\times\bigg(\frac{1}{H^2\,\eta_0^2\big(1 + q_1^2\eta_0^2\big)q_1^4}\bigg)\bigg(\frac{\widetilde{\mathcal{Q}}_3(k_1+k_2, q_1, s_L)}{(s_L+k_1+k_2)^2-q_1^2}\frac{\widetilde{\mathcal{Q}}_3(k_3+k_4, q_1, s_R)}{(s_R+k_3+k_4)^2-q_1^2}\bigg) \, .
\end{split}    
\end{equation}
Hence, the full correlator will be given by the combination of the contributions written above.

%%%%%%%%%%%%%%%%%%%%%%%%%%%%%%%%%%%%%%%%%%%%%%%%%%%%%%%%%%%%%%%%%%%%%%%%%
\subsection{$3$-site correlator in $\lambda \phi^5$ theory from the dressing rules}\label{check 3-site phi5}

We now turn to the example of a $3$-site correlator with $\phi^5$ conformally coupled interaction. Following the diagrammatic rules given in \S\ref{CC IR convergent dressing}, we work out this correlator, which is not available in the literature. We will establish the consistency of the rules by comparing the final result for the correlator with the expression that one can derive via the standard in-in principle. 

If we draw the Feynman diagram of the flat space scattering amplitude involving three $\phi^5$ interaction vertices in the bulk, there will be eleven external legs. We can label the external legs by the magnitude of three momenta, $k_1, \, k_2, \cdots, \,  k_{11}$. Hence, the sum of the external energies at three vertices will be given by $k_L=k_1+\cdots+k_4, \, k_M= k_5+\cdots k_7, \, k_R = k_8 + \cdots + k_{11}$, respectively, from left to right. The two internal legs at tree-level will be labeled by the magnitude of three momenta $p_1, p_2$, and their corresponding energies $q_1, q_2$.

\begin{figure}[h]
    \centering
   \includegraphics[width=0.95\textwidth]{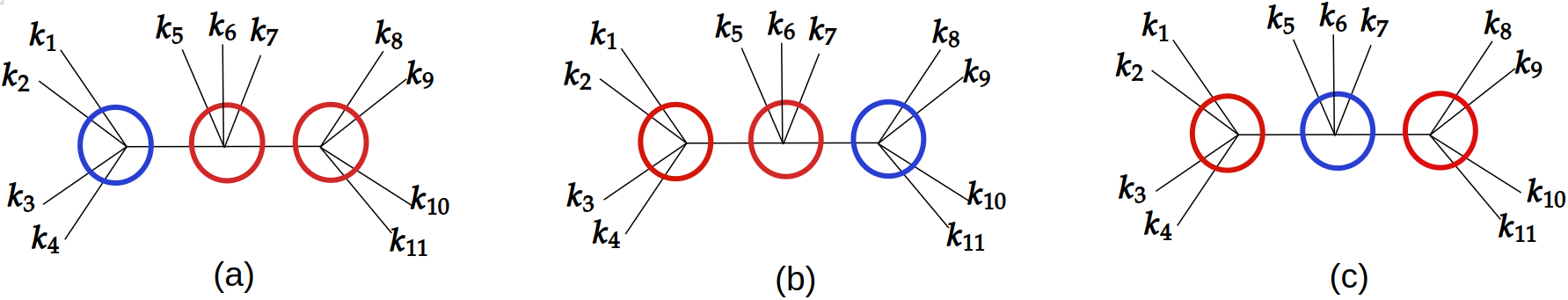}
    \caption{Diagrammatic representation of a $3$-site dS correlator for conformally coupled $\phi^5$}
    \label{dS phi5 dressing}
\end{figure}

We now apply the vertex rule to the Feynman diagram. Given that the chosen interaction is $\phi^5$,  we know that to extract the leading $\eta_0$ contribution to the correlator, we need to consider $3$ diagrams since there are $3$ vertices \footnote{Since the total number of diagrams is given by $2^{3-1}=4$. However, the diagram with all blue vertices will be subleading in $\eta_0$. Hence, we will only keep the other three diagrams that contribute to the leading $\eta_0$ case.}. We have also seen that each of the three diagrams will contain $2$ red vertices and $1$ blue vertex. Therefore, with the vertex contribution given in \S\ref{CC IR convergent dressing} along with the contribution of each internal line, we obtain the following three expressions corresponding to the diagrams shown in Fig.\ref{dS phi5 dressing}. 
\begin{equation} \label{phi5Term1}
\begin{split}
     & \text{Fig-5(a)}~~= ~ \frac{(-1)^{2+3} (-\lambda_o^c)(- i \lambda_o^\psi)^2}{\left( \prod_{i=1}^{11} 2 k_i\right)\left( H^2\eta_0^2\right)^2}  \widetilde{\mathcal{D}}_{k_M} \widetilde{\mathcal{D}}_{k_R} \left( \frac{1}{2 \pi i} \int_{-\infty}^{+\infty} \frac{dq_1}{p_1^2 - q_1^2} \right) \left( \frac{1}{2 \pi i} \int_{-\infty}^{+\infty} \frac{dq_2}{p_2^2 - q_2^2} \right) \times
     \\
     & ~~~~~~~~~~~~~~~~~~~~~~~~~~~~~~~~~~~~~~~~~~~~~~~~~~~~~~~\left( \frac{2 k_L}{q_1^2 - k_L^2} \frac{2 (q_1+q_2)}{(q_1+q_2)^2 - k_M^2} \frac{2 q_2}{q_2^2 - k_R^2} \right) \, ,
\end{split}
\end{equation}
\begin{equation}\label{phi5Term2}
\begin{split}
     & \text{Fig-5(b)}~~\equiv ~\frac{(-1)^{1+2} (- i \lambda_o^\psi)^2(-\lambda_o^c)}{\left( \prod_{i=1}^{11} 2 k_i\right)\left( H^2\eta_0^2\right)^2} \widetilde{\mathcal{D}}_{k_L} \widetilde{\mathcal{D}}_{k_M} \left( \frac{1}{2 \pi i} \int_{-\infty}^{+\infty} \frac{dq_1}{p_1^2 - q_1^2} \right) \left( \frac{1}{2 \pi i} \int_{-\infty}^{+\infty} \frac{dq_2}{p_2^2 - q_2^2} \right) \times 
     \\
     & ~~~~~~~~~~~~~~~~~~~~~~~~~~~~~~~~~~~~~~~~~~~~~~~~~~~~~~~\left( \frac{2 q_1}{q_1^2 - k_L^2} \frac{2 (q_1+q_2)}{(q_1+q_2)^2 - k_M^2} \frac{2 k_R}{q_2^2 - k_R^2} \right) \, ,
\end{split}
\end{equation}
\begin{equation}\label{phi5Term3}
\begin{split}
     & \text{Fig-5(c)}~~\equiv ~ \frac{(-1)^{1+3} (-i \lambda_o^\psi)(-\lambda_e^c) (- i \lambda_o^\psi)}{\left( \prod_{i=1}^{11} 2 k_i\right)\left( H^2\eta_0^2\right)^2} \widetilde{\mathcal{D}}_{k_L} \widetilde{\mathcal{D}}_{k_R} \left( \frac{1}{2 \pi i} \int_{-\infty}^{+\infty} \frac{dq_1}{p_1^2 - q_1^2} \right) \left( \frac{1}{2 \pi i} \int_{-\infty}^{+\infty} \frac{dq_2}{p_2^2 - q_2^2} \right) \times  \\
     & ~~~~~~~~~~~~~~~~~~~~~~~~~~~~~~~~~~~~~~~~~~~~~~~~~~~~~~~~~~~~~~~~ 
     \left( \frac{2 q_1}{q_1^2 - k_L^2} \frac{2 k_M}{(q_1+q_2)^2 - k_M^2} \frac{2 q_2}{q_2^2 - k_R^2} \right) \, .
\end{split}
\end{equation}
Note that in eq.\eqref{phi5Term1}, eq.\eqref{phi5Term2}, and eq.\eqref{phi5Term3}, we have kept track of explicit factors of $(-1)$, which is decided by the position of the red vertices, which is again consistent with our rules. 
Finally, using the residue theorem, we can evaluate the integrals in eq.\eqref{phi5Term1}, eq.\eqref{phi5Term2}, and eq.\eqref{phi5Term3}, then act with the differential operators on the resulting integrals.  Next, by summing up the three pieces, we obtain the full correlator, which agrees with the result obtained from the standard in-in formalism. This comparison is worked out in the Mathematica file \texttt{`dS_correlators.nb'}, available at - 
\href{https://github.com/DKaran98/Single-cut-to-Dressing}{(link)} \footnote{The file can be accessed at \texttt{`github.com/DKaran98/Single-cut-to-Dressing'}. The file contains the following calculations: $2$-sites with $\phi^4$, $\phi^5$, $\phi^6$ and $3$-sites with $\phi^5$, $\phi^6$ interactions.}.

%%%%%%%%%%%%%%%%%%%%%%%%%%%%%%%%%%%%%%%%%%%%%%%%%%%%%%%%%%%%%%%%
%%%%%%%%%%%%%%%%%%%%%%%%%%%%%%%%%%%%%%%%%%%%%%%%%%%%%%%%%%%%%%%%
\section{Conclusions and outlook} \label{conclusions}

 In this section, we conclude by summarising our main results and also mentioning a few future directions. Let us first briefly comment on the key highlights of our analysis. 
 
 Our primary focus has been to demonstrate that the dispersion formula acting on the discontinuities of the cosmological correlators can reconstruct them up to contact-term ambiguities. We focused on polynomial scalar interactions and restricted ourselves to tree-level diagrams. In \S\ref{Disp+Disc_reconstruction}, we have formulated an integral representation of the cosmological correlator in terms of the lowest point discontinuity data. In \S\ref{sec:rsite}, we have provided an algorithm to reconstruct a generic $r$-site correlator from its successive single cut discontinuities. The basic idea of reconstruction via dispersion integration across discontinuities is not new and has been discussed in the literature before, see \cite{Meltzer:2021zin}. However, in our analysis, we reconstructed cosmological correlators at tree level, focusing on individual channels, with an arbitrary number of vertices, which, in our opinion, represents a novel advancement in the analytical study of cosmological correlators. 
  
 Based on our general analysis, in \S\ref{GeneralDressingRules}, we have constructed a set of diagrammatic rules that, when applied to a Feynman diagram, produce the corresponding cosmological correlator. These diagrammatic dressing rules are similar to the rules derived in \cite{Chowdhury:2025ohm}, derived using a completely different method. We could present these rules more generally, without relying on the type of theory or interaction one is interested in. We derived our rules based on the discontinuity relations for cosmological correlators developed in \cite{Das:2025qsh}, following unitarity of in-in propagators. Hence, by iteratively integrating over successive discontinuities, the physical constraints of unitarity and analyticity are manifested in the color-coded Feynman diagrams through our dressing rules. Also, through our dressing rules for the Feynman diagram, the discontinuity of the corresponding correlator gets related to the off-shell intermediate exchange becoming on-shell, see related discussions in, e.g., \cite{Ansari:2026xkm, Chowdhury:2026upp}. 
 
 The diagrammatic rules we studied apply to conformally coupled and massless theories, including both IR-divergent and IR-convergent cases, which are discussed in \S\ref{ssec- CC dressing rules} and \S\ref{rules massless}. Furthermore, the applicability of our rules is verified in several examples in \S\ref{checks}. 

 In \cite{Das:2025qsh}, the discontinuity relations for correlators involved particular auxiliary objects $\widetilde{\mathcal{B}}$ which are not exactly cosmological correlators and are related to the imaginary pieces of the wave-function coefficients. It is worth noting that in our dressing rules these $\widetilde{\mathcal{B}}$ play an important role. Moreover, the role of these objects can be traced back to the corresponding rules in \cite{Chowdhury:2025ohm}. Although we do not yet have a deeper understanding of these auxiliary objects, the fact that they are crucially involved in the discontinuity of cosmological correlators is further supported by the dispersion-method-based analysis in this paper.
 We needed to employ successive single-cut discontinuities in the cosmological correlators, which differ from the simultaneous discontinuity discussed, e.g., in \cite{Melville:2021lst, Ema:2024hkj}. The specific discontinuity operation, as defined in eq.\eqref{defdisc}, and the use of successive discontinuity relations for cosmological correlators are the most essential technical novelties in our analysis. 

 The reconstruction of the cosmological correlator we have shown in this paper can be trivially applied to the flat-space correlator by first performing successive single cuts and then using the dispersion relation. In this case, the situation simplifies drastically. In particular, in the case of the flat space cosmological correlator, one can show that for any polynomial ($\phi^n$), massless interaction, it is sufficient to only consider the blue colored vertex (without any differential operator with respect to the total external energies) for all the sites in the bulk, corresponding to any conformally coupled convergent polynomial.

 Next, we will discuss some future work motivated by our investigations in this paper. Firstly, we would like to further explore the implications of our discontinuity operation. For example, one could see whether our discontinuity operation, when applied to the wave-function coefficients, reveals something interesting. 

 In this study, we derived the dressing rules for cosmological correlators by applying cutting rules to conformally coupled and massless scalar theories with polynomial interactions. A natural next step is to extend this framework to spinning correlators, such as those involving gluons and gravitons. This requires first establishing the successive single-cut discontinuity relations for spinning correlators, after which the corresponding dressing rules can be systematically extracted using dispersion relations. By expressing spinning cosmological correlators in terms of flat-space–like Feynman diagrams, one can directly inherit structures such as color–kinematics duality, providing a powerful organizing principle for their analytic properties. 

 Another important generalization concerns inflationary correlators, where one must account for the breaking of exact conformal symmetry due to the inflaton's slow roll. It would be interesting to study how inflationary correlators could retain well-defined analytic structures and discontinuities, allowing the cutting-and-dressing framework to be extended beyond exact de Sitter invariance. Deriving the single-cut relations for spinning fields in inflationary backgrounds would therefore yield dressing rules for inflationary correlators, elucidating how flat-space amplitude structures — such as color-kinematics duality — manifest themselves in inflationary observables. 

It would be interesting to see if we can understand the significance of the auxiliary objects and the $\widetilde{\text{Disc}}$ operation from a different perspective, given that now we have explored the connections between two different derivations of the dressing rules - one from the discontinuity relations for cosmological correlators, and the other from the shadow formalism \cite{Chowdhury:2025ohm}. It will be worth exploring whether the shadow formalism can shed light on the curious workings of the discontinuity relations involving the imaginary parts of the wave-function coefficients, which appear related to the auxiliary counterparts of the correlator. 

Usually, causality constraints are connected to the analyticity properties of the $S$-matrix. It would be interesting to understand more precisely how causality constraints manifest themselves in the cosmological setup. It is better understood in terms of the analytic structure of the wave-function coefficients \cite{AguiSalcedo:2023nds}. In our analysis, unitarity and analyticity (in terms of the in-in propagators) are used through the discontinuity and the dispersion relations. It would be interesting to explore the implications of causality within the framework of the dispersion and discontinuity of our analysis for cosmological correlators; see, e.g., \cite{Chowdhury:2026upp} for recent explorations in this direction.

%%%%%%%%%%%%%%%%%%%%%%%%%%%%%%%%%%
\section*{Acknowledgements}

We would like to thank Chandramouli Chowdhury for the various insightful discussions and for asking interesting questions. We are especially thankful to Suvrat Raju. We are also thankful to Arjun Bagchi, Jyotirmoy Bhattacharya, Joydeep Chakrabortty, Sabyasachi Chakraborty, Diptarka Das, Sanmay Ganguly, Sachin Jain, Apratim Kaviraj, Suvrat Raju, Ashish Shukla, and Sandip Trivedi for valuable discussions. SD would like to thank Anuj Gupta, Mudit Kumar, and Aryabrat Mahapatra for various discussions and support during this work. We thank the authors of \cite{Chowdhury:2026upp} for sharing an upcoming draft with us. DK and BK would like to thank the organizers of the 20th Asian Winter School (AWS) on Strings, Particles and Cosmology 2026 for the hospitality. DK and BK would like to thank Simon Caron-Huot, Zi-Xun Huang, Enrico Pajer, Pradipta Shankar Pathak, and Akashdeep Roy for many insightful discussions on this subject and related areas during AWS 2026. DK and BK are supported by the Department of Atomic Energy, Government of India, under project no. RTI4001. NK would like to acknowledge the warm hospitality from TIFR, Mumbai, and The Lodha Mathematical Sciences Institute (LMSI) during an academic visit while this work was in progress. NK acknowledges support from a recently concluded MATRICS research grant (MTR/2022/000794) from the Anusandhan National Research Foundation (ANRF), India.

%%%%%%%%%%%%%%%%%%%%%%%%%%%%%%%%%%%%%%%%%%%%%%%%%%%%%%%%%%%%%%%%%%%%%%%%%%%%%%%%%%%%%%%%%%%%%%%%%%%%
\appendix
%%%%%%%%%%%%%%%%%%%%%%%%%%%%%%%%%%%%%%%%%%%%%%%%%%%%%%%%%%%%%%%%%%%%%%%%%%%%%%%%%%%%%%%%%%%%%%%%%%%%%%%%%%
%%%%%%%%%%%%%%%%%%%%%%%%%%%%%%%%%%%

\section{Details on the discontinuity and the dispersion formula} \label{App_disc_disp}
In this appendix, we will review the dispersive integral technique to reconstruct a function (up to contact terms or regular pieces) from its discontinuity along a branch-cut. Let $f(z)$ be analytic on the complex plane, with its only non-analyticity arising from a branch cut along the positive real $z$-axis. Then $f(z)$ can be written as 
\begin{equation}
\label{cauchyint}
    f(z_0) = \frac{1}{2\pi i}\oint_{C_{z_0}} \frac{dz}{z-z_0}\,f(z)\,,
\end{equation}
where, $z_0$ is any point in complex z plane and $C_{z_0}$ is the contour encircling $z =z_0$ anti-clockwise.

\begin{figure}[h]
    \centering
   \includegraphics[width=0.85\textwidth]{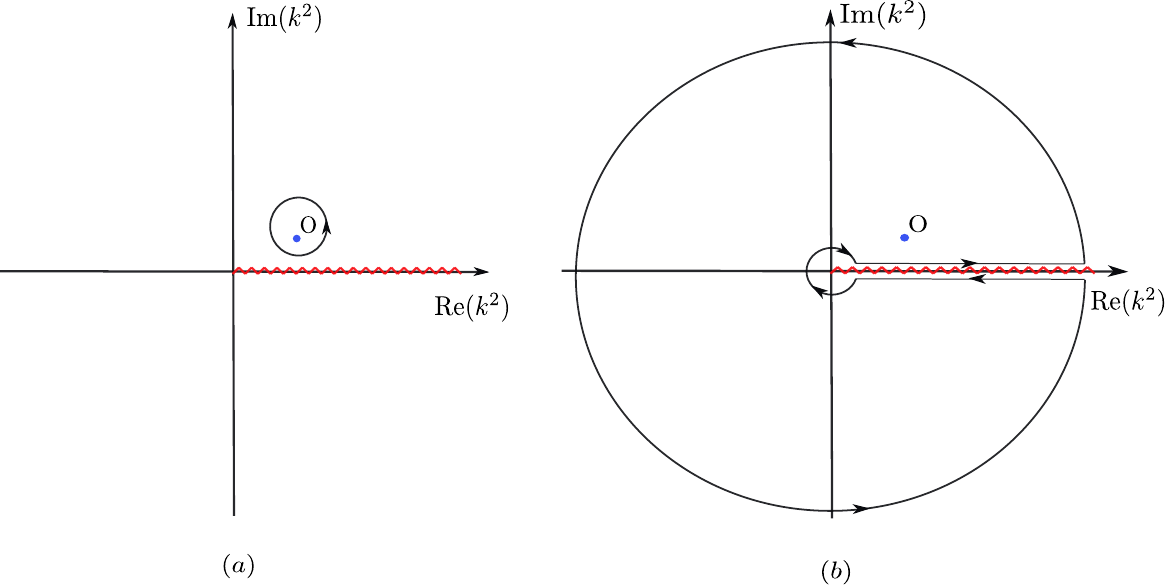}
    \caption{The left and right contours are related by contour deformation. The point $O$ is $k^2 = p^2 + i\epsilon$.}
    \label{bothcontour}
\end{figure}

Due to the presence of a branch cut along the positive real $z$-axis, we define the following
\[\lim_{\epsilon\rightarrow0^+}\,f(z+i\epsilon) = f(z)\,, \]
 for $z$ lying on the positive real axis.
Now, we can define $f(z_0)$ for real and positive values of $z_0$ as 
\begin{equation}
\label{cauchyint2}
    f(z_0) = \lim_{\epsilon\rightarrow0^+}\,f(z_0+i\epsilon)=  
    \lim_{\epsilon\rightarrow0^+}\frac{1}{2\pi i}\oint_{C_{z_{0}}} \frac{dz}{z-z_0-i\epsilon}\,f(z)\,,
\end{equation}
where, $\epsilon\rightarrow0^+$ means while taking the limit we keep $\epsilon \ge 0$. From now on, we will suppress this limit, while its meaning remains implicit. Next, we assume the following fall-off condition for the function $f(z)$ along the arc at large $|z|$, i.e., $\lim_{|z|\rightarrow \infty}\,f(z) = 0$. This condition allows us to deform our contour in eq.\eqref{cauchyint} as 
\begin{equation}
       f(z_0) = \lim_{\epsilon\rightarrow0^+}\,f(z_0+i\epsilon) = \,\frac{1}{2\pi i}\int_{0}^{\infty} \frac{dz}{z-z_0-i\epsilon}\,\textbf{Disc}_{z}f(z)\,,
\end{equation}
where, $\textbf{Disc}_{z}f(z)$ is the discontinuity of the function $f(z)$ along the branch cut, defined as
\begin{equation}
    \textbf{Disc}_{z}f(z) = \lim_{\epsilon\rightarrow 0^+}\,\bigg[f(z+i\epsilon)-f(z-i\epsilon)\bigg]\,.
\end{equation}
In our case, setting $z_0 = p^2$ and interpreting $f(p^2)$ as the cosmological correlator,
we find that the above properties hold in the following form 
\begin{itemize}
    \item For us, the cosmological correlators are functions of the magnitudes of the exchanged momenta. As functions of $k^2$ \footnote{Note, $k^2 = \vec{k}\cdot \vec{k}$. By momentum conservation at the  vertices, $\vec{k} = \sum_{\substack{\text{external}\\\text{lines}}} \vec{k}_i$.}, they have a branch cut along the positive $\mathbb{R}e(k^2)$ axis and are analytic everywhere else.
    \item Cosmological correlators vanish when the magnitude of the exchanged momentum becomes large. \footnote{This condition on the correlator follows directly from the fall-off behavior of the bulk-to-bulk propagator with Bunch-Davies initial conditions as $|k^2|\to \infty$.} In terms of $f(k^2)$, this means
\[
\lim_{|k^2|\to \infty} f(k^2) = 0.
\]
\end{itemize}
These conditions allow us to deform the contour as shown in Fig.-\ref{bothcontour}. Using this deformation, we can then write the following
\begin{equation}
\label{dispersionmomentumsp}
f(p^2) =  \frac{1}{2\pi i} \int_{0}^{\infty} \frac{dk^2}{k^2 - p^2 - i\epsilon} \, \mathbf{Disc}_{k^2} f(k^2)\,.
\end{equation}
As mentioned earlier, correlators depend only on the magnitudes of momenta. Therefore, it is natural to change variables and redefine the function as follows
\[
f(k^2) = \tilde{f}(\sqrt{k^2})\,.
\]

Here, we choose the branch cut of the square root along the positive real $k$-axis, which leads to the following simplifications:
\begin{itemize}
    \item For $k^2>0$, in the upper half of the complex $k^2$ plane:
    \[
    \lim_{\epsilon\to 0^+} f(k^2 + i\epsilon) 
    = \lim_{\epsilon\to 0^+} \tilde{f}(\sqrt{k^2 + i\epsilon})
    = \lim_{\epsilon\to 0^+} \tilde{f}\Big(k + \frac{i\epsilon}{2k}\Big)\,.
    \]
    \item For $k^2>0$, in the lower half of the complex $k^2$ plane:
    \[
    \lim_{\epsilon\to 0^+} f(k^2 - i\epsilon) 
    = \lim_{\epsilon\to 0^+} \tilde{f}(\sqrt{k^2 - i\epsilon})
    = \lim_{\epsilon\to 0^+} \tilde{f}\Big(-k + \frac{i\epsilon}{2k}\Big)\,.
    \] 
\end{itemize}
Using the above relations, we can directly write
\begin{equation}
\label{twodiscrel}
\begin{split}
\mathbf{Disc}_{k^2} f(k^2) 
&= \lim_{\epsilon \to 0^+} \big[f(k^2 + i\epsilon) - f(k^2 - i\epsilon)\big] = \lim_{\epsilon \to 0^+} \big[\tilde{f}(k + \tfrac{i\epsilon}{2k}) - \tilde{f}(-k + \tfrac{i\epsilon}{2k})\big]\,.
\end{split}
\end{equation}
In the region where $k \gg \epsilon$, we can drop the $\mathcal{O}(\epsilon)$ term from eq. \eqref{twodiscrel}. Our choice of contour in the $k^2$ plane is shown in Fig.~\ref{bothcontour} (the right one), where we make a circular deformation of radius $\delta$ such that $\epsilon \ll \delta$ (with both much smaller than unity). This allows us to write eq. \eqref{twodiscrel} as follows
\begin{equation}
\mathbf{Disc}_{k^2} f(k^2) = \text{Disc}_{k} \tilde{f}(k)\,.
\end{equation}
Now we can see how our discontinuity with respect to $k$, i.e., $\text{Disc}_k$, is related to the usual discontinuity with respect to $k^2$, i.e., $\mathbf{Disc}_{k^2}$ \footnote{We want to thank Chandramouli Chowdhury for having useful discussions with us on this issue.}. 

One can check that the integral contributions from the large and small circular arcs in Fig.~\ref{bothcontour} (the right one) vanish for cosmological correlators. We can now use eq.~\eqref{twodiscrel} in eq.~\eqref{dispersionmomentumsp} to get
\begin{equation}
\label{dispersionfromfeynman}
\tilde{f}(p) = \frac{1}{2\pi i} \int_{0}^{\infty} \frac{2k \, dk}{k^2 - p^2 - i\epsilon} \, \text{Disc}_{k} \tilde{f}(k)\,.
\end{equation}
Using the fact that $\text{Disc}_{k} \tilde{f}(k)$ is an odd function of $k$, eq.\eqref{dispersionfromfeynman} can be written as the following
\begin{equation}
\tilde{f}(p) = \frac{1}{2\pi i} \int_{-\infty}^{\infty} \frac{k \, dk}{k^2 - p^2 - i\epsilon} \, \text{Disc}_{k} \tilde{f}(k)\,, 
\end{equation}
which is same as eq.\eqref{disp+disc_relation}. One can also see that the $\widetilde{\text{Disc}}$ operation, when acting on some function, evaluates to twice the average of a function across its branch-cut.
%%%%%%%%%%%%%%%%%%%%%%%%%%%%%%%%%%%%%%%%%%%%%%%%%%%%%%%%%%%%%%%%%%%%%%%%%%%%
\section{Details for $3$-site correlator reconstruction for conformally coupled IR-convergent theories} \label{app - detail 3 site disp+disc}

In the following, we focus on conformally coupled IR-convergent theories and explicitly show how the $3$-site correlator can be reconstructed. It will be useful going forward if we break our further calculations into two cases: whether $n$ in the polynomial interaction $\lambda \phi^n$ is even or odd. We denote these two cases as $ n=2m$ and $n=2m+1$. \\

\noindent \textbf{For cases with $n=2m$:} 
From the expressions for $\mathcal{B}^{(1)}$ and $\widetilde{\mathcal{B}}^{(1)}$ written in eq.\eqref{corcompact_sec} and eq.\eqref{1 site B tilde_sec}, we have understood that for $n=2m$, $\widetilde{\mathcal{B}}^{(1)}$ can be neglected compared to $\mathcal{B}^{(1)}$ to the leading order in $\eta_0 \to 0$ limit. Therefore, among the four terms on the RHS of eq.\eqref{3sitetermstotal}, only the first term written in eq.\eqref{expterm1 final mt} will give us the leading contribution. 

The following relations can be worked out, which will be essential while computing the RHS of eq.\eqref{expterm1 final mt}-eq.\eqref{expterm4 final mt}, 
\begin{equation}
\begin{split}
 & 2 \bigg( \mathcal{B}^{(1)} (\lbrace \mathbf{k}_{M} \rbrace, q_1,  q_2)+\mathcal{B}^{(1)} (\lbrace \mathbf{k}_{M} \rbrace, -q_1,  -q_2)\bigg) \\
 & = \frac{\lambda (-1)^m (2m-4)! H^{4m-4} \eta_0^{2m}}{q_1 q_2 \lbrace 2k_i \rbrace^L} \left(  \frac{1}{(k_M+q_1+q_2)^{2m-3}} + \frac{1}{(k_M-q_1-q_2)^{2m-3}} \right)
 \\
 & = \frac{- \lambda_e^c}{q_1 q_2 \lbrace 2k_i \rbrace^{M}}  \mathcal{D}_{k_{M}} \left( \frac{2k_{M}}{(q_1+q_2)^2 - k_M^2} \right)
\end{split}
\end{equation}
in addition to the relations obtained in eq.\eqref{discleft_sec} for $\text{Disc}_q \mathcal{B}^{(1)}$. Using them, one can finally obtain eq.\eqref{B3 n=2m CC}. \\

\noindent \textbf{For cases with $n=2m+1$:}
For a $3$-site correlator with odd polynomial interaction, we need to consider all the four terms we have obtained in eq.\eqref{expterm1 final mt}-\eqref{expterm4 final mt}, unlike the even polynomial case.
We need to use the explicit forms of  $\mathcal{B}^{(1)}$ and $\widetilde{\mathcal{B}}^{(1)}$ to compute each of these four terms term for $\lambda \phi^n$ theory with $n=\text{odd}=2m+1$. 
The following expression will be needed in eq.\eqref{expterm1 final mt} which we can compute using eq.\eqref{corcompact_sec}
\begin{equation}
\begin{split}
 &  2\bigg( \mathcal{B}^{(1)} (\lbrace \mathbf{k}_{M} \rbrace, q_1,  q_2)+\mathcal{B}^{(1)} (\lbrace \mathbf{k}_{M} \rbrace, -q_1,  -q_2)\bigg) \\
 = & \frac{\lambda (-1)^{m+1} (2m-3)! H^{4m-2} (-\eta_0)^{2m+1} \eta_0 }{q_1 q_2 \lbrace 2k_i \rbrace^M } \left(\frac{1}{(k_M+q_1+q_2)^{2m-3}}    +\frac{1}{(k_M-q_1-q_2)^{2m-3}}\right)\\
 =& \frac{-\lambda_o^c}{q_1 q_2 \lbrace 2k_i \rbrace^M} \mathcal{D}_{k_M} \bigg( \frac{2k_M}{(q_1+q_2)^2-k_M^2}  \bigg) \, ,
\end{split}
\end{equation}
where $\mathcal{D}_{k_M} = \partial_{k_M}^{2m-4},~ \lambda_o^{c} = \lambda (2m-3)(-1)^{m} H^{4m-2} \eta_0^{2m+2}$. 
We also compute the following expression using eq.\eqref{1 site B tilde_sec}, which will be needed in eq.\eqref{expterm2 final mt}
\begin{equation}
\begin{split}
    & 2 \bigg( \widetilde{\mathcal{B}}^{(1)} (\lbrace \mathbf{k}_{M} \rbrace, q_1,  q_2)-  \widetilde{\mathcal{B}}^{(1)} ( \lbrace \mathbf{k}_{M} \rbrace, -q_1,  -q_2)\bigg)
    \\
    = &  \frac{i\lambda (-1)^m (2m-3)! H^{4m-2} \eta_0^{2m+1}}{q_1 q_2 \lbrace 2k_i \rbrace^M} \left( \frac{1}{(k_M +q_1 + q_2)^{2m-2}} -  \frac{1}{(k_M - q_1 -q_2)^{2m-2}}\right)
    \\
    = & \frac{-i \lambda_o^\psi}{q_1 q_2 \lbrace 2 k_i \rbrace^M} \widetilde{\mathcal{D}}_{k_M} \left(\frac{2( q_1 +q_2)}{(q_1 +q_2)^2 - k_M^2} \right) \, ,
\end{split}
\end{equation}
 where $\widetilde{\mathcal{D}}_{k_M} \equiv \partial_{k_M}^{2m-3},~ \lambda_o^{\psi} = \lambda (-1)^{m} H^{4m-2} \eta_0^{2m+1}$. 
We also need to compute the following quantities that we will use 
\begin{equation}\label{discBodd}
 \begin{split}
     \text{Disc}_q \mathcal{B}^{(1)}( \lbrace \mathbf{k}_{L} \rbrace ,q) &= \frac{\lambda (-1)^m (2m-3)! H^{4m-2} \eta_0^{2m+2}}{q \lbrace 2k_i \rbrace^L} \left( \frac{1}{(k_L+q)^{2m-3}} + \frac{1}{(k_L-q)^{2m-3}}  \right)
     \\
     & = \frac{-\lambda_o^c}{q \lbrace 2k_i \rbrace^L} \mathcal{D}_{k_L} \bigg( \frac{2k_L}{q^2-k_L^2}  \bigg) \, ,
 \end{split}   
\end{equation}
where $\mathcal{D}_{k_L} = \partial_{k_L}^{2m-4},~ \lambda_o^{c} = \lambda (2m-3)(-1)^{m} H^{4m-2} \eta_0^{2m+2}$, 
and recall that we previously computed 
\begin{equation}
\begin{split}
\widetilde{\text{Disc}}_p  \widetilde{\mathcal{B}}^{(1)}( \lbrace \mathbf{k}_{L} \rbrace ,p)=& \frac{-i \lambda_o^{\psi}}{q \lbrace 2k_i \rbrace^L} \widetilde{\mathcal{D}}_{k_L} \left( \frac{2q}{q^2-k_L^2} \right)\, ,
\end{split}    
\end{equation}
where $\widetilde{\mathcal{D}}_{k_L} \equiv \partial_{k_L}^{2m-3},~~~\lambda_o^{\psi} = \lambda (-1)^{m} H^{4m-2} \eta_0^{2m+1}$. 
Using all the relations written above, we can straightforwardly obtain eq.\eqref{expterm1 final mt CC} - \eqref{expterm4 final mt CC}.
%%%%%%%%%%%%%%%%%%%%%%%%%%%%%%%%%%%%%%%%%%%%%%%%%%%%%%%%%%%%%%%%%%%%%%%%%

%%%%%%%%%%%%%%%%%%%%%%%%%%%%%%%%%%%%%%%%%%%%%%%%%%%%%%%%%%%%%%%%%%%%%%%%%
%%%%%%%%%%%%%%%%%%%%%%%%%%%%%%%%%%%%%%%%%%%%%%%%%%%%%%%%%%%%%%%%%
%%%%%%%%%%%%%%%%%%%%%%%%%%%%%%%%%%%%%%%%%%%%%%%%%%%%%%%%%%%%
\section{Details for the $r$-site correlator reconstruction} \label{app_r_site}
\subsection{Deriving the identity in eq.\eqref{single cut auxiliary correlator def2}}\label{app_r_site_subsec1}
In this appendix, our aim is to derive the identity relating ${\text {Disc}}_{p_{1}} \widetilde{\mathcal{B}}^{(r)}$ to ${\text {Disc}}_{p_{1}} \mathcal{B}^{(r)}$, written in eq.\eqref{single cut auxiliary correlator def2} and eq.\eqref{single cut auxiliary correlator def3}. In \S \ref{sec:3site} and in \S \ref{sec:rsite}, this identity was used to algebraically manipulate the successive discontinuity relation for the $3$-site and $r$-site correlator. 

Let us start by writing the auxiliary counterpart $\widetilde{\mathcal{B}}^{(r)}$ for of the $r$-site correlator, from its definition given in eq.(4.39) in \cite{Das:2025qsh}
\begin{equation}\label{rsiteauxinin}
\begin{split}
    &\widetilde{\mathcal{B}}^{(r)}\left( \lbrace \mathbf{k}_L,...,\mathbf{k}_R \rbrace ; \lbrace p_1,..,p_{r-1} \rbrace \right) = (i\lambda)^2 \int d\eta_r \int d\eta_{r-1} \times \\
    & \qquad \bigg[ \mathcal{V}_{(r-2)}^+ \mathcal{K}_{\eta_0 \eta_{r-1}}^+(\lbrace \mathbf{k}_{M_{r-2}} \rbrace) G^{++}_{p_{r-1}}(\eta_{r-1},\eta_r) \mathcal{K}_{\eta_0 \eta_r}^+(\lbrace \mathbf{k}_R \rbrace) 
    \\& \qquad 
    +\mathcal{V}_{(r-2)}^+ \mathcal{K}_{\eta_0 \eta_{r-1}}^+(\lbrace \mathbf{k}_{M_{r-2}} \rbrace) G^{+-}_{p_{r-1}}(\eta_{r-1},\eta_r) \mathcal{K}_{\eta_0 \eta_r}^-(\lbrace \mathbf{k}_R \rbrace) 
    \\& \qquad 
    -\mathcal{V}_{(r-2)}^- \mathcal{K}_{\eta_0 \eta_{r-1}}^-(\lbrace \mathbf{k}_{M_{r-2}} \rbrace) G^{-+}_{p_{r-1}}(\eta_{r-1},\eta_r) \mathcal{K}_{\eta_0 \eta_r}^+(\lbrace \mathbf{k}_ R\rbrace)
    \\& \qquad 
    - \mathcal{V}_{(r-2)}^- \mathcal{K}_{\eta_0 \eta_{r-1}}^-(\lbrace \mathbf{k}_{M_{r-2}} \rbrace) G^{--}_{p_{r-1}}(\eta_{r-1},\eta_r) \mathcal{K}_{\eta_0 \eta_r}^-(\lbrace \mathbf{k}_R \rbrace) \bigg] \, ,
\end{split}
\end{equation}
such that $\mathcal{V}_{(r-2)}^{\pm}$ has the following expression
\begin{equation}
\begin{split}
    &\mathcal{V}_{(r-2)}^{\sigma \in \pm} \left( \lbrace \eta_1, \cdots, \eta_{r-1} \rbrace \right)=  (i \lambda)^{r-2} \int^{(r-2)} d\eta ~ \sum_{\lbrace \sigma_i \rbrace_{i=1}^{r-2} \in \pm } \Bigg[ \left( \prod_{j=1}^{r-2} \sigma_j \right)  \times \\
    &\left( \mathcal{K}^{\sigma_1}_{\eta_0 \eta_1}\left( \lbrace \mathbf{k}_L \rbrace \right) \prod_{j=2}^{r-2}  \mathcal{K}^{\sigma_{j}}_{\eta_0\eta_j} \left(\lbrace \mathbf{k}_{M_{j-1}} \rbrace \right)\right) 
     \left( \prod_{j=1}^{r-3} G_{p_j}^{\sigma_j \sigma_{j+1}} ( \eta_j, \eta_{j+1} )\right) G_{p_{r-2}}^{\sigma_{r-2}\sigma} ( \eta_{r-2}, \eta_{r-1} ) \Bigg].
\end{split}
\end{equation}
Next, applying discontinuity operation ${\text {Disc}}_{p_{r-1}}$ on $\widetilde{\mathcal{B}}_r$, we get two contributions 
\begin{equation}\label{Br disc step2}
\begin{split}
      &{\text {Disc}}_{p_{r-1}} \widetilde{\mathcal{B}}^{(r)}( \lbrace \mathbf{k}_L,..., \mathbf{k}_R \rbrace ; \lbrace p_1, ..., p_{r-1} \rbrace ) = {\text {Disc}}_{p_{r-1}} \widetilde{\mathcal{B}}^{(r)} \bigg|_{\text{Term 1}} + {\text {Disc}}_{p_{r-1}} \widetilde{\mathcal{B}}^{(r)} \bigg|_{\text{Term 2}} \, ,
\end{split}
\end{equation}
such that 
\begin{equation}\label{Br disc step2 term1}
\begin{split}
      &{\text {Disc}}_{p_{r-1}} \widetilde{\mathcal{B}}^{(r)}\bigg|_{\text{Term 1}} 
     = \frac{(i\lambda)^2}{2 P_{p_{r-1}} (\eta_0)}\times\prod_{i = r}^{r+1}\int d\eta_i\Bigg[\bigg( \mathcal{V}_{(r-2)}^+ \mathcal{K}_{\eta_0 \eta_{r-1}}^+(\lbrace \mathbf{k}_{M_{r-2}} \rbrace)\, K_{p_{r-1}}^+(\eta_0, \eta_{r-1})  \\
     &- \mathcal{V}_{(r-2)}^- \mathcal{K}_{\eta_0 \eta_{r-1}}^-(\lbrace \mathbf{k}_{M_{r-2}} \rbrace)\,K_{p_{r-1}}^-(\eta_0, \eta_{r-1}) \bigg) - (p_{r-1} \leftrightarrow -p_{r-1})\Bigg] \times 
     \\
     & \Bigg[\Big( K_{p_{r-1}}^+(\eta_0, \eta_r)\, \mathcal{K}_{\eta_0 \eta_r}^+(\lbrace \mathbf{k}_R \rbrace) + K_{p_{r-1}}^-(\eta_0, \eta_r)\, \mathcal{K}_{\eta_0 \eta_r}^-(\lbrace \mathbf{k}_R \rbrace)\Big) - (p_{r-1} \leftrightarrow -p_{r-1}) \Bigg] \, ,
\end{split}
\end{equation}
and 
\begin{equation}\label{Br disc step2 term2}
\begin{split}
      &{\text {Disc}}_{p_{r-1}} \widetilde{\mathcal{B}}^{(r)}\bigg|_{\text{Term 2}}
     = \frac{(i\lambda)^2}{2 P_{p_{r-1}} (\eta_0)}
     \prod_{i = r}^{r+1}\int d\eta_i\Bigg[\bigg( -\mathcal{V}_{(r-2)}^+ \mathcal{K}_{\eta_0 \eta_{r-1}}^+(\lbrace \mathbf{k}_{M_{r-2}} \rbrace)\, K_{p_{r-1}}^+(\eta_0, \eta_{r-1})  \\
      & - \mathcal{V}_{(r-2)}^- \mathcal{K}_{\eta_0 \eta_{r-1}}^-(\lbrace \mathbf{k}_{M_{r-2}} \rbrace)\,K_{p_{r-1}}^-(\eta_0, \eta_{r-1}) \bigg) + (p_{r-1} \leftrightarrow -p_{r-1})\Bigg] \times 
     \\
     & \Bigg[\Big( K_{p_{r-1}}^+(\eta_0, \eta_r)\, \mathcal{K}_{\eta_0 \eta_r}^+(\lbrace \mathbf{k}_R \rbrace) - K_{p_{r-1}}^-(\eta_0, \eta_r)\, \mathcal{K}_{\eta_0 \eta_r}^-(\lbrace \mathbf{k}_R \rbrace)\Big) + (p_{r-1} \leftrightarrow -p_{r-1})\Bigg] \, .
\end{split}
\end{equation}
One can further simplify eq.\eqref{Br disc step2}, with eq.\eqref{Br disc step2 term1} and eq.\eqref{Br disc step2 term2}, as the following
\begin{equation}
\label{single cut r-site auxiliary correlator app}
\begin{split}
    &{\text {Disc}}_{p_{r-1}} \widetilde{\mathcal{B}}^{(r)}( \lbrace \mathbf{k}_L,..., \mathbf{k}_R \rbrace ; \lbrace p_1, ..., p_{r-1} \rbrace )= \frac{1}{2 P_{p_{r-1}} (\eta_0)} \times  \\
    & \bigg[\text{Disc}_{p_{r-1}} \mathcal{B}^{(r-1)}\left(  \lbrace \mathbf{k}_L,...,\mathbf{k}_{M_{r-2}} \rbrace, p_{r-1} ; \lbrace p_1,..,p_{r-2} \rbrace \right) \text{Disc}_{p_{r-1}} \widetilde{\mathcal{B}}^{(1)}( \lbrace \mathbf{k}_{R} \rbrace, p_{r-1})\\
    &-  \widetilde{\text{Disc}}_{p_{r-1}} \widetilde{\mathcal{B}}^{(r-1)}\left(  \lbrace \mathbf{k}_L,...,\mathbf{k}_{M_{r-2}} \rbrace, p_{r-1} ; \lbrace p_1,..,p_{r-2} \rbrace \right) \widetilde{\text{Disc}}_{p_{r-1}} \mathcal{B}^{(1)}( \lbrace \mathbf{k}_{R} \rbrace, p_{r-1}) \bigg]\,.
\end{split}
\end{equation}
In deriving eq.\eqref{single cut r-site auxiliary correlator app}, we have to use various relations which were presented in \S (4.2) of \cite{Das:2025qsh}. Actually, one must follow similar steps that were used to derive eq.(4.44) starting from eq.(4.26) in \cite{Das:2025qsh}. 

Finally, to get our desired identity, we must compare eq.\eqref{single cut r-site auxiliary correlator app} with ${\text {Disc}}_{p_{r-1}} \mathcal{B}^{(r)}$, which was written in eq.\eqref{single cut r-site correlator}. Consequently, we can easily check that the following relation holds 
 \begin{equation}
 \label{single cut auxiliary correlator def2  app} 
 \begin{split}
     &{\text {Disc}}_{p_{r-1}} \widetilde{\mathcal{B}}^{(r)}( \lbrace \mathbf{k}_L,..., \mathbf{k}_R \rbrace ; \lbrace p_1, ..., p_{r-1} \rbrace )  \\&= {\text {Disc}}_{p_{r-1}} \mathcal{B}^{(r)}( \lbrace \mathbf{k}_L,..., \mathbf{k}_R \rbrace ; \lbrace p_1, ..., p_{r-1} \rbrace ) \Bigg|_{\mathcal{B}^{(1)}( \lbrace \mathbf{k}_{R} \rbrace, p_{r-1})\leftrightarrow\widetilde{\mathcal{B}}^{(1)}( \lbrace \mathbf{k}_{R} \rbrace, p_{r-1})}\,.
\end{split}
 \end{equation}
 which is our desired identity appearing before as eq.\eqref{single cut auxiliary correlator def2} and eq.\eqref{single cut auxiliary correlator def3}.
 
%%%%%%%%%%%%%%%%%%%%%%%%%%%%%%%%%%%%%%%%%%%%%%%%%%%%%%%%%%%%%%%
 \subsection{Details for the induction method for $r$-site correlator reconstruction} \label{app induction_details r site}
 
In this subsection, we explain in some more detail why the apparent problems in executing the induction method for reconstructing the $r$-site correlator, mentioned after eq.\eqref{r-site successive dis with r-1 site repeat}, can be resolved. 

By assumption of the inducion method, $\widehat{\mathcal{I}}_{(r-1)}\left(  \lbrace \mathbf{k}_L,...,\mathbf{k}_{M_{r-2}} \rbrace, p_{r-1} ; \lbrace p_1,..,p_{r-2} \rbrace \right)$ is consistent with the algorithm presented in \S~\ref{sec:rsite}, and hence it will allow for a reorganisation same as eq.\eqref{r-1-site successive disc}. In terms of $\widehat{\mathcal{I}}_{(r-1)}$, the successive discontinuity of the $r$-site correlator, i.e., $\widehat{\mathcal{I}}_{(r)}$, is given in eq.\eqref{r-site successive dis with r-1 site repeat}. Our goal in this Appendix is, therefore, to argue that the RHS of eq.\eqref{r-site successive dis with r-1 site repeat} is consistent with our algorithm for an $r$-site correlator. 

The contribution from the right-most vertex (the $(r-1)$-th vertex from the perspective of the $(r-1)$-site correlator) will contribute 
\begin{equation} \label{app C term1}
\begin{split}
    & \text{either }~~ \text{Disc}_{p_{r-2}}\mathcal{B}^{(1)}( \lbrace \mathbf{k}_{M_{r-2}} \rbrace, p_{r-2},p_{r-1}) \, , \quad  \text{or }~~ \widetilde{\text{Disc}}_{p_{r-2}}\widetilde{\mathcal{B}}^{(1)}( \lbrace \mathbf{k}_{M_{r-2}} \rbrace, p_{r-2},p_{r-1}) \, .
\end{split}
\end{equation}
But from the perspective of the $r$-site correlator, this vertex is one of the middle ones (adjacent to the right edge vertex), so if $\widehat{\mathcal{I}}_{(r)}$ has to be consistent with the algorithm, we expect a contribution from this $(r-1)$-th vertex as 
\begin{equation} \label{app C term2}
\begin{split}
    & \text{either }~~ \widetilde{\text{Disc}}_{p_{r-2},p_{r-1}}\mathcal{B}^{(1)}( \lbrace \mathbf{k}_{M_{r-2}} \rbrace, p_{r-2},p_{r-1}) \, , \quad  \text{or }~~ \text{Disc}_{p_{r-2},p_{r-1}}\widetilde{\mathcal{B}}^{(1)}( \lbrace \mathbf{k}_{M_{r-2}} \rbrace, p_{r-2},p_{r-1}) \, .
\end{split}
\end{equation}
It can be verified that the terms in eq.\eqref{app C term1} are contained in $\mathcal{I}_{(r-1)}$, i.e. the first term on the RHS of eq.\eqref{r-1-site successive disc}. Whereas, the terms in eq.\eqref{app C term2} are contained in the second term on the RHS of eq.\eqref{r-1-site successive disc}. Therefore, it is possible to make the swapping between these terms within the first and the second terms on the RHS of eq.\eqref{r-1-site successive disc}. This can be expressed schematically as subjecting $\mathcal{I}_{(r-1)}\left(  \lbrace \mathbf{k}_L,...,\mathbf{k}_{M_{r-2}} \rbrace, p_{r-1} ; \lbrace p_1,..,p_{r-2} \rbrace \right)$ to 
\begin{equation}\label{app C term2a}
\begin{split}
    &\text{Disc}_{p_{r-2}}\mathcal{B}^{(1)}( \lbrace \mathbf{k}_{M_{r-2}} \rbrace, p_{r-2},p_{r-1})\leftrightarrow\widetilde{\text{Disc}}_{p_{r-2},p_{r-1}}\mathcal{B}^{(1)}( \lbrace \mathbf{k}_{M_{r-2}} \rbrace, p_{r-2},p_{r-1})\, ,\\ & \qquad \qquad \qquad \qquad \text{or}\\ 
    &\widetilde{\text{Disc}}_{p_{r-2}}\widetilde{\mathcal{B}}^{(1)}( \lbrace \mathbf{k}_{M_{r-2}} \rbrace, p_{r-2},p_{r-1})\leftrightarrow\text{Disc}_{p_{r-2},p_{r-1}}\widetilde{\mathcal{B}}^{(1)}( \lbrace \mathbf{k}_{M_{r-2}} \rbrace, p_{r-2},p_{r-1}) \, .
\end{split}
\end{equation}
This establishes that the first term on the RHS of eq.\eqref{r-site successive dis with r-1 site repeat} is consistent with our algorithm. 

Similarly, for the second term on the RHS of eq.\eqref{r-site successive dis with r-1 site repeat}, from \begin{equation} \label{app C term3a}
    \widehat{\mathcal{I}}_{(r-1)}\left(  \lbrace \mathbf{k}_L,...,\mathbf{k}_{M_{r-2}} \rbrace, p_{r-1} ; \lbrace p_1,..,p_{r-2} \rbrace \right) \bigg|_{\mathcal{B}^{(1)}( \lbrace \mathbf{k}_{M_{r-2}} \rbrace, p_{r-2},p_{r-1})\leftrightarrow\widetilde{\mathcal{B}}^{(1)}( \lbrace \mathbf{k}_{M_{r-2}} \rbrace, p_{r-2},p_{r-1})} \, ,
\end{equation}
we will obtain 
\begin{equation} \label{app C term3}
\begin{split}
    & \text{either }~~  \text{Disc}_{p_{r-2}}\widetilde{\mathcal{B}}^{(1)}( \lbrace \mathbf{k}_{M_{r-2}} \rbrace, p_{r-2},p_{r-1})\, , \quad  \text{or }~~  \widetilde{\text{Disc}}_{p_{r-2}}\mathcal{B}^{(1)}( \lbrace \mathbf{k}_{M_{r-2}} \rbrace, p_{r-2},p_{r-1})\, ,
\end{split}
\end{equation}
as contribution from the $(r-1)$-th vertex from the perspective of the $(r-1)$-site correlator $\widehat{\mathcal{I}}_{(r-1)}$ in  eq.\eqref{app C term3a}. But from the $r$-site correlator, we expect contributions from this last of the middle vertices, as 
\begin{equation} \label{app C term4}
\begin{split}
    & \text{either }~~  \text{Disc}_{p_{r-2},p_{r-1}}\widetilde{\mathcal{B}}^{(1)}( \lbrace \mathbf{k}_{M_{r-2}} \rbrace, p_{r-2},p_{r-1})\, , ~~ \text{or }~~  \widetilde{\text{Disc}}_{p_{r-2},p_{r-1}}\mathcal{B}^{(1)}( \lbrace \mathbf{k}_{M_{r-2}} \rbrace, p_{r-2},p_{r-1})\, .
\end{split}
\end{equation}
Again, one can verify that, from eq.\eqref{r-1-site successive disc}, the resolution is to subject the term in eq.\eqref{app C term3a} to the swapping 
\begin{equation}\label{app C term4a}
\begin{split}
    &\text{Disc}_{p_{r-2}}\widetilde{\mathcal{B}}^{(1)}( \lbrace \mathbf{k}_{M_{r-2}} \rbrace, p_{r-2},p_{r-1})\leftrightarrow\text{Disc}_{p_{r-2},p_{r-1}}\widetilde{\mathcal{B}}^{(1)}( \lbrace \mathbf{k}_{M_{r-2}} \rbrace, p_{r-2},p_{r-1})\, ,\\ & \qquad \qquad \qquad \qquad \text{or}\\ 
    &\widetilde{\text{Disc}}_{p_{r-2}}\mathcal{B}^{(1)}( \lbrace \mathbf{k}_{M_{r-2}} \rbrace, p_{r-2},p_{r-1})\leftrightarrow \widetilde{\text{Disc}}_{p_{r-2},p_{r-1}}\mathcal{B}^{(1)}( \lbrace \mathbf{k}_{M_{r-2}} \rbrace, p_{r-2},p_{r-1}) \, .
\end{split}
\end{equation}
This establishes that the second term on the RHS of eq.\eqref{r-site successive dis with r-1 site repeat} is also consistent with our algorithm. 

Let us now discuss the resolution of the second problem, which was the sign issue for the second term in the RHS of eq.\eqref{r-site successive dis with r-1 site repeat}. We have already argued how to obtain the correct repositioning of $(\text{Disc}, \, \widetilde{\text{Disc}})$ in  $\mathcal{I}_{(r-1)}$, through the interchanges mentioned in eq.\eqref{app C term2a} and \eqref{app C term4a}, it should be noted that on the RHS of eq.\eqref{r-site successive dis with r-1 site repeat}, the additional factors of $\text{Disc}_{p_{r-1}} \mathcal{B}^{(1)}$ and $\widetilde{\text{Disc}}_{p_{r-1}} \widetilde{\mathcal{B}}^{(1)}( \lbrace \mathbf{k}_{ R} \rbrace, p_{r-1})$ are properly placed according to the algorithm as the right edge vertex from the perspective of the $r$-site correlator. 

Now, if we assume the $(r-1)$-th  vertex of $\mathcal{ I} _{(r-1)}\left(  \lbrace \mathbf{k}_L,...,\mathbf{k}_{M_{r-2}} \rbrace, p_{r-1} ; \lbrace p_1,..,p_{r-2} \rbrace \right)$ is $\mathcal{B}^{(1)}$, after exchanging with $\widetilde{\mathcal{B}}^{(1)}$, there is a new auxiliary object at position $(r-1)$ and there is also a new $\widetilde{\mathcal{B}}^{(1)}$ leg attached at the $r$-th vertex position. Apart from this two new insertions of $\widetilde{\mathcal{B}}^{(1)}$ everything was consistent with the sign rule since $\mathcal{ I} _{(r-1)}\left(  \lbrace \mathbf{k}_L,...,\mathbf{k}_{M_{r-2}} \rbrace, p_{r-1} ; \lbrace p_1,..,p_{r-2} \rbrace \right)$, by assumption, follows the algorithm. Now, if we apply our sign rule for the last two vertices, we get
\[(-1)^{r-1}(-1)^r = -1  \, , \]
which is the relative minus sign present in the second term on the RHS of eq.\eqref{r-site successive dis with r-1 site repeat}.

Alternatively, if the $(r-1)$-th vertex of $\mathcal{ I} _{(r-1)}\left(  \lbrace \mathbf{k}_L,...,\mathbf{k}_{M_{r-2}} \rbrace, p_{r-1}; \lbrace p_1,..,p_{r-2} \rbrace \right)$ is $\widetilde{\mathcal{B}}^{(1)}$, after exchanging with $\mathcal{B}^{(1)}$, there will be an excess sign factor of $(-1)^{r-1}$ due to our sign rule. Now, multiplying $(-1)^{r-1}$ with the relative minus sign before the second term of RHS of eq.\eqref{r-site successive dis with r-1 site repeat}, we get $(-1)^r$, which we can attach to the new $r$-th vertex with one $\widetilde{\mathcal{B}}^{(1)}$. This will be consistent with our sign rule, which holds for any $r$-site correlator, provided it holds for the $(r-1)$- site correlator.
%%%%%%%%%%%%%%%%%%%%%%%%%%%%%%%%%%%%%%%%%%%%%%%%%%%%%%%%%%%%%%%

%%%%%%%%%%%%%%%%%%%%%%%%%%%%%%%%%%%%%%%%%%%%%%%%%%%%%%%%%%%%%%%
%%%%%%%%%%%%%%%%%%%%%%%%%%%%%%%%%%%%%%%%%%%%%%%%%%%%%%%%%%%%%%%
\section{Details of the diagrammatic dressing rules} \label{appCC_IRdiv}

Here, we will give a detailed derivation of the vertex contribution for a conformally coupled IR divergent and massless scalar polynomial interaction.
%%%%%%%%%%%%%%%%%%%%%%%%%%%%%%%%%%%%%%%%%%%%%%%%%%%%%%%%%
\subsection{IR divergent $\phi^3$ interaction}\label{detail vertex IR div cc}

Here we will focus on deriving the vertex contribution for the conformally coupled $\phi^3$ interaction, which is IR divergent. For this case, the contact wave-function coefficient can be easily obtained, and it has the following expression
\begin{equation}
\begin{split}
       \psi^{(1)}(k_1,k_2,p)=\frac{i \lambda}{H^4 \eta_0^3} \log \left((k_1+k_2+p) \eta_0 \right)+\frac{\pi \lambda}{2 H^4 \eta_0^3} \, .
\end{split}
\end{equation}
Therefore, the $1$-site correlator $\mathcal{B}^{(1)}$, and the $1$-site auxiliary object $\widetilde{\mathcal{B}}^{(1)}$ are obtained directly from their definition as follows
\begin{equation} \label{BBtilphi3}
 \begin{split}
     \mathcal{B}^{(1)}(k_1,k_2,p)=\frac{\pi \lambda H^2 \eta_0^3}{8k_1 k_2 p} \, , \quad \widetilde{\mathcal{B}}^{(1)} (k_1, k_2, p) = \frac{i \lambda H^2 \eta_0^3}{4 k_1 k_2 p} \log((k_1+k_2+p)\eta_0) \, .
 \end{split}   
\end{equation}
Using the explicit form of $\mathcal{B}^{(1)}$ and  $\widetilde{\mathcal{B}}^{(1)}$ as given in eq.\eqref{BBtilphi3}, and using the definitions of $\text{Disc}_q$ and $\widetilde{\text{Disc}_q}$ from eq.\eqref{defdisc}, we get the following expressions for the left and right most blue vertices
\begin{equation}\label{DIscB}
\begin{split}
    \text{Disc}_{q_1} \mathcal{B}^{(1)}(\lbrace \mathbf{k}_L \rbrace ,q_1) = &\frac{\pi \lambda H^2 \eta_0^3}{\lbrace 2 k_i \rbrace^L q_1}\, , \quad 
     \text{Disc}_{q_{r-1}} \mathcal{B}^{(1)}(\lbrace \mathbf{k}_R \rbrace ,q_{r-1}) =\frac{\pi \lambda H^2 \eta_0^3}{\lbrace 2 k_i \rbrace^R q_{r-1}} \, ,
\end{split}
  \end{equation}
and similarly, the contribution of the left and right-most red vertices can also be obtained through the following simple calculation
\begin{equation}\label{DIsctildeB}
\begin{split}
     & \widetilde{\text{Disc}}_{q_1} \widetilde{\mathcal{B}}^{(1)}( \lbrace \mathbf{k}_L \rbrace ,q_1)= \frac{i \lambda H^2 \eta_0^3}{\lbrace 2 k_i \rbrace^L q_1} \log\bigg(\frac{k_L+q_1}{k_L-q_1}  \bigg) = \frac{i \lambda H^2 \eta_0^3}{\lbrace 2 k_i \rbrace^L q_1}  \int_{0}^{\infty} ds_1 \frac{2q_1}{(s_1+k_L)^2-q_1^2} \, ,\\
     & \widetilde{\text{Disc}}_{q_{r-1}} \widetilde{\mathcal{B}}^{(1)}( \lbrace \mathbf{k}_R \rbrace ,q_{r-1})= \frac{i \lambda H^2 \eta_0^3}{\lbrace 2 k_i \rbrace^R q_{r-1}}  \int_{0}^{\infty} ds_2 \frac{2q_{r-1}}{(s_2+k_R)^2-q_{r-1}^2} \, .
\end{split}
\end{equation}
Now we calculate the contribution of the middle vertices of both color codes. For the blue middle vertex, we have the following contribution 
\begin{equation}
\begin{split}
    2~\widetilde{\text{Disc}}_{q_{j+1},q_j}\mathcal{B}^{(1)}(k_{M_{j}}, q_{j+1}, q_{j}) =  \frac{\pi \lambda H^2 \eta_0^3}{\lbrace 2k_i \rbrace^{M_j} q_j q_{j+1}} \, ,
\end{split}
\end{equation}
and for the red middle vertex, the contribution is the following 
\begin{equation}
\begin{split}
   2 ~ \text{Disc}_{q_j,q_{j+1}}\, \widetilde{\mathcal{B}}^{(1)} (\lbrace \mathbf{k}_{M_j} \rbrace, q_j,  q_{j+1}) = \frac{i \lambda H^2 \eta_0^3}{\lbrace 2 k_i \rbrace^{M_j}  q_{j} q_{j+1}}  \int_{0}^{\infty} ds_j \frac{2(q_j+q_{j+1})}{(s_j+k_{M_{j}})^2-(q_j+q_{j+1})^2} \, .
\end{split}
\end{equation}
Note that for the $\phi^3$ interaction, each internal site has just one bulk to boundary line. As a result, the above notation $\lbrace 2 k_i \rbrace^{M_j}$ in principle means $2k_{M_j}$ where $j$ is the position of sites counted from the left-most vertex. In other words, for an $r$-site tree level diagram, $j$ can take integer values from $1$ to $r-2$. From here, one can directly obtain eq.\eqref{discleftphi3cc} and eq.\eqref{discrightphi3cc}. 

%%%%%%%%%%%%%%%%%%%%%%%%%%%%%%%%%%%%%%%%%%%%%%%%%%%
\subsection{Massless $\chi^3$ interaction} \label{app-massless-details}

In this sub-section, we will focus on massless scalar fields (denoted by $\chi$) with a cubic interaction $ \lambda \chi^3$ and determine the contributions from the dressed vertex factors (red and blue), as seen in eq.\eqref{firsteq of diag blue} to eq.\eqref{secondeq of diag red}, for the color-coded Feynman diagrams. For this, we need to know $\mathcal{B}^{(1)}$ and  $\widetilde{\mathcal{B}}^{(1)}$. They will be obtained from the $1$-site wave function coefficient for $\chi^3$ theory given by 
\begin{equation}
\begin{split}
    \psi_3(\vec{k_1},\vec{k_2},\vec{k_3}) = i\lambda\int_{-\infty}^{0}\frac{d\eta}{H^4\eta^{4}}\prod_{j= 1}^{3}(1-i\,k_j\,\eta)\times e^{i(k_1+k_2+k_3)\eta}\,.
\end{split}
\end{equation}
Note that the wave-function coefficient is IR divergent due to the presence of the following kind of integration
\begin{equation}
\label{IRdivint1}
    \int_{-\infty}^{0}\frac{d\eta}{\eta^{\alpha}}e^{ik\eta}\,,\quad \text{for}~\alpha\geq 1 \, , \text{ from } \eta \to 0\,.
\end{equation}
To handle such divergent contributions, we will use dimensional regularization, so that the contact wave-function coefficient becomes the following \footnote{We have analytically continued $k \rightarrow k(1 -i \beta )$ with $\beta$ being a small positive number, which introduces a damping factor ensuring that the integrand falls off as $\eta \rightarrow -\infty$.}
\begin{equation}
\begin{split}
    \psi_3(\vec{k_1},\vec{k_2},\vec{k_3}) = i\lambda\int_{-\infty}^{0}\frac{d\eta}{H^4\eta^{(4-\epsilon)}}\prod_{j= 1}^{3}(1-i\,k_j\,\eta)\times e^{i(k_1+k_2+k_3 - i \, \beta)\eta}\, . 
\end{split}
\end{equation}
Next, the real and imaginary pieces of $\psi^{(1)}$ can be obtained as follows
\begin{equation}
\begin{split}
    \mathbb{R}e\bigg(\psi^{(1)}(k_1,k_2,k_3)\bigg) =\frac{\lambda}{H^4}\int_{0}^{\infty}ds\bigg(\frac{s^{(3-\epsilon)}}{\Gamma(4-\epsilon)}+\frac{s^{(2-\epsilon)}\,k_{123}}{\Gamma(3-\epsilon)}+\frac{s^{(1-\epsilon)}\,b_{12}}{\Gamma(2-\epsilon)}+\frac{s^{-\epsilon}\,c_{123}}{\Gamma(1-\epsilon)}\bigg)\frac{1}{(s+k_{123})} \, ,
\end{split}
\end{equation}
and
\begin{equation}
\begin{split}
    \mathbb{I}m\bigg(\psi^{(1)}(k_1,k_2,k_3)\bigg) = \alpha \,   \mathbb{R}e\bigg(\psi^{(1)}(k_1,k_2,k_3)\bigg) \, ,
\end{split}
\end{equation}
where we have defined the following notations
$b_{12} = \sum_{\substack{1 \le j < i \le 3}} k_i k_j\, , \quad c_{123} = k_1k_2k_3$.
In deriving the above relations, we have used the following very useful identity
\begin{equation}
    (-i\eta)^{-a}e^{ik\eta} = \frac{1}{\Gamma(a)}\int_{0}^{\infty} ds \,s^{(a-1)}\,e^{i(s+k)\eta}\,,\quad(\text{for Re(a)}>0) \, .
\end{equation}
For matching the regularized answer with \cite{Chowdhury:2025ohm}, we choose $\alpha = \frac{\pi}{2}\epsilon$. In general, our answer should be regularization-parameter-independent, but to show this, we need proper renormalization, which we are not doing here. Hence, we have the following form of contact correlator, $\mathcal{B}^{(1)}$ and auxiliary object, $\widetilde{\mathcal{B}}^{(1)}$
\begin{equation}
\label{masslesscontactdata_app}
\begin{split}
    &\mathcal{B}^{(1)}(k_1,k_2,k_3) \\
    & =\lim_{\epsilon\rightarrow 0^+}\frac{\lambda\,H^2}{8k_1^3k_2^3k_3^3}\int_{0}^{\infty}ds\bigg(\frac{s^{(3-\epsilon)}}{\Gamma(4-\epsilon)}+\frac{s^{(2-\epsilon)}\,k_{123}}{\Gamma(3-\epsilon)}+\frac{s^{(1-\epsilon)}\,b_{12}}{\Gamma(2-\epsilon)}+\frac{s^{-\epsilon}\,c_{123}}{\Gamma(1-\epsilon)}\bigg)\frac{\cos(\pi\epsilon/2)}{(s+k_{123})}\,,\\&
   \widetilde{\mathcal{B}}^{(1)}(k_1,k_2,k_3) \\
    & =\lim_{\epsilon\rightarrow 0^+}\frac{\lambda\,H^2}{8k_1^3k_2^3k_3^3}\int_{0}^{\infty}ds\bigg(\frac{s^{(3-\epsilon)}}{\Gamma(4-\epsilon)}+\frac{s^{(2-\epsilon)}\,k_{123}}{\Gamma(3-\epsilon)}+\frac{s^{(1-\epsilon)}\,b_{12}}{\Gamma(2-\epsilon)}+\frac{s^{-\epsilon}\,c_{123}}{\Gamma(1-\epsilon)}\bigg)\frac{\sin(\pi\epsilon/2)}{(s+k_{123})}\,.
\end{split}
\end{equation}
Here, we have used the following identities
$\lim_{\epsilon\rightarrow0} \cos(\epsilon) = 1\, ,~ \lim_{\epsilon\rightarrow0} \sin(\epsilon) = \epsilon\,$. 
Using eq.\eqref{masslesscontactdata_app}, we get following 
\begin{equation}
\begin{split}
    &\text{Disc}_{p_1}\mathcal{B}^{(1)}(k_1,k_2,p_1) = \lim_{\epsilon\rightarrow 0^+}\frac{\lambda\,H^2}{4k_1^3k_2^3p_1^3}\int_{0}^{\infty}ds\bigg(\frac{s^{(3-\epsilon)}(k_{12}+s)}{\Gamma(4-\epsilon)}+\frac{s^{(2-\epsilon)}\,\left(k_{12}(k_{12}+s)-p_1^2\right)}{\Gamma(3-\epsilon)}\\&
    +\frac{s^{(1-\epsilon)}\,\left(k_1k_2(k_{12}+s)-p_1^2k_{12}\right)}{\Gamma(2-\epsilon)}-\frac{s^{-\epsilon}\,k_1k_2p_1^2}{\Gamma(1-\epsilon)}\bigg)\frac{\cos(\pi\epsilon/2)}{\left((k_{12}+s)^2-p_1^2\right)} \, .
\end{split}
\end{equation}
The above equation can be written more compactly as 
\begin{equation}
\label{masslessdiscb}
  \text{Disc}_{p_1}\mathcal{B}^{(1)}(k_1,k_2,p_1) =   \lim_{\epsilon\rightarrow 0^+}\frac{\lambda\,H^2\,\cos(\pi\epsilon/2)}{4k_1^3k_2^3p_1^3}\int_{0}^{\infty}ds\frac{\mathcal{Q}_3(k_{12},p_1,s)}{(s+k_{12})^2-p_1^2}\,,
\end{equation}
where, 
\begin{equation}
\label{defQ3_app}
\mathcal{Q}_3(k_{12},p_1,s) = s^{-\epsilon}\sum_{m,n = 0}^{m+n = 3}\frac{a_jb_is^{3-m-n}}{\Gamma(4-m-n-\epsilon)}\left((s+k_{12})\cos(\frac{\pi}{2}n)+i p_1\sin(\frac{\pi}{2}n)\right)\, ,
\end{equation}
such that only the following coefficients are nonzero: $a_0 = 1\, ,~a_1 = i p_1\,,~b_0 = 1\, ,~b_1 = k_1+k_2\, , ~b_2 = k_1k_2$. Similarly, using eq.\eqref{masslesscontactdata_app}, we get following 
\begin{equation}
\begin{split}
   &\widetilde{ \text{Disc}}_{p_1}\widetilde{\mathcal{B}}^{(1)}(k_1,k_2,p_1) = \lim_{\epsilon\rightarrow 0^+}\frac{\lambda\,H^2}{4k_1^3k_2^3p_1^3}\int_{0}^{\infty}ds\bigg(\frac{-s^{(3-\epsilon)}p_1}{\Gamma(4-\epsilon)}+\frac{s^{(2-\epsilon)}\,sp_1}{\Gamma(3-\epsilon)}\\&
    +\frac{s^{(1-\epsilon)}\,p_1\left(k_1^2+k_2^2+k_1k_2+sk_{12}\right)}{\Gamma(2-\epsilon)}+\frac{s^{-\epsilon}\,k_1k_2p_1(k_{12}+s)}{\Gamma(1-\epsilon)}\bigg)\frac{\sin(\pi\epsilon/2)}{\left((k_{12}+s)^2-p_1^2\right)} \, ,
\end{split}
\end{equation}
The above equation can be written more compactly as follows
\begin{equation}
  \label{masslessdisctbt}\widetilde{\text{Disc}}_{p_1}\widetilde{\mathcal{B}}^{(1)}(k_1,k_2,p_1) =   \lim_{\epsilon\rightarrow 0^+}\frac{i \lambda\,H^2\,\sin(\pi\epsilon/2)}{4k_1^3k_2^3p_1^3}\int_{0}^{\infty}ds\frac{\widetilde{\mathcal{Q}}_3(k_{12},p_1,s)}{(s+k_{12})^2-p_1^2}\,,
\end{equation}
where,
\begin{equation}
\label{defQ3tilde_app}
\widetilde{\mathcal{Q}}_3(k_{12},p_1,s) = s^{-\epsilon}\sum_{m,n = 0}^{m+n = 3}\frac{a_jb_is^{3-m-n}}{\Gamma(4-m-n-\epsilon)}\left(i p_1\cos(\frac{\pi}{2}n)-(s+k_{12})\sin(\frac{\pi}{2}n)\right)\,. 
\end{equation}
This completes the derivation of the vertex factors in \S \ref{rules massless}. 
%%%%%%%%%%%%%%%%%%%%%%%%%%%%%%%%%%%%%%%%%%%%%%%%%%%%%%%%%%%%%%%%%%%%%%%

%%%%%%%%%%%%%%%%%%%%%%%%%%%%%%%%%%%%%%%%%%%%%%%%%%%%%
%%%%%%%%%%%%%%%%%%%%%%%%%%%%%%%%%%%%%%%%%%%%%%%%%

%%%%%%%%%%%%%%%%%%%%%%%%%%%%%%%%%%%%%%%%%%%%%%%%%%%%%%
\bibliographystyle{JHEP}

\bibliography{dressingrules.bib}

%%%%%%%%%%%%%%%%%%%%%%%%%%%%%%%%%%%%%%%%%%%%%%%%%%%

%%%%%%%%%%%%%%%%%%%%%%%%%%%%%%%%%%%%%%%%%%%%%%%%%%%%%%
\end{document}